\documentclass[preprint,showpacs,preprintnumbers,floatfix,superscriptaddress,nofootinbib,showkeys]{revtex4}

\usepackage[reqno]{amsmath}
\usepackage{amsthm}
\usepackage{amssymb}

\usepackage{graphicx}
\usepackage{lscape}
\usepackage{longtable}
\usepackage{afterpage}

\newcommand{\beq}{\begin{equation}}
\newcommand{\eeq}{\end{equation}}
\newcommand{\bea}{\begin{eqnarray}}
\newcommand{\eea}{\end{eqnarray}}
\begin{document}

\preprint{MSUCL-1378, NSF-KITP-08-110}

\title{
Symmetry Energy I: Semi-Infinite Matter
}

\author{Pawe\l~Danielewicz\email{danielewicz@nscl.msu.edu}}

\affiliation{National Superconducting Cyclotron Laboratory and\\
Department of Physics and Astronomy, Michigan State University,
\\
East Lansing, Michigan 48824, USA\\
}

\affiliation{Kavli Institute for Theoretical Physics\\
University of California,
Santa Barbara, CA 93106
}

\author{Jenny Lee\email{lee@nscl.msu.edu}}

\affiliation{National Superconducting Cyclotron Laboratory and\\
Department of Physics and Astronomy, Michigan State University,
\\
East Lansing, Michigan 48824, USA\\
}


\begin{abstract}
Energy for a nucleus is considered in macroscopic limit, in terms of nucleon numbers.  Further considered for a nuclear system is the Hohenberg-Kohn energy functional, in terms of
proton and neutron densities. Finally, Skyrme-Hartree-Fock calculations are carried out
for a half-infinite particle-stable nuclear-matter.  In each case, the attention is focused on the role of neutron-proton asymmetry and on the nuclear symmetry energy.  We extend the considerations on the symmetry term from an energy formula to the respective term in the Hohenberg-Kohn functional.  We show, in particular, that in the limit of an analytic functional, and subject to possible Coulomb corrections, it is possible to construct isoscalar and isovector densities out of the proton and neutron densities, that retain a universal relation to each other, approximately independent of asymmetry.  In the so-called local approximation, the isovector density is inversely proportional to the symmetry energy in uniform matter at the local isoscalar density.  Generalized symmetry coefficient of a nuclear system is related, in the analytic limit of a functional, to an integral of the isovector density.  We test the relations, inferred from the Hohenberg-Kohn functional, in the Skyrme-Hartree-Fock calculations of half-infinite matter.  Within the calculations, we obtain surface symmetry coefficients and parameters characterizing the densities, for the majority of Skyrme parameterizations proposed in the literature.  The volume-to-surface symmetry-coefficient ratio and the displacement of nuclear isovector relative to isoscalar surfaces both strongly increase as the slope of symmetry energy in the vicinity of normal density increases.

\end{abstract}




\maketitle


\section{Introduction}

Nuclear symmetry energy ties different areas of nuclear physics, including structure of ground-state nuclei,
dynamics of central nuclear reactions, physics of giant collective excitations and physics of neutron stars.
Recently, the interest in symmetry energy has been stirred up by novel
astrophysical observations and by
the availability of exotic beams in accelerators, with greater span of asymmetries for given nuclear mass than for stable beams.  Particularly important in the different areas, and simultaneously uncertain, is the density dependence of symmetry
energy in uniform matter.  As situation in the literature progresses, an increasing wider range of theoretical
conclusions is getting made on that density dependence, as well as on some associated nuclear characteristics.
In that situation, not unfamiliar from other contexts, rather than increasing claim statistics, it may be useful to take a step back
and examine in detail the chain of implications following from the symmetry energy, to better understand the connections and to see how firmly any conclusions can be reached.
We hope to make such a progress here within nuclear structure.

Two main directions are advanced in our paper.  One is in extending and combining symmetry-energy considerations from nuclear energy formula and from uniform nuclear matter, within the nuclear energy functional of the Hohenberg-Kohn type~\cite{hohenberg-1964}.
Another direction is in carrying calculations of semi-infinite nuclear-matter within the Skyrme-Hartree-Fock approach~\cite{Reinhard:1991}.  The framework developed in considering the energy functional is employed in analyzing the Hartree-Fock calculations.  In the past, an analysis pertaining to the
 symmetry-energy, in the context of Hohenberg-Kohn functional, has been carried out by Farine \cite{Fari85}.
Semi-infinite asymmetric matter has been examined before in the
Hartree-Fock approach, in particular by Kohler \cite{Kohler:1969,Kohler:1976} and by Pearson {\em et al.}~\cite{Cote:1978,Farine:1980,PhysRevC.24.303}, and in the relativistic Hartree approach by Del Estal {\em et al.} \cite{DelEstal99}.  Also, a number of Thomas-Fermi calculations for the half-infinite asymmetric matter have been done~\cite{Brack85,Mye85,Kol85,Trei86,Cen98,PhysRevC.63.034318,Steiner:2004fi}.
 Since the latest detailed Hartree-Fock studies by Pearson {\em et al}.~\cite{PhysRevC.24.303} have been done over two decades ago, there is potential for computational progress in these calculations, aside from our differing strategy in analyzing the results.

In the following, Sec.~\ref{sec:enfo} reviews the basic nuclear energy formula, emphasizing the symmetry energy. Section \ref{sec:enfu} extends the symmetry-energy considerations to the context of a continuous nuclear Hohenberg-Kohn energy-functional.  The limit of uniform matter is discussed there too.  In Sec.~\ref{sec:nucal}, details of our numerical calculations of half-infinite matter are presented and the results analyzed.  Overall conclusions are presented in Sec.~\ref{sec:conc}.

\section{Energy Formula}
\label{sec:enfo}

\subsection{Elementary Formula}
The first basic context within which the nuclear symmetry energy is encountered is the empirical nuclear energy formula.  In its most
elementary textbook form, the formula contains just four macroscopic terms:
\bea
E(N,Z) & = & E_\text{mac} + E_\text{mic}= E_V + E_S + E_a + E_C + E_\text{mic} \nonumber \\[.5ex]
& =  & -a_V \, A + a_S \, A^{2/3} + a_a \, \frac{(N-Z)^2}{A} + a_C \, \frac{Z^2}{A^{1/3}} + E_\text{mic} \, .
\label{eq:enz}
\eea
Here, $N$ and $Z$ are the neutron and proton numbers, respectively, $A=N+Z$ is the mass number and $a_V$, $a_S$, $a_a$ and
$a_C$ are the volume, surface, symmetry and Coulomb coefficients.  The microscopic energy $E_\text{mic}$ contains shell and pairing energy
corrections.  In spite of its simplicity, the energy formula is extremely powerful both in describing nuclear masses and in giving access
the nature of nuclear interactions.  When the microscopic corrections are disregarded, and the macroscopic coefficients are fitted to the data set
on energies of over 3100 $A \ge 10$ nuclei, to yield $a_V = 15.3$~MeV, $a_S=16.1$~MeV,
$a_a = 22.5$~MeV and $a_C=0.69$~MeV,
the rms deviation of the formula from data turns to be just 3.6~MeV, to be compared to the span of energies for the nuclei of
over 2000~MeV.  The nuclear energies are dominated by the volume term,
responsible for nuclear binding, with binding energies reduced by the surface, symmetry and Coulomb
terms.

Regarding the nature of nuclear interactions, the dominance in $E$ of the negative volume term $E_V$, proportional to $A$,
tells that the attractive nuclear
interactions are short-range.  The invariance of the symmetry term $E_a$ with respect to neutron-proton interchange demonstrates that
the nuclear interactions possess such an interchange symmetry, commonly termed charge symmetry.
The positive value of $a_a$ suggests that neutron-proton interactions are more attractive than like-nucleon interactions;
the Pauli principle effects contribute positively to that coefficient as well.  The surface term $E_S$ scales with $A$ as in proportion to the surface of a nucleus with
fixed volume per nucleon, represented as $4 \pi \, r_0^3/3$, or, equivalently, with fixed density $\rho_0 = 3/(4\pi r_0^3)$.
The reduction in the binding proportional to the surface may be thought
of as one associated with the fact that nucleons at the surface experience less attraction than the nucleons in interior;
the curvature of nuclear wavefunction at the surface also, in fact, contributes to that reduction.  The form of the Coulomb term $E_C$ is such as expected
for a~uniformly charged sphere, specifically $(3/5) \, (Ze)^2/4\pi \epsilon_0 \, R$, where $R=r_0 \, A^{1/3}$.

Quantitatively, the fitted Coulomb parameter provides an estimate of the nuclear radius parameter,
$r_0 \simeq (3/5) \, e^2/(4\pi \epsilon_0 \, a_c) = 1.25$~fm.  With necessarily enhanced error,
due to the raising of $r_0$ to cube power, the parameter further implies an estimate for
the nuclear density~$\rho_0$.  Under changing nuclear density $\rho$, the nuclear energy should minimize at $\rho_0$, reaching
the value of $-a_V$ per nucleon
for $N=Z$ and in absence of Coulomb interactions or boundaries.
Given the surface term, one can estimate the nuclear surface tension, which is the energetic cost of creating the surface per unit area,
or tension,
$\sigma = \partial E/ \partial \Sigma = a_S / 4 \pi r_0^2 \simeq 0.8$~MeV/fm$^2$, again with a deteriorated expected accuracy due
to the power of radius and due to combining of parameters.  In the tension,
$\Sigma$ is nuclear surface area taking on the value of $4\pi  r_0^2 \, A^{2/3}$ for a spherical shape.
It is apparent that the energy formula can provide a wealth of information on nuclei.
Naturally, this can be expected to extend to the changes in nuclear properties with
changes in the relative
neutron-proton asymmetry, $\eta = (N-Z)/A$.

The success of the energy formula may be extended by modifying terms and incorporating new ones, with or without new parameters, including the
microscopic terms.  The terms may be guided by physics or {\em ad hoc}.  The formula success can be perceived in at least two different
ways.  One is in terms of a better predictive power of the formula for net energy.  In fact, by adding different terms and
parameters, the rms deviation from fitted nuclear can be reduced by a factor of order of~5 \cite{Moller:1993ed,koura2000,pomorski-2003,royer:067302,mendoza-2008}.  Another measure of success is in enhanced
access to fundamental nuclear properties.  The advances in the two directions do not necessarily go hand in hand.  Thus, obviously, adding
an {\em ad hoc} term may throw off the value of a parameter that has a physical meaning, while helping though to reduce the rms error~\cite{royer:067302}.  Even
when a term of genuine physical meaning is added, problems can occur when many terms and parameters are present~\cite{Kir08}.  As an example, one can
envision that the range of variation of $N$ and $Z$ for measured nuclei may not be suffice to make two terms in an energy formula distinct, with
different parameter combinations yielding comparable agreement with data and a parameter expected {\em a priori} to be obscure, eating into a major
parameter.  Another example might be a term introduced to improve the description of one $N$-$Z$ region (e.g.\ low $A$)
that gets its parameters determined,
in an unguided fit, by the prevalence of nuclei in another $N$-$Z$ region (such as high $A$).  The problems with multiparametric fits are
known from other areas and the developed remedy is to introduce as many prejudices~\cite{tikhonov-1963} into the fit as possible.  The prejudices may, in fact, include
the qualitative features of fitted functions, following from physical considerations, common-sense limits on region
for parameter determination and the use of constraints on the parameters from auxiliary investigations.  The latter constraints, in the case
of an energy formula, could e.g.\ involve information on the proton and neutron densities, $\rho_p$ and~$\rho_n$~\cite{Danielewicz:2003dd,royer:067302,Diep07}.

\subsection{Surface Symmetry Energy}
\label{ssec:ssenergy}

Examining the symmetry term in the basic formula (\ref{eq:enz}), one can notice that this term has a volume character, i.e.\ it changes in proportion
to $A$, like $E_V$, when $N$ and $Z$ are changed by same factor.  One question which arises, when addressing the symmetry energy,
is whether there is a conceptual need for adding
a {\em surface} symmetry term to the energy formula.  Another question is how such term should enter that formula.  In the literature, there are at least two ways
of introducing such a term,
with a surface symmetry parameter having different meanings in the two formulations.

Regarding the first question, the volume energy becomes less negative, when magnitude of nuclear asymmetry increases from zero, at a fixed $A$.
With this, less work should be required for developing a surface, since there is less lost binding then to compensate for compared to zero asymmetry.
Thus the surface tension should drop with increase in asymmetry magnitude.  The quantity associated
with asymmetry, shared by systems in contact in macroscopic equilibrium, such as the surface region and interior, is the asymmetric chemical potential
\beq
\mu_a = \frac{\partial E}{ \partial (N-Z)} = \frac{1}{2} \big( \mu_n - \mu_p \big) \, .
\eeq
Under charge symmetry of nuclear interactions, with Coulomb ignored for the moment, the nuclear tension should depend only on square of
$\mu_a$ and, following the above, behave for small asymmetries as
\beq
\label{eq:sig=}
\sigma = \frac{\partial E_S}{\partial \Sigma} = \sigma_0 - \nu \, \mu_a^2 \, ,
\eeq
where $\sigma_0 = a_S/4\pi r_0^2$ is the tension for symmetric nuclear matter and $\nu$ is some positive constant.  When the tension
depends on asymmetry, so must the surface energy $E_S$, cf.~\eqref{eq:sig=}.  The~next question is of the form of an energy formula which incorporates that dependence
and, of course, of the consequences of that dependence.

\begin{figure}
\centerline{\includegraphics[width=.4\linewidth]{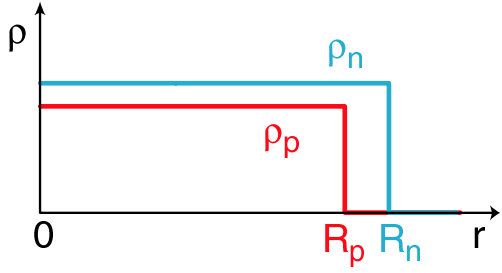}}
\caption{Surface asymmetry can be understood in terms of different radii for proton and neutron distributions.
}
\label{fig:droplet}
\end{figure}

The inverse Legendre transformation for asymmetry, from $\mu_a$ to $N-Z$, is given by
\beq
\label{eq:nz=}
N - Z = \frac{\partial \Phi}{\partial \mu_a } \, ,
\eeq
where
\beq
\label{eq:Fi=}
\Phi = \mu_a (N - Z) - E = \mu_a (N - Z) - E_V - E_S  \, .
\eeq
Upon inserting of (\ref{eq:Fi=}) into (\ref{eq:nz=}), we find
\beq
\label{eq:dNZ}
\frac{\partial (N - Z)}{\partial \mu_a} = 2 \, \frac{\partial E}{\partial \mu_a^2} =
2 \left( \frac{\partial E_V}{\partial \mu_a^2} +  \frac{\partial E_S}{\partial \mu_a^2}      \right)
\, .
\eeq
In the equation above, $N-Z$, $E$, $E_V$ and $E_S$ are all extensive, i.e.\ they grow, in a characteristic fashion, as the system size grows.
The equation implies that, when the surface energy depends on $\mu_a$, the
net asymmetry partitions into volume and surface portions, $N_V - Z_V$ and $N_S - Z_S$, satisfying respectively
\beq
\label{eq:dNZs}
\frac{\partial (N - Z)_{V,S}}{\partial \mu_a} = 2 \, \frac{\partial E_{V,S}}{\partial \mu_a^2} \, .
\eeq
The surface asymmetry may be transcribed onto a difference of radii for neutron and proton distributions, which is the basis of
the droplet model (to be discussed) of nuclei and is schematically indicated in Fig.~\ref{fig:droplet}.

Given that, at low asymmetries, due to charge symmetry, the energies must depend quadratically on asymmetry,
as does the tension in Eq.~(\ref{eq:sig=}),
the second derivatives of energy in Eqs.~(\ref{eq:dNZ}) and~(\ref{eq:dNZs}) can only depend on $A$. In consequence, at low asymmetries, also
the first derivatives of $N-Z$ in those equations must also depend solely on $A$.
Accounting for the dimensions and for scalings with $A$, we can write
\beq
\label{eq:nvs}
N_V - Z_V = \frac{A}{2 \, a_a^V} \, \mu_a \hspace*{2em}\text{and}\hspace*{2em} N_S - Z_S = \frac{A^{2/3}}{2 \, a_a^S}  \, \mu_a \, ,
\eeq
where $a_a^V$ and $a_a^S$ are constants characterizing the nuclear volume and surface, respectively, with the dimension of energy.
Sum of the two asymmetries gives us the relation between $\mu_a$ and net asymmetry:
\beq
\label{eq:nzv}
N-Z = N_V - Z_V + N_S - Z_S = \frac{\mu_a}{2} \left( \frac{A}{a_a^V} +  \frac{A^{2/3}}{a_a^S}   \right) \, .
\eeq
From Eqs.\ (\ref{eq:dNZs}) and (\ref{eq:nvs}), we further find
\beq
\label{eq:ev=}
E_V = E_V^0 + \frac{A}{4 \, a_a^V} \, \mu_a^2 = -a_V \, A +  a_a^V \, \frac{(N_V - Z_V)^2}{A} \, ,
\eeq
and
\beq
\label{eq:es=}
E_S = E_S^0 + \frac{A^{2/3}}{4 \, a_a^S} \, \mu_a^2 = a_S \, A^{2/3} +  a_a^S \, \frac{(N_S - Z_S)^2}{A^{2/3}} \, ,
\eeq
where $E_{V,S}^0$ are the volume and surface energies for symmetric $N=Z$ matter.  It should be mentioned that there is no
error in the sign difference in symmetry terms between Eqs.\ (\ref{eq:es=}) and (\ref{eq:sig=}).  Whereas the volume and surface energies increase with asymmetry, the surface
tension decreases.  The derivative of surface energy with respect to surface area, for the surface tension $\sigma$,
must be taken at constant $N_S - Z_S$, which yields $\nu^{-1}= 16 \, \pi \, r_0^2 \, a_a^S$.

On adding up the energy contributions (\ref{eq:ev=}) and (\ref{eq:es=}), we get for the net energy
\beq
\label{eq:e=}
E = E^0 + \frac{\mu_a^2}{4} \left( \frac{A}{a_a^V} + \frac{A^{2/3}}{a_a^S}   \right)
= -a_V \, A + a_S \, A^{2/3} + \frac{(N-Z)^2}{\frac{A}{a_a^V} + \frac{A^{2/3}}{a_a^S}} \, ,
\eeq
where we have eliminated $\mu_a$ in favor of $N-Z$ using (\ref{eq:nz=}).
The nuclear energies above, quadratic in asymmetry, exhibit
analogy to the capacitor energy in terms of electric charge.  While asymmetry is analogous to the electric charge within this analogy,
the chemical potential is analogous to the electric potential.  The coefficients of proportionality between volume and surface
asymmetries and chemical potential in (\ref{eq:nvs}) are analogs of capacitance, with the two capacitances proportional, respectively, to the volume
and surface.  For connected capacitors, the potentials are equal.  The net symmetry partitions itself in (\ref{eq:nzv}), between
the volume and surface, in proportion to the capacitances.  Finally, net energy for connected capacitors can be represented in (\ref{eq:e=})
either in terms of net potential squared multiplied by net capacitance or in terms of net asymmetry squared
divided by the net capacitance \footnote{Beyond the quadratic approximation for the energy in asymmetry, the capacitance may be defined \cite{kubis:025801} in terms of the derivative of asymmetry with respect to the chemical potential, i.e.\ the l.h.s.\ of \eqref{eq:dNZ}.}.

Upon adding the Coulomb and microscopic contributions, we now arrive at the energy formula with a mass-dependent symmetry
coefficient
\beq
E(N,Z)
 =   -a_V \, A + a_S \, A^{2/3} + \frac{a_a(A)}{A} \, (N-Z)^2 + a_C \, \frac{Z^2}{A^{1/3}} + E_\text{mic} \, ,
\label{eq:enza}
\eeq
where
\beq
\label{eq:aa}
a_a(A) = \frac{a_a^V}{1 + {a_a^V}/({a_a^S} \, A^{1/3})} \, .
\eeq
In the limit of large $A$, the asymmetry gets primarily stored within the volume and then the coefficient $a_a(A)$ tends towards $a_a^V$.
On the other hand, in the limit of small $A$, the storage of asymmetry gets shifted to the surface and then the ratio $a_a(A)/A$
approaches $a_a^S/A^{2/3}$.

Although the surface symmetry effects can be important in learning on symmetry energy,
as far as the net nuclear energy is concerned, those effects just correct the symmetry energy which itself is a correction to the leading term  $-a_V \, A$ term in an nuclear energy.
Additionally, since most of the measured asymmetric nuclei are heavy, those nuclei can provide little variation
for $A^{1/3}$ in the symmetry-energy term.
In consequence, the basic energy formula~\eqref{eq:enz} can provide a fit to the measured nuclear energies that is quite satisfactory without
any surface symmetry
modification.

The coefficient $a_a^S$ of surface symmetry energy is generally of interest as a fundamental quantity and because,
in the context of the changing density at nuclear surface, it can provide access to the density dependence
of symmetry energy in bulk nuclear matter.  However, obviously, since $a_a^S$ appears within a
correction to a correction
in the energy formula, it can be difficult to learn about that coefficient
by fitting nuclear energy data with a formula
where further types of secondary corrections may need to be included.  The latter corrections may compete against each other and against the surface symmetry energy \cite{Diep07,Kir08}.
This can be especially true for blind fits.
The potential secondary corrections include those to the asymmetry-{\em dependent} Coulomb term and those to the asymmetry-{\em independent} part of the formula.
The Coulomb energy e.g.\ also affects the surface asymmetry.  Corrections to the asymmetry-independent part of the formula are important, possibly against expectations,
because the grouping of nuclei around the line of stability, in the $(N,Z)$-plane, introduces mass-asymmetry correlations for the fitted data.

In this and in a subsequent paper, we shall try to understand the interplay between the dependence of bulk symmetry energy on density,
density distributions in nuclei and
the symmetry coefficient $a_a(A)$.  On one hand, we shall try understand a connection between~$a_a^S$ and the density dependence of symmetry
energy.  On the other hand, we shall try to determine $a_a(A)$ from nuclear data,
refraining from a blind fit and without insisting on validity of Eq.~(\ref{eq:aa}).
Finally, we shall seek additional information on symmetry energy within density distributions.

\subsection{Droplet Model}
\label{ssec:droplet}

A result of the form (\ref{eq:aa}), for the symmetry coefficient, has first appeared in the literature in the droplet model \cite{Myers:1969} by Myers and
Swiatecki.
The droplet model relies on the assumption that the neutron and proton surfaces shift relative to each other as nuclear asymmetry changes.
The surface energy per the elementary area of $4 \pi \, r_0^2$ depends in the model on that shift, $D_{np}$,
and on the relative asymmetry in nuclear interior, $\eta_V$, according to:
\beq
\label{eq:dromoES}
\frac{4\pi \, r_0^2 \, E_S}{\Sigma} = a_S + H \, \left( \frac{D_{np}}{r_0} \right)^2 + 2 P \, \eta_V \,  \frac{D_{np}}{r_0}  - G \, \eta_V^2 \, .
\eeq
In the above, $H$, $P$ and $G$ are model constants and we have dropped a higher-order curvature term employed in \cite{Myers:1969}.  (Strictly, the surface energy
density ${E_S}/{\Sigma}$ is identified in \cite{Myers:1969} as the surface tension $\sigma$, but these two are different when they depend on asymmetry, cf.\ Subsection~\ref{ssec:ssenergy}; Eq.~(\ref{eq:dromoES}) is consistent with the {\em use} of the formula by the authors of \cite{Myers:1969}.)
When a~higher-order Coulomb correction included
in \cite{Myers:1969} is disregarded, the symmetry coefficient emerges such as given by Eq.~(\ref{eq:aa}), with the surface symmetry coefficient given
by
\beq
\label{eq:aaSQ}
a_a^S = \frac{4}{9} \, Q = \frac{8}{27} \, \frac{H^2 \, G}{P \, a_a^V} \, ,
\eeq
where $Q$ is an auxiliary constant.

Following \cite{Danielewicz:2003dd} and Subsection \ref{ssec:ssenergy}, Eq.~(\ref{eq:aa}) {\em must} emerge in the macroscopic limit at large~$A$, for an energy
analytic in asymmetry, provided underlying model
assumptions represent short-range interactions obeying charge symmetry.  Later in the paper, we shall encounter a~greater richness of changes with asymmetry in the nuclear
surface region for different Skyrme interactions, in the Hartree-Fock calculations, than envisioned in the droplet model.  However, all those calculations will be consistent with the symmetry energy being
quadratic in asymmetry in the macroscopic limit, with the symmetry coefficient conforming with (\ref{eq:aa}) at large~$A$.

The droplet-model energy formula eventually acquired \cite{Moller:1993ed} as many as 38 independent parameters that either got set by fitting nuclear energies or
using auxiliary information.  The~general difficulty in settling on the value of $a_a^S$ in an energy formula is
illustrated by the fact that this coefficient has risen by over a factor of 2
within the history of the droplet model, from $a_a^S=7.1$~MeV~\cite{Myers:1969} to 15.7~MeV~\cite{Myers:2000ry}.  At the same time, the volume symmetry constant
of the model has varied within the range of only about 25\%, from $a_a^V=28.1$~MeV~\cite{Myers:1969} to $a_a^V=36.5$~MeV \cite{Myers:1974}.
A~still wider range of surface symmetry coefficients is found when considering other formulas in the literature.
Thus, e.g.\ the fits by Pomorski and Dudek \cite{pomorski-2003}, who have emphasized asymmetry effects, give rise to the coefficient values
of $a_a^S = (\kappa_\text{vol} \, b_\text{vol})^2/
(\kappa_\text{surf} \,  b_\text{surf}) = 21.3$~MeV and~50.1~MeV, in terms of these authors' notation, depending on the variant of the drop formula they
employ in describing nuclear ground-state energies and fission barriers.  Depending on the formula variant, the fits, at the same time,
produce the values for the volume symmetry coefficient of either  $a_a^V = - \kappa_\text{vol} \, b_\text{vol} = 28.8$~MeV or 25.4~MeV.
Notably, the droplet formula~\cite{Moller:1993ed} and
the Lublin-Strasbourg formula~\cite{pomorski-2003} produce comparable rms deviations from the measured energies and barriers.
Similarly wide variations of the surface symmetry coefficient, of nearly an order of magnitude, will be found in Sec.~\ref{sec:nucal} for the effective Skyrme interactions
employed in the literature.  The volume symmetry coefficients will, again, exhibit relatively less variation.  The values of $a_a^S$ will be found to be
correlated with the density dependence of symmetry energy.

\section{Energy Functional}
\label{sec:enfu}

\subsection{Hohenberg-Kohn Functional}

Here, we shall consider the nuclear energy functional in terms of neutron $\rho_n( {\pmb r})$ and proton~$\rho_p( {\pmb r})$ densities.
Given a system hamiltonian in terms of nuclear and Coulomb components,
\beq
\hat{H} = \hat{H}_\text{nucl} + \hat{H}_C \, ,
\eeq
different densities $\rho_n( {\pmb r})$ and proton~$\rho_p( {\pmb r})$ and different values of the functional $E(\rho_n, \rho_p) $ will result, when minimizing the expectation value of the hamiltonian in the presence of
different external potentials \cite{hohenberg-1964},
$V_n({\pmb r})$ and $V_p({\pmb r})$:
\beq
\langle \hat{H} + \hat{V}_n + \hat{V}_p \rangle
= E(\rho_n, \rho_p) + \int \text{d} {\pmb r} \left( V_n({\pmb r}) \, \rho_n( {\pmb r}) +  V_p({\pmb r}) \, \rho_p( {\pmb r})  \right) \, .
\eeq
The functional $E$, defined in this way, is specified for integer nucleon numbers, $N$ and $Z$, only.  In finding the ground-state densities and energy, the functional
$E$ is minimized with respect to variations in $\rho_n$ and $\rho_p$.
In the context of an energy formula, we will actually be interested in the functional $E$ averaged out and smoothed
out across discrete $N$ and $Z$, defined then also for fractional nucleon numbers and made analytic in the nucleon densities.

The smoothed-out functional $\overline{E}$ will be partitioned into the parts associated with nuclear and with Coulomb interactions.  The nuclear part will
be further partitioned into the energy of symmetric matter and a correction associated with asymmetry:
\beq
\overline{E}(\rho_n, \rho_p) =  E_\text{nucl} (\rho, \rho_{np}) + E_C (\rho, \rho_{np})
= E_0 (\rho) + E_a ( \rho, \rho_{np} ) + E_C (\rho, \rho_{np}) \, ,
\eeq
where $\rho=\rho_n+\rho_p$, $\rho_{np}=\rho_n-\rho_p$ and $E_0 (\rho) \equiv E_\text{nucl} (\rho, 0)$.

At $\rho_{np}=0$, the symmetry energy $E_a$ vanishes by definition.
Given that energies of nuclei, corrected for microscopic and Coulomb contributions, minimize at a lowest asymmetry, the energy $E_a$ must be quadratic
in asymmetry at low asymmetries:
\beq
\label{eq:EaS}
E_a (\rho , \rho_{np}) = \int \text{d} {\pmb r}_1 \, \text{d} {\pmb r}_2 \, \rho_{np} ({\pmb r}_1) \, {\mathcal S}( \rho, {\pmb r}_1, {\pmb r}_2) \,
\rho_{np} ({\pmb r}_2) + {\mathcal O}(\eta^4) \, .
\eeq
Because of the minimum, ${\mathcal S}$ must be a positive-definite symmetric bilinear integral operator, with regard to its arguments ${\pmb r}_1$ and ${\pmb r}_2$.  This operator
represents an analog of the potential matrix in electrostatics, dependent on conductor described in terms of~$\rho$.    For~the purposes of assessing later the accuracy of our considerations, we have indicated the principal presence of the subsequent terms of fourth order in asymmetry and higher.  At~other times, we may omit underscoring the presence or order of any subsequent terms in an expansion, beyond the directly considered order.

Minimal energy $\overline{E}$ at given $N$ and $Z$ will be achieved for densities $\rho$ and $\rho_{np}$ which satisfy the equations
\beq
\label{eq:dErho}
\frac{\delta \overline{E}}{\delta \rho({\pmb r})} = \mu
\eeq
and
\beq
\label{eq:dErhonp}
\frac{\delta \overline{E}}{\delta \rho_{np}({\pmb r})} = \mu_a
\, ,
\eeq
where $\mu$ and $\mu_a$ are Lagrange multipliers associated with the constraints
\beq
\label{eq:=A}
\int \text{d} {\pmb r} \, \rho({\pmb r}) = A \, ,
\eeq
and
\beq
\int \text{d} {\pmb r} \, \rho_{np}({\pmb r}) = N - Z \, ,
\eeq
respectively.
The Lagrange multipliers coincide with the corresponding chemical potentials, which are derivatives of energy $\overline{E}$ that gets minimized,
$\mu = \partial \overline{E}/\partial A$ and $\mu_a = \partial \overline{E}/\partial (N-Z)$,
hence the notation for multipliers.  Thus,
e.g.\ upon differentiating the energy with respect to mass number,
\beq
\frac{\partial \overline{E}}{\partial A} = \int \text{d} {\pmb r} \,
\frac{\delta \overline{E}}{\delta \rho({\pmb r})} \, \frac{\partial \rho({\pmb r})}{\partial A} =
\mu \int \text{d} {\pmb r} \, \frac{\partial \rho({\pmb r})}{\partial A} = \mu \, ,
\eeq
we find that the l.h.s.\ coincides with $\mu$ introduced as a Lagrange multiplier in~\eqref{eq:dErho}.
Since the nucleonic chemical potentials are equal to
\beq
\mu_{n,p} = \mu \pm \mu_a \, ,
\eeq
the system can spontaneously emit nucleons, either neutrons or protons, if
\beq
|\mu_a| > - \mu \, .
\eeq

If we consider the symmetric term $E_0$ alone, we can expand it around the density $\rho_0({\pmb r})$ that minimizes $E_0$ amongst the densities that
meet the condition (\ref{eq:=A}) at a prescribed $A$:
\beq
\begin{split}
\label{eq:E0exp}
E_0(\rho) & =  E_0(\rho_0) + \mu_0 \int \text{d} {\pmb r} \, \Delta \rho ({\pmb r}) + \int \text{d} {\pmb r}_1 \, \text{d} {\pmb r}_2 \,
\Delta \rho ({\pmb r}_1) \, {\mathcal K} ({\pmb r}_1 , {\pmb r}_2) \, \Delta \rho ({\pmb r}_2) + {\mathcal O}\left((\Delta \rho)^3\right)  \\
& = E_0(A) + \int \text{d} {\pmb r}_1 \, \text{d} {\pmb r}_2  \,
\Delta \rho ({\pmb r}_1) \, {\mathcal K} ({\pmb r}_1 , {\pmb r}_2) \, \Delta \rho ({\pmb r}_2) + {\mathcal O}\left((\Delta \rho)^3\right)   \, ,
\end{split}
\eeq
where $\Delta \rho({\pmb r}) = \rho({\pmb r}) - \rho_0({\pmb r})$, $\mu_0 = \text{d} E_0/\text{d}A$,
${\mathcal K}$ is a positive-definite operator and $\int \text{d} {\pmb r} \, \Delta \rho = 0$ holds amongst the densities that meet the condition (\ref{eq:=A}).

Note that, depending on context, we may use the symbol $\rho_0$ to denote either the function of position that minimizes the energy of a finite or semi-finite symmetric system, in the absence of Coulomb interactions, or to denote single density value that minimizes the energy of symmetric uniform matter.  Use of a position argument for the quantity will obviously imply the function.  The distinction may be, otherwise, made by directly naming the quantity.

\subsection{Densities and Generalized Symmetry Coefficient}
In the following, we shall suppress the Coulomb term $E_C$ and we shall explore what consequences the discussed features of the functional $\overline{E}$ may have on the asymmetry dependence
of nuclear densities and of nuclear energy.

Thus, at given $N$ and $Z$, for low asymmetry $|N-Z|$, the minimal energy may be generally represented
as
\beq
\label{eq:enz1}
\overline{E}(N,Z) = E_0(A) + \frac{a_a(A)}{A} \, (N-Z)^2  + {\mathcal O}(\eta^4)  \, ,
\eeq
where $a_a(A)$ is a general
$A$-dependent symmetry coefficient.  From (\ref{eq:enz1}), the asymmetry chemical potential, in terms of the coefficient, is
\beq
\label{eq:mua=}
\mu_a = \frac{2 a_a(A)}{A} \, (N-Z)  + {\mathcal O}(\eta^3)  \, ,
\eeq
while the symmetric chemical potential from (\ref{eq:enz1}) is
\beq
\label{eq:mu=}
\mu =  \mu_0 + (N-Z)^2 \, \frac{d}{d A} \, \frac{a_a(A)}{A}  + {\mathcal O}(\eta^4) \, .
\eeq

When the Coulomb term is suppressed, Eqs.~(\ref{eq:dErho}) and \eqref{eq:dErhonp} for the densities produce
\beq
\label{eq:dErhoa}
\mu_a = 2 \int \text{d} {\pmb r}_1 \, {\mathcal S}( \rho, {\pmb r}, {\pmb r}_1) \,
\rho_{np} ({\pmb r}_1) + {\mathcal O}(\eta^3)  \, ,
\eeq
and
\beq
\label{eq:dErho1}
\mu   =
\frac{\delta E_0}{\delta \rho({\pmb r})} + \int \text{d} {\pmb r}_1 \, \text{d} {\pmb r}_2 \, \rho_{np} ({\pmb r}_1) \,
\frac{\delta {\mathcal S}(\rho, {\pmb r}_1, {\pmb r}_2)}{\delta \rho({\pmb r})} \,
\rho_{np} ({\pmb r}_2) + {\mathcal O}(\eta^4)
 \, .
\eeq
An operator ${\mathcal S}^{-1}$ inverse to ${\mathcal S}$,
\beq
\label{eq:SS-1}
\int \text{d} {\pmb r}_2 \, {\mathcal S}(\rho, {\pmb r}_1, {\pmb r}_2) \,
{\mathcal S}^{-1}(\rho, {\pmb r}_2, {\pmb r}_3) =
\int \text{d} {\pmb r}_2 \, {\mathcal S}^{-1}(\rho, {\pmb r}_1, {\pmb r}_2) \,
{\mathcal S}(\rho, {\pmb r}_2, {\pmb r}_3) = \delta ({\pmb r}_1 - {\pmb r}_3 ) \, ,
\eeq
should exist as ${\mathcal S}$ must have positive eigenvalues only.  Upon applying the inverse operator to both sides of \eqref{eq:dErhoa}, we can formally solve that equation to get
\beq
\label{eq:rhonpS}
\rho_{np} ({\pmb r}) = \frac{\mu_a}{2} \int \text{d} {\pmb r}_1 \, {\mathcal S}^{-1}( \rho, {\pmb r}, {\pmb r}_1) + {\mathcal O}(\eta^3) \, .
\eeq
Upon integrating both sides of \eqref{eq:rhonpS} over space, we find for the asymmetry chemical potential
\beq
\label{eq:muaS}
\mu_a = \frac{2(N-Z)}{\int \text{d} {\pmb r}_1 \, \text{d} {\pmb r}_2 \, {\mathcal S}^{-1}( \rho, {\pmb r}_1, {\pmb r}_2)} + {\mathcal O}(\eta^3) \, .
\eeq
As a consequence, the nucleon density difference from \eqref{eq:rhonpS} can be represented as
\beq
\label{eq:rhonpNZ}
\rho_{np} ({\pmb r}) = (N-Z) \, \frac{ \int \text{d} {\pmb r}_1 \, {\mathcal S}^{-1}( \rho, {\pmb r}, {\pmb r}_1)}{\int \text{d} {\pmb r}_1 \, \text{d} {\pmb r}_2 \, {\mathcal S}^{-1}( \rho, {\pmb r}_1, {\pmb r}_2)} + {\mathcal O}(\eta^3) \, .
\eeq

Upon substituting \eqref{eq:rhonpNZ} into the symmetry energy \eqref{eq:EaS}, we get
\beq
\label{eq:EaS-1}
E_a = \frac{(N-Z)^2}{\int \text{d} {\pmb r} \, \text{d} {\pmb r}_1 \, {\mathcal S}^{-1}( \rho, {\pmb r}, {\pmb r}_1)} + {\mathcal O}(\eta^4) \, .
\eeq
Substitutions into the equation \eqref{eq:dErho1} for the net density, next yield
\beq
\label{eq:mufrho}
\mu_0 = \frac{\delta E_0}{\delta \rho({\pmb r})} + (N-Z)^2 \, \frac{\delta}{\delta \rho({\pmb r})}
\left( \frac{1}{\int \text{d} {\pmb r}_1 \, \text{d} {\pmb r}_2 \, {\mathcal S}^{-1}( \rho, {\pmb r}_1, {\pmb r}_2)} -  \frac{a_a(A)}{A} \right)  + {\mathcal O}(\eta^4) \, ,
\eeq
where we have used $\delta A/\delta \, \rho({\pmb r}) = 1$ and the identity
\beq
\int \text{d} {\pmb r}_2 \, \left( {\mathcal S}(\rho, {\pmb r}_1, {\pmb r}_2) \,
\frac{\delta {\mathcal S}^{-1}(\rho, {\pmb r}_2, {\pmb r}_3)}{\delta \rho({\pmb r})}
+  \frac{\delta {\mathcal S}(\rho, {\pmb r}_1, {\pmb r}_2)}{\delta \rho({\pmb r})} \,
{\mathcal S}^{-1}(\rho, {\pmb r}_2, {\pmb r}_3) \right) = 0 \, ,
\eeq
following from differentiating Eq.~\eqref{eq:SS-1} side-by-side .
With $\rho_0({\pmb r})$ representing the solution of
\beq
\mu_0 = \frac{\delta E_0}{\delta \rho({\pmb r})} \, ,
\eeq
we find from \eqref{eq:mufrho} that the net density for a finite asymmetry differs from that for symmetric matter only up to second order in asymmetry,
\beq
\label{eq:rhoeta2}
\rho({\pmb r}) = \rho_0 ({\pmb r}) + {\mathcal O}(\eta^2) \, ,
\eeq
with the second order brought in by the second term on the r.h.s.\ of \eqref{eq:mufrho}.  Upon comparing the expanded energy in \eqref{eq:enz1} to the combination of energies $E_0$ from \eqref{eq:E0exp} and $E_a$ from~\eqref{eq:EaS-1}, with the density $\rho_0({\pmb r})$ from \eqref{eq:rhoeta2}, we arrive at a result for the generalized symmetry coefficient in terms of the operator~${\mathcal S}$,
\beq
\label{eq:aaS}
\frac{A}{a_a(A)} = \int \text{d} {\pmb r} \, \text{d} {\pmb r}_1 \, {\mathcal S}^{-1}( \rho_0, {\pmb r}, {\pmb r}_1 ) \, .
\eeq

Given the weak dependence on asymmetry for the net density, stated in Eq.~\eqref{eq:rhoeta2}, we conclude, from \eqref{eq:rhonpS} or \eqref{eq:rhonpNZ}, a similarly weak dependence on asymmetry for the density difference $\rho_{np}$ scaled by $\mu_a$ or $(N-Z)$ (or another quantity odd in asymmetry).  In that context, we introduce an isovector density changing weakly with asymmetry, that will be normalized either locally or globally.
Normalized locally, in terms of the intense chemical potential, the isovector density is defined as
\beq
\label{eq:rhoa=}
\rho_a ({\pmb r}) = \frac{2 a_a^V }{\mu_a} \, \rho_{np}({\pmb r}) =   \frac{a_a^V }{ a_a(A) }  \, \frac{ \rho_{np}({\pmb r})}{ \eta} + {\mathcal O}(\eta^2)
 \, ,
\eeq
where the r.h.s.\ follows from \eqref{eq:mua=}.
For an extended system at low asymmetry, the dimensionless ratio normalizing $\rho_{np}$ on the
r.h.s.\ of \eqref{eq:rhoa=} can be identified with volume asymmetry~$\eta_V$, cf.\ Eq.\ \eqref{eq:nvs}.  The normalized chemical potential $\mu_a/2 a_a^V$ differs from that asymmetry only by higher order terms and may be termed effective volume asymmetry:
\beq
\eta_V' \equiv \frac{\mu_a}{2 a_a^V} =  \frac{a_a^V }{ a_a(A) }  \, \eta + {\mathcal O}(\eta^3)
=  \eta_V  + {\mathcal O}(\eta^3)
\, .
\eeq
In the limit of symmetric matter, the isovector density becomes
\beq
\label{eq:rhoadef}
\rho_a^0 ({\pmb r}) = 2 a_a^V  \left. \frac{\partial \rho_{np}({\pmb r})}{\partial \mu_a} \right|_{\mu_a =0} = \left. \frac{\partial \rho_{np}({\pmb r})}{\partial \eta_V} \right|_{\eta=0} \, .
\eeq
Normalized globally, to integrate to a mass number $A$, the isovector density $\rho_A$ is defined as
\beq
\label{eq:rhoA}
\rho_A({\pmb r}) =  \frac{A}{N-Z}  \, \rho_{np}({\pmb r}) \equiv \frac{\eta_V'}{\eta} \, \rho_a({\pmb r}) = \frac{a_a(A)}{a_a^V} \, \rho_a({\pmb r})  + {\mathcal O}(\eta^2)  \, .
\eeq
In the limit of symmetric matter, the density normalized that way becomes
\beq
\rho_A^0({\pmb r}) = \frac{\partial \rho_{np}({\pmb r})}{\partial \eta} \, .
\eeq
From \eqref{eq:rhonpS}, it follows that the isovector density $\rho_a$ may be expressed in terms of the operator~${\mathcal S}$ as
\beq
\label{eq:rhoaexp}
\rho_a({\pmb r}) = a_a^V \int \text{d} {\pmb r}_1 \, {\mathcal S}^{-1}( \rho, {\pmb r}, {\pmb r}_1) + {\mathcal O}(\eta^2) = \rho_a^0({\pmb r}) + {\mathcal O}(\eta^2) \, ,
\eeq
with the $\eta=0$ limit of
\beq
\rho_a^0({\pmb r}) = a_a^V \int \text{d} {\pmb r}_1 \, {\mathcal S}^{-1}( \rho^0, {\pmb r}, {\pmb r}_1)
\, .
\eeq
From \eqref{eq:aaS}, it follows that the symmetry coefficient is related to the locally normalized density with
\beq
\label{eq:aaaa}
\frac{A}{a_a(A)} = \frac{1}{a_a^V}
\int \text{d} {\pmb r} \rho_{a}^0 ({\pmb r}) \, .
\eeq

The result \eqref{eq:rhoaexp} for the isovector density is a counterpart of the result \eqref{eq:rhoeta2} for the isoscalar net density.  The obvious important implication of those results is that the proton and neutron densities should be describable, up to second-order in asymmetry and up to Coulomb corrections discussed in the next paper, in terms of two densities universal across an isobar chain,
\beq
\label{eq:rhonp}
\rho_{n,p} ({\pmb r}) = \frac{1}{2} \left( \rho ({\pmb r}) \pm \eta \, \rho_A ({\pmb r}) \right)
= \frac{1}{2} \left( \rho ({\pmb r}) \pm \eta_V' \, \rho_a ({\pmb r}) \right) =
\frac{1}{2} \left( \rho_0 ({\pmb r}) \pm \eta_V \, \rho_a^0 ({\pmb r}) \right) + {\mathcal O}(\eta^2) \, .
\eeq
As is apparent, the isovector density is tied to the symmetry coefficient.  We shall investigate quantitatively the relation between the densities $\rho$ and $\rho_a$ for different interactions in half-infinite matter in the next section.  It should be mentioned that, while $\rho_a$ represents an isovector form factor, it is, at this point, an isoscalar with regard to its transformation properties in isobar space.  Using analogy with spin, that density might further be termed an isospin susceptibility.

One issue that Eq.~\eqref{eq:rhonp} raises is that of the importance of higher-order terms in asymmetry expansion.  That importance needs to be considered not just globally but locally as well.  In~the next section, we shall see that $\rho_a$ is typically larger than $\rho$ outside of the nuclear surface.  Since nucleon densities must be positive, the higher-order terms must become significant, at a given~$\eta_V$, at distances outside of the surface where
\beq
\label{eq:break}
|\eta_V | \, \rho_a^0 ({\pmb r}) > \rho_0 ({\pmb r}) \, .
\eeq

\subsection{Uniform Matter}

In uniform matter, the energy is an integral of constant energy per unit volume $e$,
\beq
E = \int \text{d} {\pmb r} \, e(\rho_n, \rho_p) = V e  \, , \hspace*{1.5em} \text{with}  \hspace*{1.em}
e = \frac{E}{V} = \rho \, \frac{E}{A} \, ,
\eeq
and
\beq
\label{eq:EA}
\frac{E}{A} = \frac{E_0}{A} (\rho) + \frac{E_a}{A} (\rho , \eta) \simeq \frac{E_0}{A} (\rho) + S(\rho) \, \eta^2 \, ,
\eeq
where $S$ is a density-dependent symmetry coefficient.  That last coefficient is on its own commonly termed the symmetry energy.
An obvious test of the usefulness of the quadratic expansion of symmetry energy is whether the approximate equality,
\beq
\frac{E_a}{A}(\rho,1) \stackrel{?}{\simeq} S(\rho) \, ,
\eeq
holds at different densities, particularly at $\rho=\rho_0$.

To the extend that the approximation of the r.h.s.\ of (\ref{eq:EA}) holds \cite{PhysRevC.60.024605}, the functions $\frac{E_0}{A}(\rho)$ and
$S(\rho)$ allow for the determination, at any asymmetry in uniform matter, of nuclear energy and nuclear pressure,
\beq
P = \rho^2 \, \frac{\text{d}}{\text{d} \rho} \left( \frac{E}{A} \right) \simeq
\rho^2 \, \frac{\text{d}}{\text{d} \rho} \left( \frac{E_0}{A} \right) + \eta^2 \, \rho^2 \, \frac{\text{d} S}{\text{d} \rho} \, ,
\eeq
and nucleonic chemical potentials,
\beq
\label{eq:munp}
\mu_{n,p} = \rho \, \frac{\partial }{\partial \rho_{n,p}} \left( \frac{E}{A} \right) + \frac{E}{A}
=  \mu \pm \mu_a
 \, ,
\eeq
where
\beq
\label{eq:mu}
\mu = \rho \, \frac{\partial }{\partial \rho} \left( \frac{E}{A} \right) + \frac{E}{A} =
\frac{P}{\rho} + \frac{E}{A}
 \, ,
\eeq
and
\beq
\label{eq:mua}
\mu_a = \frac{\partial }{\partial \eta } \left( \frac{E}{A} \right) \simeq
2 \eta \, S(\rho) \, .
\eeq
The pressure and chemical potentials are important in calculating neutron-star properties and in simulating supernova
explosions  \cite{Lattimer:2006xb}.

Around $\rho_0$, it is common to expand both $E_0/A$,
\beq
\label{eq:E0Aexp}
\frac{E_0}{A}(\rho) = -a_V + \frac{K}{18 \rho_0^2} \, \left( \rho - \rho_0 \right)^2 + \ldots \, ,
\eeq
and $S$,
\beq
\label{eq:Sexp}
S(\rho) = a_a^V + \frac{L}{3 \rho_0} \, \left( \rho - \rho_0 \right) + \frac{K_\text{sym}}{18 \rho_0^2} \, \left( \rho - \rho_0 \right)^2 + \ldots \, .
\eeq
Notation for the constant terms in the expansion is consistent with that in the energy formula, Sec.~\ref{sec:enfo}.  Upon employing
the expansions (\ref{eq:E0Aexp}) and (\ref{eq:Sexp}), the energy per nucleon (\ref{eq:EA}) may be rewritten, around its minimum, in the form
\beq
\frac{E}{A} = -a_V + a_a^V \, \eta^2 + \frac{K_\text{eff}}{18 \rho_\text{min}^2} \, \left( \rho - \rho_\text{min} \right)^2 + \ldots \, ,
\eeq
where we retain an accuracy of the order of $\mathcal{O}(\eta^4)$ in the coefficients.  Within that accuracy the density which minimizes the energy is
\beq
\label{eq:rhomin}
\rho_\text{min} (\eta) \simeq \rho_0 - \frac{3 L}{ K} \, \rho_0 \, \eta^2 \, ,
\eeq
and the effective incompressibility is
\beq
K_\text{eff} (\eta) = K + \left(K_\text{sym} - 6 L \right) \, \eta^2 \, .
\eeq

In self-bound matter, on one hand, the pressure vanishes, $P=0$, and, on the other, the nucleonic potentials are negative.  When assuming a density close to normal, from \eqref{eq:munp}-\eqref{eq:rhomin}, one finds that the magnitude of bulk drip asymmetry~$\eta_V^d$, above which nuclear matter emits nucleons, is approximately
\beq
\label{eq:etaVd}
\eta_V^d \simeq \frac{a_V}{2 a_a^V} \, .
\eeq
In gravitationally bound neutron-star matter, due to the fact that the energy of symmetric matter minimizes at $\rho_0$, the constant $L$ from the symmetry-energy expansion \eqref{eq:Sexp},
largely determines the nuclear contribution to pressure
in neutron-star matter around $\rho_0$,
\beq
P \simeq
\eta^2 \, \rho^2 \, \frac{\text{d} S}{\text{d} \rho} \simeq \frac{L}{3} \, \rho_0 \, \eta^2 \, .
\eeq

Table \ref{tab:skypro} shows a selection of bulk nuclear parameters for the majority of Skyrme interactions employed in the literature
in nuclear mean-field calculations.  The force constants for most of those interactions
have been compiled
by Jirina Stone~\cite{PhysRevC.68.034324,Jirina:priv}.  Basic information on the determination of $E_0/A$ and $S$ of uniform matter for those interactions will be provided later in this section; more thorough explorations of the properties of uniform matter may be found elsewhere, e.g.~in~\cite{PhysRevC.68.034324}.  On the other hand, procedures for the determination of surface parameters and values of those parameters for the interactions will be discussed in the following section.

Variations in the bulk parameter values in Table \ref{tab:skypro} can serve as illustration of the status of the specific parameters.  Thus,
there is a general consensus in the literature regarding $a_V$ and~$\rho_0$, and the respective values in the table,
across the listed interactions,
are $a_V = 15.9 \pm 0.3\,\text{MeV}$ and $\rho_0 = 0.158 \pm 0.005\, \text{fm}^\text{-3}$.
Regarding variation of the energy of symmetric matter with density, recent analyses of
the energies of giant isoscalar monopole resonances point \cite{colo-2004-70} to nuclear incompressibilities within the range of
$K = (230-250)\, \text{MeV}$, and fewer than 30\% of all the interactions in Table \ref{tab:skypro} give rise to
incompressibilities definitely outside of
that range.  Some of the interactions characterized by high $K$-values outside of the above range have been, incidentally,
purposely constructed to explore the effects of high $K$-values on properties
of finite nuclei.

Regarding the dependence of $E/A$ on $\eta$, the values of expansion coefficients for normal $\rho_0$-matter in Table \ref{tab:skypro},
$a_a^V$, are lower by just $\sim 1 \, \text{MeV}$ than the values of symmetry energy
per nucleon at $\eta=1$, $E_a/A(\rho_0,1)$.  This indicates that the quadratic expansion of energy in asymmetry is fairly accurate
near $\rho_0$ for the Skyrme interactions.  The next
quartic coefficients in the expansion of $E/A$ with respect to asymmetry must be positive for the Skyrme interactions and be of the order of 1~MeV.
The inferred magnitude of quartic terms indicates that a term breaking charge invariance,
that might be potentially introduced into~$E/A$, could be more important around
$\rho_0$ than the quartic term, for typical $\eta$ of interest.

\begin{figure}
\centerline{\includegraphics[width=.70\linewidth]{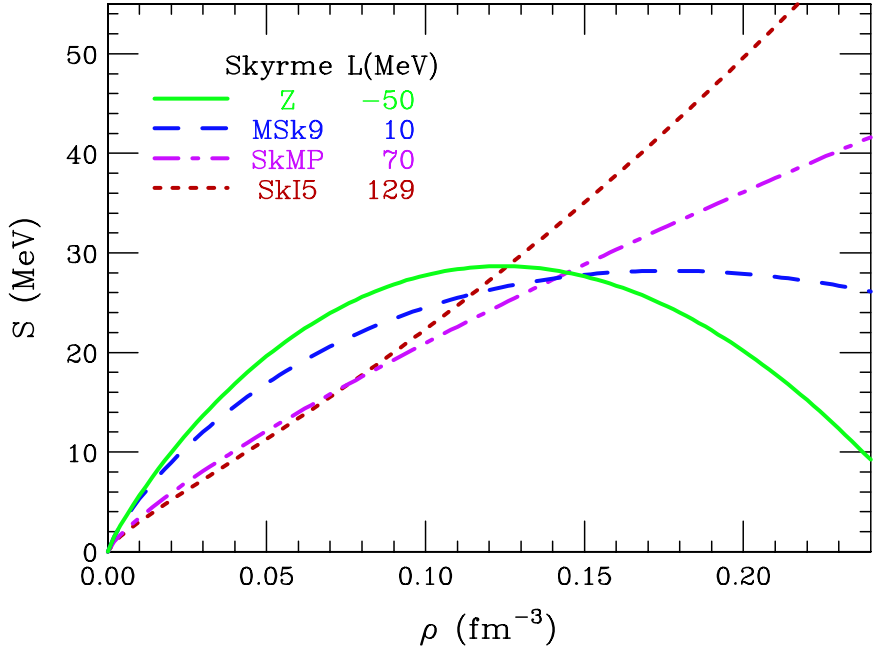}}
\caption{Symmetry energy $S$ as a function of density $\rho$, for sample Skyrme
interactions characterized by widely varying derivative-parameters $L$.
}
\label{fig:syme}
\end{figure}

The relative variation across Table \ref{tab:skypro} is far greater for $a_a^V \equiv S(\rho_0)$ than for $a_V$.
The~lowest~$a_a^V$ values, of the order of 23~MeV,
are close to the $a_a$-value inferred from the basic energy formula (\ref{eq:enz}) from Sec.~\ref{sec:enfo}, while some of the high $a_a^V$ values
in the table actually exceed 37~MeV.
An even greater degree of variation is found in the table
for the constants $L$ and $K_\text{sym}$, characterizing the $\rho$-dependence of~$S$.  Thus, the derivative-constant $L$ takes on
positive and negative values in the table, within principally as much as 3 orders of magnitude; the average value of $L$ there is $35 \pm 33 \, \text{MeV}$.
The situation is found to be similar for $K_\text{sym}$ that takes on both negative and
positive values within 2 orders of magnitude.  Functions $S(\rho)$ for the Skyrme interactions representing two more typical and
two more extreme $L$-values are
shown in Fig.~\ref{fig:syme}.

\begin{figure}
\centerline{\includegraphics[width=.70\linewidth]{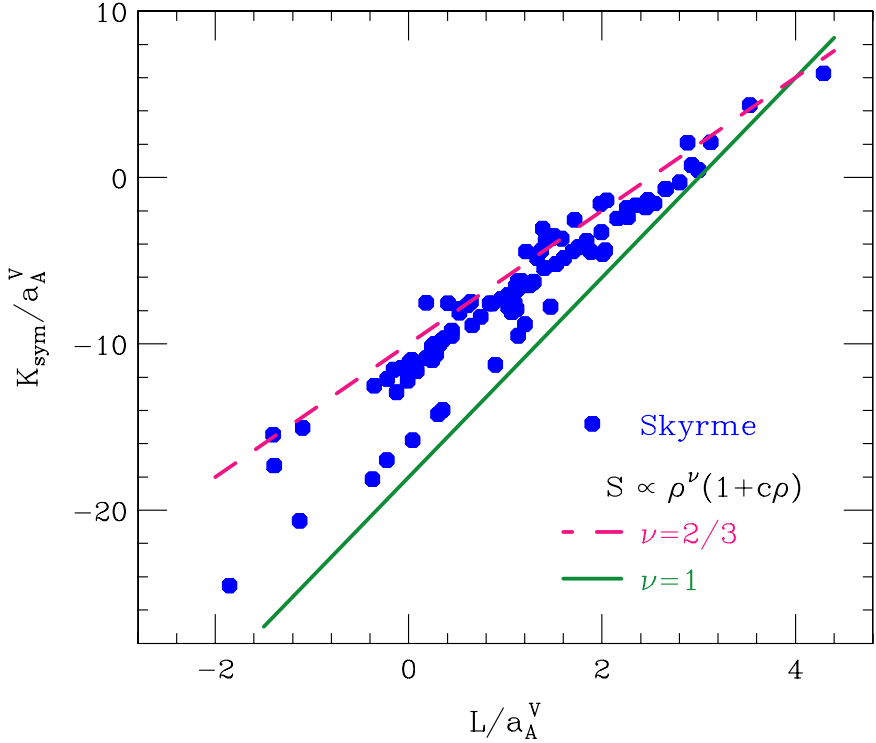}}
\caption{
Correlation between the symmetry-energy parameter-ratios $K_\text{sym}/a_a^V$ and $L/a_a^V$.
The~circles represent Skyrme interactions, while the lines
represent the modified power parameterization $S(\rho) \propto \rho^\nu (1+ c \rho)$.
}
\label{fig:lka}
\end{figure}

While the parameters $L$ and $K_\text{sym}$ vary widely between interactions, they do so in a~fairly correlated manner.
The relatively tight correlation is illustrated in Fig.~\ref{fig:lka} displaying results for the Skyrme interactions, in the plane of those parameters scaled by $a_a^V$.
The~correlation may be understood as due to the fact that just one physical scale of $\rho_0$ governs
the low-density variation of nuclear energy, limiting possible independent variations in the first and second derivatives of~$S$ at~$\rho_0$.
If~we were to approximate $S(\rho)$-functions for the Skyrme interactions,
such as those shown in Fig.~\ref{fig:syme}, with parabolas, $S(\rho) \propto \rho (1 + c \, \rho)$, i.e.\ extending the
expansion~(\ref{eq:Sexp}) to the whole subnormal region of densities, we would arrive at the following relation between the scaled parameters:
\beq
\label{eq:KL1}
\frac{K_\text{sym}}{a_a^V} \approx 6 \, \frac{L}{a_a^V} - 18 \, .
\eeq
That relation,
represented by the solid line in Fig.~\ref{fig:lka}, qualitatively explains the correlation for the interactions in the figure.
However, a parabola produces an incorrect linear behavior of $S$ at low~$\rho$, where $S$ is expected to be dominated by the Fermi-energy
contribution proportional to~$\rho^{2/3}$.  An attempt to improve the approximation to $S$, using $S(\rho) \propto \rho^\nu (1 + c \, \rho)$,
leads to the relation
\beq
\label{eq:KL2}
\frac{K_\text{sym}}{a_a^V} \approx 6 \nu \, \frac{L}{a_a^V} - 9 \nu (\nu+1) \, ,
\eeq
represented by the dashed line in Fig.~\ref{fig:lka}, for $\nu=2/3$.  It is apparent that the $\nu=1$ and $\nu=2/3$ relations practically bracket
the correlation between the $L$ and $K_\text{sym}$ parameters, in~spite of the fact that $S$ for the Skyrme interactions differs from either of
the forms used in the approximations.  Practical implication of the observed correlation is that any constraining of either of the two
parameter will produce constraints
on the likely values of the other parameter.

With $a_V \simeq 16 \, \text{MeV}$, and
with the average volume symmetry parameter for the Skyrme interactions being $a_a^V \sim 30 \, \text{MeV}$, the magnitude of drip asymmetry expected for the Skyrme interactions from the simple expression \eqref{eq:etaVd} is $\sim 0.27$, which turns out to be close to the typical magnitude of $\eta_V^d$ in Table \ref{tab:skypro}.  The simple formula implies further an anticorrelation between the bulk asymmetry coefficient and the drip asymmetry.  Indeed, the lowest and the highest drip asymmetry values in Table \ref{tab:skypro} are associated with some of the highest and the lowest, respectively, volume coefficient values~$a_a^V$ in the table.  Overall, the magnitudes of $\eta_V^d$ in the table may seem to be low.   However, these values pertain to nuclear interiors.  For a given Skyrme interaction, due to asymmetry effects in the surface, nuclei may generally have a net asymmetry $\eta$ larger in magnitude than $\eta_V^d$ and still be stable.


\newpage

\pagestyle{empty}
\begin{landscape}
\begin{center}
\setlength{\LTcapwidth}{7.9in}
\begin{longtable}
               {l|c c c c c c c c c c c c c c c c}

\caption[Bulk nuclear properties for different Skyrme interactions with force constants given in the indicated references.]
{Bulk nuclear properties for different Skyrme interactions with force constants given in the indicated references.}
\label{tab:skypro}\\

\hline
Name & $\rho_0$ & $a_V$ & $m^*/m$ & $K$  & $a_a^V$ & $L$ & $K_\text{sym}$ & $\frac{E_a}{A}(\rho_0,1)$ & $\eta_V^d $  & $a_S$ & $a_a^S$ & $d_0$ & $\Delta R^0$ & $\Delta_e R$ & $d_a^0$ & Ref. \\
& (fm$^\text{-3}$) & (MeV) & & (MeV) & (MeV) & (MeV) & (MeV) & (MeV) &     &  (MeV) & (MeV) & (fm) & (fm) & (fm) & (fm) &    \\
\hline
\endfirsthead

\multicolumn{17}{l}{{\hspace*{7em}\tablename} \thetable{} -- Continued}\\
\hline
Name & $\rho_0$ & $a_V$ & $m^*/m$ & $K$  & $a_a^V$ & $L$ & $K_\text{sym}$ & $\frac{E_a}{A}(\rho_0,1)$ & $\eta_V^d $  & $a_S$ & $a_a^S$ & $d_0$ & $\Delta R^0$ & $\Delta_e R$ & $d_a^0$ & Ref. \\
& (fm$^\text{-3}$) & (MeV) & & (MeV) & (MeV) & (MeV) & (MeV) & (MeV) &     &  (MeV) & (MeV) & (fm) & (fm) & (fm) & (fm) &    \\
\hline
\endhead

\multicolumn{17}{l}{\hspace*{7em}Table Continued on Next Page\ldots}
\endfoot

\hline
\endlastfoot

SI        & 0.1553 & -15.99 & 0.911 & 370.3 & 29.24 &    1.2 & -461.8 & 30.27 & 0.325 & 17.36 &  26.00 &  0.438 &  0.432 &  0.433 &  0.461 & \cite{PhysRevC.5.626}\\
SII       & 0.1482 & -15.96 & 0.580 & 340.8 & 34.14 &   50.1 & -265.1 & 35.75 & 0.274 & 19.38 &  17.54 &  0.517 &  0.718 &  0.761 &  0.439 & \cite{PhysRevC.5.626}\\
SIII      & 0.1453 & -15.85 & 0.763 & 355.3 & 28.16 &    9.9 & -393.7 & 29.38 & 0.337 & 18.54 &  21.77 &  0.486 &  0.518 &  0.509 &  0.474 & \cite{Beiner:1974gc}\\
SIIIs     & 0.1507 & -16.57 & 0.789 & 371.8 & 32.66 &   29.3 & -367.7 & 33.60 & 0.299 & 20.06 &  17.70 &  0.492 &  0.653 &  0.717 &  0.412 & \cite{PhysRevC.21.2076}\\
SIV       & 0.1509 & -15.96 & 0.471 & 324.5 & 31.22 &   63.5 & -136.7 & 33.22 & 0.312 & 18.87 &  13.39 &  0.529 &  0.779 &  0.906 &  0.404 & \cite{Beiner:1974gc}\\
SV        & 0.1551 & -16.05 & 0.383 & 305.7 & 32.83 &   96.1 &   24.2 & 35.32 & 0.312 & 19.16 &  10.27 &  0.553 &  0.941 &  1.231 &  0.368 & \cite{Beiner:1974gc}\\
SVI       & 0.1435 & -15.75 & 0.949 & 363.6 & 26.88 &   -7.3 & -471.3 & 27.86 & 0.354 & 18.10 &  26.27 &  0.466 &  0.428 &  0.404 &  0.496 & \cite{Beiner:1974gc}\\
SVII      & 0.1434 & -15.79 & 1.001 & 366.4 & 26.96 &  -10.1 & -488.9 & 27.89 & 0.354 & 18.11 &  27.43 &  0.462 &  0.415 &  0.388 &  0.500 & \cite{PhysRevC.21.2076}\\
SkT       & 0.1476 & -15.40 & 0.602 & 333.3 & 24.89 &   28.2 & -236.7 & 26.25 & 0.385 & 14.21 &  17.58 &  0.431 &  0.477 &  0.554 &  0.393 & \cite{1974NuPhA.236..269K}\\
SkT1      & 0.1610 & -15.98 & 1.000 & 236.2 & 32.02 &   56.2 & -134.9 & 32.73 & 0.305 & 18.26 &  14.60 &  0.556 &  0.799 &  0.834 &  0.450 & \cite{1984NuPhA.420..297T}\\
SkT2      & 0.1610 & -15.94 & 1.000 & 235.8 & 32.00 &   56.2 & -134.7 & 32.71 & 0.305 & 18.01 &  14.71 &  0.552 &  0.794 &  0.828 &  0.450 & \cite{1984NuPhA.420..297T}\\
SkT3      & 0.1610 & -15.94 & 1.000 & 235.8 & 31.50 &   55.3 & -132.1 & 32.21 & 0.311 & 17.78 &  15.33 &  0.547 &  0.776 &  0.782 &  0.468 & \cite{1984NuPhA.420..297T}\\
SkT4      & 0.1590 & -15.95 & 1.000 & 235.5 & 35.46 &   94.1 &  -24.5 & 36.16 & 0.284 & 18.17 &  11.57 &  0.558 &  0.986 &  1.171 &  0.390 & \cite{1984NuPhA.420..297T}\\
SkT5      & 0.1640 & -16.00 & 1.000 & 201.7 & 37.01 &   98.5 &  -25.0 & 37.72 & 0.274 & 18.13 &  10.91 &  0.604 &  1.084 &  1.283 &  0.400 & \cite{1984NuPhA.420..297T}\\
SkT6      & 0.1609 & -15.96 & 1.000 & 236.0 & 29.97 &   30.9 & -211.6 & 30.67 & 0.321 & 18.22 &  18.10 &  0.555 &  0.658 &  0.630 &  0.508 & \cite{1984NuPhA.420..297T}\\
SkT7      & 0.1606 & -15.94 & 0.833 & 235.7 & 29.52 &   31.1 & -209.9 & 30.72 & 0.326 & 18.12 &  18.21 &  0.559 &  0.650 &  0.617 &  0.516 & \cite{1984NuPhA.420..297T}\\
SkT8      & 0.1607 & -15.94 & 0.833 & 235.7 & 29.92 &   33.7 & -187.6 & 30.77 & 0.322 & 18.13 &  17.88 &  0.559 &  0.657 &  0.637 &  0.509 & \cite{1984NuPhA.420..297T}\\
SkT9      & 0.1603 & -15.88 & 0.833 & 234.9 & 29.76 &   33.8 & -185.7 & 30.60 & 0.323 & 17.79 &  17.81 &  0.555 &  0.654 &  0.636 &  0.508 & \cite{1984NuPhA.420..297T}\\
SkTK      & 0.1681 & -16.70 & 0.611 & 253.3 & 35.57 &   41.6 & -221.9 & 37.35 & 0.275 & 19.78 &  19.46 &  0.579 &  0.696 &  0.685 &  0.512 & \cite{1976JPhG....2..285T}\\
SkM       & 0.1603 & -15.77 & 0.789 & 216.7 & 30.75 &   49.4 & -148.9 & 32.07 & 0.314 & 16.85 &  14.84 &  0.558 &  0.756 &  0.789 &  0.460 & \cite{krivine:1980}\\
SkM1      & 0.1603 & -15.77 & 0.789 & 216.7 & 25.17 &  -35.3 & -389.0 & 26.53 & 0.414 & 17.46 &  59.67 &  0.575 &  0.180 &  0.161 &  0.729 & \cite{gomez:1995}\\
SkMP      & 0.1570 & -15.56 & 0.654 & 230.9 & 29.89 &   70.3 &  -49.8 & 31.26 & 0.336 & 16.58 &  12.06 &  0.549 &  0.850 &  0.950 &  0.423 & \cite{PhysRevC.40.2834}\\
SkMs      & 0.1603 & -15.77 & 0.789 & 216.7 & 30.03 &   45.8 & -156.0 & 31.39 & 0.322 & 17.46 &  14.48 &  0.575 &  0.765 &  0.790 &  0.470 & \cite{1982NuPhA.386...79B}\\
SKa       & 0.1554 & -15.99 & 0.608 & 263.2 & 32.91 &   74.6 &  -78.5 & 34.56 & 0.301 & 18.77 &  12.51 &  0.561 &  0.877 &  1.013 &  0.409 & \cite{Kohler:1976}\\
SKb       & 0.1554 & -15.99 & 0.608 & 263.2 & 23.88 &   47.6 &  -78.5 & 25.53 & 0.460 & 18.77 &  10.27 &  0.561 &  0.768 &  0.895 &  0.408 & \cite{Kohler:1976}\\
SGI       & 0.1544 & -15.89 & 0.608 & 261.8 & 28.33 &   63.9 &  -52.0 & 29.62 & 0.365 & 17.48 &  12.76 &  0.531 &  0.770 &  0.856 &  0.424 & \cite{1981PhLB..106..379V}\\
SGII      & 0.1583 & -15.59 & 0.786 & 214.7 & 26.83 &   37.6 & -146.0 & 28.10 & 0.365 & 16.10 &  15.11 &  0.538 &  0.660 &  0.679 &  0.464 & \cite{1981PhLB..106..379V}\\
RATP      & 0.1598 & -16.05 & 0.667 & 239.6 & 29.26 &   32.4 & -191.3 & 30.80 & 0.332 & 18.68 &  17.95 &  0.579 &  0.658 &  0.621 &  0.534 & \cite{1982AA...116..183R}\\
SkP       & 0.1625 & -15.95 & 1.000 & 201.0 & 30.00 &   19.7 & -266.7 & 31.33 & 0.316 & 18.18 &  18.02 &  0.607 &  0.670 &  0.631 &  0.533 & \cite{1984NuPhA.422..103D}\\
E         & 0.1591 & -16.13 & 0.868 & 333.4 & 27.66 &  -31.2 & -570.8 & 28.83 & 0.359 & 18.94 &  37.62 &  0.485 &  0.321 &  0.280 &  0.576 & \cite{PhysRevC.33.335}\\
Es        & 0.1628 & -16.02 & 0.839 & 248.6 & 26.44 &  -36.8 & -457.9 & 27.72 & 0.390 & 18.21 &  52.37 &  0.543 &  0.222 &  0.191 &  0.689 & \cite{PhysRevC.33.335}\\
Gs        & 0.1576 & -15.59 & 0.784 & 237.3 & 31.37 &   94.0 &   14.0 & 32.61 & 0.336 & 16.07 &  10.10 &  0.512 &  0.929 &  1.191 &  0.357 & \cite{PhysRevC.33.335}\\
Rs        & 0.1577 & -15.59 & 0.783 & 237.4 & 30.58 &   85.7 &   -9.1 & 31.82 & 0.340 & 16.06 &  10.59 &  0.512 &  0.888 &  1.106 &  0.366 & \cite{PhysRevC.33.335}\\
T         & 0.1613 & -15.93 & 1.000 & 235.7 & 28.35 &   27.2 & -206.8 & 29.06 & 0.343 & 17.74 &  22.64 &  0.547 &  0.587 &  0.476 &  0.604 & \cite{PhysRevC.33.335}\\
Z         & 0.1589 & -15.97 & 0.842 & 330.3 & 26.82 &  -49.7 & -657.9 & 27.97 & 0.391 & 17.74 &  51.51 &  0.465 &  0.213 &  0.199 &  0.588 & \cite{PhysRevC.33.335}\\
Zs        & 0.1630 & -15.88 & 0.783 & 233.4 & 26.69 &  -29.3 & -401.6 & 28.01 & 0.371 & 17.00 &  46.62 &  0.534 &  0.233 &  0.217 &  0.650 & \cite{PhysRevC.33.335}\\
Zss       & 0.1625 & -15.96 & 0.775 & 234.9 & 28.80 &   -4.5 & -332.7 & 30.14 & 0.330 & 17.32 &  29.34 &  0.541 &  0.406 &  0.372 &  0.594 & \cite{PhysRevC.33.335}\\
SkSC1     & 0.1607 & -15.85 & 1.000 & 234.6 & 28.10 &    0.2 & -312.1 & 28.81 & 0.338 & 17.28 &  26.64 &  0.536 &  0.448 &  0.401 &  0.589 & \cite{1991NuPhA.528....1P}\\
SkSC2     & 0.1607 & -15.90 & 1.000 & 235.2 & 24.74 &   11.0 & -228.3 & 25.44 & 0.401 & 17.45 &  19.73 &  0.540 &  0.515 &  0.477 &  0.535 & \cite{1991NuPhA.528....1P}\\
SkSC3     & 0.1606 & -15.85 & 1.000 & 234.5 & 27.01 &    0.8 & -296.3 & 27.71 & 0.355 & 16.88 &  27.39 &  0.529 &  0.434 &  0.375 &  0.605 & \cite{1991NuPhA.528....1P}\\
SkSC4     & 0.1606 & -15.86 & 1.000 & 234.8 & 28.80 &   -2.1 & -329.6 & 29.51 & 0.328 & 17.37 &  28.15 &  0.538 &  0.439 &  0.389 &  0.602 & \cite{1992NuPhA.549..155A}\\
SkSC4o    & 0.1606 & -15.84 & 1.000 & 234.5 & 26.98 &   -9.6 & -337.8 & 27.69 & 0.357 & 17.35 &  30.46 &  0.538 &  0.385 &  0.337 &  0.619 & \cite{pearson:2000}\\
SkSC5     & 0.1606 & -15.85 & 1.000 & 234.5 & 30.99 &   -6.9 & -375.2 & 31.69 & 0.300 & 17.28 &  32.03 &  0.536 &  0.415 &  0.368 &  0.627 & \cite{PhysRevC.50.460}\\
SkSC6     & 0.1607 & -15.92 & 1.000 & 235.5 & 24.57 &   11.0 & -226.3 & 25.28 & 0.406 & 17.59 &  19.56 &  0.543 &  0.518 &  0.478 &  0.535 & \cite{PhysRevC.50.460}\\
SkSC10    & 0.1607 & -15.96 & 1.000 & 235.9 & 22.83 &   19.1 & -172.8 & 23.53 & 0.457 & 17.75 &  15.61 &  0.546 &  0.577 &  0.556 &  0.497 & \cite{PhysRevC.50.460}\\
SkSC11    & 0.1606 & -15.86 & 1.000 & 234.8 & 28.80 &   -2.1 & -329.6 & 29.51 & 0.328 & 17.37 &  28.15 &  0.538 &  0.439 &  0.389 &  0.602 & \cite{PhysRevC.52.2254}\\
SkSC14    & 0.1607 & -15.92 & 1.000 & 235.5 & 30.00 &   33.2 & -202.9 & 30.71 & 0.320 & 17.60 &  17.94 &  0.543 &  0.656 &  0.636 &  0.497 & \cite{pearson:2000}\\
SkSC15    & 0.1607 & -15.88 & 1.000 & 235.0 & 28.00 &    6.7 & -284.6 & 28.71 & 0.341 & 17.43 &  23.93 &  0.539 &  0.492 &  0.445 &  0.568 & \cite{pearson:2000}\\
Skyrme1p  & 0.1553 & -15.99 & 0.911 & 370.3 & 29.35 &   35.3 & -259.2 & 30.38 & 0.329 & 17.36 &  19.06 &  0.438 &  0.548 &  0.592 &  0.407 & \cite{1995NuPhA.584..675P}\\
MSkA      & 0.1535 & -15.99 & 0.794 & 313.3 & 30.35 &   57.2 & -135.4 & 31.60 & 0.324 & 19.00 &  14.60 &  0.508 &  0.731 &  0.803 &  0.408 & \cite{PhysRevLett.74.3744}\\
SkI1      & 0.1604 & -15.95 & 0.693 & 242.8 & 37.52 &  161.0 &  234.7 & 38.20 & 0.331 & 17.45 &  11.42 &  0.541 &  1.126 &  1.253 &  0.443 & \cite{1995NuPhA.584..467R}\\
SkI2      & 0.1575 & -15.77 & 0.685 & 241.0 & 33.38 &  104.3 &   70.7 & 34.02 & 0.316 & 17.03 &  11.68 &  0.537 &  0.951 &  1.096 &  0.412 & \cite{1995NuPhA.584..467R}\\
SkI3      & 0.1577 & -15.98 & 0.577 & 258.2 & 34.83 &  100.5 &   73.0 & 35.20 & 0.293 & 17.77 &  12.77 &  0.542 &  0.908 &  1.045 &  0.413 & \cite{1995NuPhA.584..467R}\\
SkI4      & 0.1601 & -15.94 & 0.649 & 248.0 & 29.50 &   60.4 &  -40.6 & 30.08 & 0.344 & 17.48 &  20.83 &  0.539 &  0.665 &  0.540 &  0.582 & \cite{1995NuPhA.584..467R}\\
SkI5      & 0.1558 & -15.85 & 0.579 & 255.8 & 36.64 &  129.3 &  159.5 & 37.01 & 0.288 & 17.31 &  10.71 &  0.536 &  1.034 &  1.316 &  0.374 & \cite{1995NuPhA.584..467R}\\
SkI6      & 0.1591 & -15.92 & 0.640 & 248.6 & 30.09 &   59.7 &  -47.3 & 30.64 & 0.333 & 17.42 &  19.07 &  0.539 &  0.683 &  0.602 &  0.542 & \cite{PhysRevC.53.740}\\
SLy0      & 0.1603 & -15.97 & 0.698 & 229.7 & 31.98 &   47.1 & -116.3 & 32.67 & 0.301 & 18.21 &  16.75 &  0.580 &  0.713 &  0.727 &  0.500 & \cite{ChabanatPhd1995}\\
SLy1      & 0.1603 & -15.99 & 0.698 & 229.9 & 31.99 &   47.1 & -116.6 & 32.68 & 0.301 & 18.31 &  16.19 &  0.582 &  0.725 &  0.752 &  0.484 & \cite{ChabanatPhd1995}\\
SLy2      & 0.1605 & -15.99 & 0.698 & 230.0 & 32.00 &   47.5 & -115.2 & 32.69 & 0.301 & 18.20 &  17.28 &  0.579 &  0.706 &  0.705 &  0.516 & \cite{ChabanatPhd1995}\\
SLy230a   & 0.1600 & -15.99 & 0.697 & 229.9 & 31.99 &   44.3 &  -98.3 & 32.23 & 0.301 & 18.20 &  18.19 &  0.580 &  0.668 &  0.670 &  0.525 & \cite{1997NuPhA.627..710C}\\
SLy3      & 0.1604 & -15.97 & 0.696 & 229.9 & 31.99 &   45.3 & -122.2 & 32.68 & 0.300 & 18.21 &  16.74 &  0.579 &  0.706 &  0.728 &  0.490 & \cite{ChabanatPhd1995}\\
SLy4      & 0.1595 & -15.97 & 0.695 & 230.0 & 32.00 &   46.0 & -119.8 & 32.68 & 0.301 & 18.24 &  16.60 &  0.581 &  0.712 &  0.735 &  0.490 & \cite{1998NuPhA.635..231C}\\
SLy5      & 0.1606 & -15.98 & 0.698 & 230.0 & 32.01 &   48.2 & -112.8 & 32.70 & 0.302 & 18.28 &  16.06 &  0.581 &  0.728 &  0.759 &  0.481 & \cite{1998NuPhA.635..231C}\\
SLy6      & 0.1590 & -15.92 & 0.690 & 229.9 & 31.96 &   47.5 & -112.8 & 32.63 & 0.300 & 17.53 &  17.12 &  0.562 &  0.689 &  0.713 &  0.487 & \cite{1998NuPhA.635..231C}\\
SLy7      & 0.1583 & -15.90 & 0.688 & 229.8 & 31.99 &   47.0 & -114.4 & 32.65 & 0.299 & 17.35 &  17.45 &  0.559 &  0.680 &  0.701 &  0.490 & \cite{1998NuPhA.635..231C}\\
SLy8      & 0.1603 & -15.97 & 0.696 & 229.9 & 32.00 &   47.2 & -115.7 & 32.68 & 0.301 & 18.18 &  16.40 &  0.578 &  0.717 &  0.743 &  0.485 & \cite{ChabanatPhd1995}\\
SLy9      & 0.1512 & -15.79 & 0.666 & 229.9 & 31.98 &   54.9 &  -81.5 & 32.58 & 0.300 & 17.59 &  15.93 &  0.574 &  0.742 &  0.780 &  0.486 & \cite{ChabanatPhd1995}\\
SLy10     & 0.1556 & -15.90 & 0.683 & 229.7 & 31.98 &   38.8 & -142.3 & 32.62 & 0.297 & 17.05 &  20.23 &  0.549 &  0.607 &  0.608 &  0.518 & \cite{ChabanatPhd1995}\\
SkX       & 0.1554 & -16.05 & 0.993 & 271.1 & 31.10 &   33.2 & -252.2 & 32.36 & 0.307 & 16.04 &  19.31 &  0.476 &  0.581 &  0.620 &  0.440 & \cite{PhysRevC.58.220}\\
SkXce     & 0.1553 & -15.86 & 1.006 & 268.2 & 30.20 &   33.7 & -238.5 & 31.46 & 0.314 & 16.10 &  18.34 &  0.480 &  0.592 &  0.634 &  0.437 & \cite{PhysRevC.58.220}\\
SkXm      & 0.1589 & -16.05 & 0.966 & 238.2 & 31.20 &   32.1 & -242.9 & 32.45 & 0.306 & 15.69 &  19.56 &  0.497 &  0.597 &  0.609 &  0.469 & \cite{PhysRevC.58.220}\\
SkO       & 0.1605 & -15.83 & 0.896 & 223.4 & 31.97 &   79.1 &  -43.2 & 32.86 & 0.320 & 17.14 &  14.94 &  0.546 &  0.856 &  0.815 &  0.485 & \cite{PhysRevC.60.014316}\\
SkOp      & 0.1602 & -15.75 & 0.896 & 222.4 & 31.95 &   68.9 &  -78.8 & 32.79 & 0.309 & 16.07 &  15.74 &  0.524 &  0.802 &  0.773 &  0.453 & \cite{PhysRevC.60.014316}\\
SKRA      & 0.1595 & -15.78 & 0.748 & 217.0 & 31.32 &   53.1 & -139.3 & 32.71 & 0.308 & 16.71 &  14.79 &  0.556 &  0.766 &  0.808 &  0.455 & \cite{2000MPLA...15.1287R}\\
MSk1      & 0.1575 & -15.83 & 1.000 & 233.8 & 30.00 &   33.9 & -200.1 & 30.70 & 0.318 & 17.21 &  18.08 &  0.536 &  0.651 &  0.636 &  0.492 & \cite{PhysRevC.62.024308}\\
MSk2      & 0.1575 & -15.83 & 1.050 & 231.7 & 30.00 &   33.4 & -203.5 & 30.66 & 0.318 & 17.22 &  18.06 &  0.538 &  0.654 &  0.636 &  0.494 & \cite{PhysRevC.62.024308}\\
MSk3      & 0.1578 & -15.82 & 1.001 & 233.6 & 27.99 &    6.9 & -284.4 & 28.69 & 0.339 & 17.08 &  24.37 &  0.533 &  0.485 &  0.440 &  0.562 & \cite{PhysRevC.62.024308}\\
MSk4      & 0.1575 & -15.79 & 1.050 & 231.2 & 28.00 &    7.2 & -284.1 & 28.66 & 0.339 & 17.06 &  24.04 &  0.534 &  0.494 &  0.446 &  0.563 & \cite{PhysRevC.62.024308}\\
MSk5      & 0.1575 & -15.79 & 1.050 & 231.2 & 28.00 &    7.6 & -282.6 & 28.66 & 0.339 & 17.07 &  23.90 &  0.535 &  0.496 &  0.449 &  0.561 & \cite{PhysRevC.62.024308}\\
MSk5s     & 0.1561 & -15.78 & 0.800 & 243.8 & 28.00 &    7.0 & -290.7 & 29.17 & 0.337 & 17.25 &  24.15 &  0.533 &  0.481 &  0.445 &  0.549 & \cite{PhysRevC.64.027301}\\
MSk6      & 0.1575 & -15.79 & 1.050 & 231.2 & 28.00 &    9.7 & -274.4 & 28.66 & 0.339 & 17.09 &  23.10 &  0.536 &  0.510 &  0.464 &  0.556 & \cite{PhysRevC.62.024308}\\
MSk7      & 0.1575 & -15.79 & 1.050 & 231.3 & 27.95 &    9.4 & -274.7 & 28.61 & 0.340 & 17.11 &  23.12 &  0.536 &  0.509 &  0.463 &  0.557 & \cite{2001ADNDT..77..311G}\\
MSk8      & 0.1575 & -15.79 & 1.100 & 229.4 & 27.93 &    8.3 & -280.1 & 28.56 & 0.340 & 17.12 &  23.38 &  0.537 &  0.507 &  0.458 &  0.561 & \cite{2001NuPhA.688..349G}\\
MSk9      & 0.1575 & -15.80 & 1.000 & 233.4 & 28.00 &   10.4 & -270.3 & 28.70 & 0.339 & 17.15 &  23.02 &  0.536 &  0.510 &  0.466 &  0.553 & \cite{2001NuPhA.688..349G}\\
v070      & 0.1575 & -15.80 & 1.050 & 231.3 & 28.00 &   -3.5 & -361.7 & 29.49 & 0.337 & 17.41 &  24.61 &  0.541 &  0.466 &  0.436 &  0.547 & \cite{PhysRevC.64.027301}\\
v075      & 0.1575 & -15.80 & 1.050 & 231.3 & 28.00 &   -0.3 & -342.0 & 29.32 & 0.337 & 17.31 &  24.40 &  0.539 &  0.477 &  0.439 &  0.552 & \cite{PhysRevC.64.027301}\\
v080      & 0.1575 & -15.79 & 1.050 & 231.2 & 28.00 &    2.3 & -325.7 & 29.18 & 0.337 & 17.18 &  24.30 &  0.536 &  0.483 &  0.441 &  0.555 & \cite{PhysRevC.64.027301}\\
v090      & 0.1575 & -15.79 & 1.050 & 231.2 & 28.00 &    5.1 & -304.3 & 28.94 & 0.338 & 17.08 &  24.22 &  0.535 &  0.489 &  0.443 &  0.561 & \cite{PhysRevC.64.027301}\\
v100      & 0.1575 & -15.79 & 1.050 & 231.2 & 28.00 &    8.8 & -281.5 & 28.75 & 0.339 & 17.02 &  23.68 &  0.534 &  0.502 &  0.453 &  0.560 & \cite{PhysRevC.64.027301}\\
v105      & 0.1575 & -15.79 & 1.050 & 231.2 & 28.00 &    7.1 & -284.6 & 28.66 & 0.339 & 17.05 &  24.11 &  0.534 &  0.492 &  0.445 &  0.563 & \cite{PhysRevC.64.027301}\\
v110      & 0.1575 & -15.79 & 1.050 & 231.2 & 28.00 &    7.5 & -279.7 & 28.59 & 0.339 & 17.11 &  23.78 &  0.535 &  0.496 &  0.451 &  0.559 & \cite{PhysRevC.64.027301}\\
SKz1      & 0.1600 & -16.01 & 0.700 & 230.1 & 32.01 &   27.7 & -242.5 & 33.60 & 0.294 & 17.35 &  19.03 &  0.554 &  0.616 &  0.641 &  0.490 & \cite{PhysRevC.66.014303}\\
SKz2      & 0.1600 & -16.00 & 0.700 & 230.1 & 32.01 &   16.8 & -259.8 & 33.27 & 0.293 & 17.34 &  23.95 &  0.554 &  0.523 &  0.509 &  0.549 & \cite{PhysRevC.66.014303}\\
SKz3      & 0.1600 & -16.01 & 0.700 & 230.1 & 32.01 &   13.0 & -242.0 & 32.77 & 0.293 & 17.35 &  28.05 &  0.554 &  0.465 &  0.435 &  0.596 & \cite{PhysRevC.66.014303}\\
SKz4      & 0.1600 & -16.01 & 0.700 & 230.1 & 32.01 &    5.8 & -241.0 & 32.35 & 0.293 & 17.35 &  34.04 &  0.554 &  0.393 &  0.358 &  0.643 & \cite{PhysRevC.66.014303}\\
BSk1      & 0.1573 & -15.80 & 1.050 & 231.4 & 27.81 &    7.2 & -281.9 & 28.47 & 0.342 & 17.22 &  23.72 &  0.539 &  0.499 &  0.449 &  0.566 & \cite{2002NuPhA.700..142S}\\
BSk2      & 0.1575 & -15.79 & 1.042 & 233.7 & 28.00 &    8.0 & -297.1 & 29.02 & 0.338 & 17.14 &  23.15 &  0.534 &  0.509 &  0.463 &  0.552 & \cite{PhysRevC.66.024326}\\
BSk2p     & 0.1575 & -15.79 & 1.049 & 233.4 & 28.00 &    7.8 & -298.1 & 29.02 & 0.338 & 17.11 &  23.20 &  0.534 &  0.508 &  0.462 &  0.553 & \cite{PhysRevC.66.024326}\\
BSk3      & 0.1575 & -15.80 & 1.123 & 234.8 & 27.94 &    6.8 & -307.0 & 28.95 & 0.339 & 17.16 &  23.37 &  0.531 &  0.506 &  0.458 &  0.552 & \cite{PhysRevC.68.054325}\\
BSk4      & 0.1575 & -15.77 & 0.920 & 236.9 & 27.99 &   12.5 & -266.0 & 28.90 & 0.339 & 16.91 &  23.01 &  0.530 &  0.511 &  0.466 &  0.551 & \cite{PhysRevC.68.054325}\\
BSk5      & 0.1575 & -15.80 & 0.920 & 237.2 & 28.70 &   21.4 & -240.4 & 29.65 & 0.331 & 17.04 &  20.65 &  0.533 &  0.570 &  0.532 &  0.528 & \cite{PhysRevC.68.054325}\\
BSk6      & 0.1575 & -15.75 & 0.800 & 229.2 & 28.00 &   16.9 & -215.3 & 28.72 & 0.339 & 16.74 &  22.73 &  0.542 &  0.511 &  0.472 &  0.564 & \cite{PhysRevC.68.054325}\\
BSk7      & 0.1575 & -15.76 & 0.800 & 229.3 & 28.00 &   18.0 & -209.4 & 28.69 & 0.340 & 16.70 &  22.81 &  0.541 &  0.514 &  0.470 &  0.568 & \cite{PhysRevC.68.054325}\\
BSk8      & 0.1589 & -15.82 & 0.800 & 230.3 & 28.00 &   14.9 & -221.0 & 28.70 & 0.341 & 17.12 &  23.13 &  0.549 &  0.506 &  0.462 &  0.578 & \cite{samyn:044309}\\
BSk9      & 0.1589 & -15.92 & 0.800 & 231.5 & 30.00 &   39.9 & -145.4 & 30.62 & 0.323 & 17.42 &  17.83 &  0.556 &  0.666 &  0.643 &  0.520 & \cite{samyn:044309}\\
BSk10     & 0.1593 & -15.91 & 0.920 & 238.9 & 30.00 &   37.3 & -195.0 & 31.01 & 0.321 & 17.51 &  17.48 &  0.540 &  0.671 &  0.655 &  0.491 & \cite{goriely:2006}\\
BSk11     & 0.1586 & -15.86 & 0.920 & 238.1 & 30.00 &   38.4 & -189.9 & 30.99 & 0.320 & 17.32 &  17.42 &  0.536 &  0.671 &  0.658 &  0.486 & \cite{goriely:2006}\\
BSk12     & 0.1586 & -15.86 & 0.920 & 238.1 & 30.00 &   38.0 & -191.4 & 31.00 & 0.320 & 17.30 &  17.51 &  0.535 &  0.668 &  0.655 &  0.487 & \cite{goriely:2006}\\
BSk13     & 0.1586 & -15.86 & 0.920 & 238.1 & 30.00 &   38.8 & -188.0 & 30.99 & 0.320 & 17.34 &  17.34 &  0.536 &  0.673 &  0.661 &  0.485 & \cite{goriely:2006}\\
BSk14     & 0.1586 & -15.85 & 0.800 & 239.4 & 30.00 &   43.9 & -152.1 & 30.92 & 0.323 & 17.17 &  17.22 &  0.537 &  0.680 &  0.666 &  0.492 & \cite{goriely:064312}\\
SK255     & 0.1573 & -16.33 & 0.797 & 255.0 & 37.40 &   95.1 &  -58.3 & 38.77 & 0.268 & 19.28 &  12.44 &  0.561 &  0.967 &  1.153 &  0.389 & \cite{PhysRevC.68.031304}\\
SK272     & 0.1553 & -16.28 & 0.773 & 271.5 & 37.40 &   91.7 &  -67.8 & 38.71 & 0.264 & 19.21 &  13.48 &  0.545 &  0.916 &  1.068 &  0.395 & \cite{PhysRevC.68.031304}\\
QMC1      & 0.1373 & -14.00 & 0.926 & 328.7 & 29.68 &   -6.7 & -504.2 & 31.03 & 0.272 & 23.72 &  21.44 &  0.663 &  0.708 &  0.555 &  0.680 & \cite{guichon:132502}\\
QMC2      & 0.1403 & -14.29 & 0.834 & 330.1 & 28.70 &    8.7 & -408.4 & 29.84 & 0.290 & 18.84 &  21.00 &  0.526 &  0.586 &  0.544 &  0.520 & \cite{guichon:132502}\\
QMC3      & 0.1607 & -15.98 & 0.825 & 366.9 & 45.78 &   91.8 & -211.0 & 47.07 & 0.197 & 21.17 &  18.92 &  0.508 &  0.847 &  0.921 &  0.418 & \cite{guichon:132502}\\
KDE0v     & 0.1608 & -16.10 & 0.717 & 228.8 & 32.98 &   45.2 & -144.9 & 33.80 & 0.291 & 17.27 &  18.22 &  0.552 &  0.672 &  0.689 &  0.488 & \cite{agrawal:014310}\\
KDE0v1    & 0.1646 & -16.23 & 0.744 & 227.6 & 34.58 &   54.7 & -127.2 & 35.34 & 0.281 & 17.66 &  17.25 &  0.558 &  0.730 &  0.757 &  0.474 & \cite{agrawal:014310}\\
LNS       & 0.1746 & -15.31 & 0.826 & 210.8 & 33.43 &   61.5 & -127.4 & 34.63 & 0.276 & 15.77 &  14.10 &  0.517 &  0.765 &  0.878 &  0.404 & \cite{cao:014313}\\

\end{longtable}
\setlength{\LTcapwidth}{4in}
\end{center}

\end{landscape}
\pagestyle{plain}

\subsection{Local Approximation}
\label{ssec:local}

We now consider
the Hohenberg-Kohn (HK) functional in the context of uniform and near-uniform matter.  For a uniform $\rho$, given the requirement of translation invariance, we may notice that ${\mathcal S}$
in Eq.~(\ref{eq:EaS}) must be a symmetric function of the difference of its arguments,
\beq
{\mathcal S}( \rho, {\pmb r}, {\pmb r}_1) = {\mathcal S}( \rho, {\pmb r} - {\pmb r}_1) = {\mathcal S}( \rho, {\pmb r}_1 - {\pmb r}) \, .
\eeq
Upon comparing (\ref{eq:EaS}) and (\ref{eq:EA}), we arrive at the constraint in uniform matter
\beq
\label{eq:Srho}
\int \text{d} {\pmb r} \, {\mathcal S}( \rho, {\pmb r})
= \int \text{d} {\pmb r} \, {\mathcal S}( \rho, {\pmb r} - {\pmb r}_1 ) = \frac{S(\rho)}{\rho} \, .
\eeq
Given the evidence for a local nature of nuclear energy functional in the energy formula and the above constraint,
it should be possible to approximate
the operator ${\mathcal S}$ for uniform $\rho$ by
\beq
{\mathcal S}( \rho, {\pmb r}) \simeq \frac{S(\rho)}{\rho} \, \delta ({\pmb r}) \, ,
\eeq
in the action of this operator on sufficiently uniform densities $\rho_{np}$.
Also, such an approximation should be valid for densities $\rho$ that are not
uniform generally, but are sufficiently uniform locally, with the operator then given by
\beq
\label{eq:Slocal}
{\mathcal S}( \rho, {\pmb r} , {\pmb r}_1 ) \simeq \frac{S(\rho ({\pmb r}))}{\rho ({\pmb r})} \, \delta ( {\pmb r} - {\pmb r}_1 ) \, .
\eeq
On applying an inverse operator to ${\mathcal S}$ for uniform $\rho$, to the middle and r.h.s.\ expressions in~(\ref{eq:Srho}), we find
a constraint on the inverse operator,
\beq
\label{eq:Srho1}
\int \text{d} {\pmb r} \, {\mathcal S}^{-1}( \rho, {\pmb r})
=  \frac{\rho}{S(\rho)} \, .
\eeq
A consistent approximation for the inverse operator, in matter sufficiently uniform locally,
is
\beq
{\mathcal S}^{-1} ( \rho, {\pmb r} , {\pmb r}_1 ) \simeq \frac{\rho ({\pmb r})}{S(\rho ({\pmb r}))} \, \delta ( {\pmb r} - {\pmb r}_1 ) \, .
\eeq

From \eqref{eq:rhoaexp} and \eqref{eq:Srho1}, it follows that the isovector density in uniform matter can be represented as
\beq
\label{eq:rhoaS}
\rho_a = \frac{a_a^V}{S(\rho)} \, \rho + {\mathcal O} (\eta^2) \, .
\eeq
The isovector density is then seen normalized locally in such a manner that it approaches the  normal density $\rho_0$ when isoscalar density itself approaches $\rho_0$.
Equation (\ref{eq:rhoaS}) is further expected to hold in nonuniform matter that is sufficiently uniform on the scale of nonlocality
in the energy functional.  The contributions from such regions to $A/a_a(A)$ in (\ref{eq:aaaa}) are then
\beq
\label{eq:AaaA}
\frac{A}{a_a(A)} =
\int \text{d} {\pmb r} \, \frac{\rho({\pmb r})}{S(\rho({\pmb r}))} + \ldots \, ,
\eeq
where dots represent contributions from other regions.  In (\ref{eq:AaaA}), the contributions from locally uniform regions
combine as capacitances of independent
capacitors in electrostatics, with $\rho/2S(\rho)$ playing a role of local density of capacitance and $A/2 a_a(A)$ playing the role of the net capacitance, cf.~Eqs.~\eqref{eq:enz1} and~\eqref{eq:mua=}.

One extreme limit that helps to develop an intuition, following the local approximation, is that of a constant symmetry energy, $S(\rho) \equiv a_a^V$.  We find in that limit, as common sense suggests, that $a_a(A) \equiv a_a^V$, cf.~\eqref{eq:AaaA}.  In the context of Eq.~\eqref{eq:aa}, we can see that $a_a(A)$ and~$a_a^V$  are expected to coincide when $a_a^S$ approaches infinity.  As a further implication of $S \equiv a_a^V$, we find, cf.~\eqref{eq:mufrho}, that the net nuclear density becomes identical to the density for symmetric matter up to the 4$^\text{th}$ order in asymmetry not just 2$^\text{nd}$, $\rho({\pmb r}) = \rho_0({\pmb r}) + {\mathcal O}(\eta^4)$.  For the isovector density, cf.~\eqref{eq:rhoaS}, we find $\rho_a^0({\pmb r}) \equiv \rho_0({\pmb r})$ and $\rho_a({\pmb r}) = \rho({\pmb r}) + {\mathcal O}(\eta^2) = \rho_0({\pmb r}) + {\mathcal O}(\eta^2)$.  Since the densities coincide for symmetric matter, $\rho_a^0({\pmb r}) \equiv \rho_0({\pmb r})$, the condition \eqref{eq:break}, signaling a~breakdown of the quadratic approximation to the symmetry energy, is never satisfied.

These $S \equiv a_a^V$ results suggest that, for interactions for which the function $S$ varies weakly with $\rho$, the isoscalar density should exhibit a~weaker variation with asymmetry, than the isovector density.  The result \eqref{eq:rhoaS} suggests that the stronger variation of $S(\rho)$ with~$\rho$, the more pronounced should be the differences between the isovector and isoscalar densities.  Moreover, Eq.~\eqref{eq:AaaA} suggests that the stronger that variation the larger should be the difference between $a_a(A)$ and $a_a^V$.  If that difference were attributed to the nuclear surface, a larger difference would correspond to a reduced value of surface symmetry coefficient $a_a^S$, cf.~\eqref{eq:AaaA}. Finally, the stronger the variation of $S$ with density, the sooner the invariance of the isoscalar density, and then of isovector density as well, should break down.  Indeed, in a~region where the local approximation holds, we find, from \eqref{eq:mufrho},
\beq
\label{eq:rhofmu}
\frac{\delta E_0}{\delta \rho({\pmb r})} = \mu_0 + \eta^2 \, a_a(A) \left[ \frac{A}{a_a(A)} \frac{\text{d} a_a}{\text{d}A} + \frac{a_a(A)}{S(\rho_0({\pmb r}))} \left(1 - \frac{\rho_0({\pmb r})}{S(\rho_0({\pmb r}))}\frac{\text{d}S}{\text{d}\rho}\right) - 1 \right] + {\mathcal O}(\eta^4) \, .
\eeq
From \eqref{eq:rhofmu},
it appears that, simultaneously, a deviation of the local symmetry coefficient $S$ from the global coefficient $a_a$, local dependence of $S$ on $\rho$ and the dependence of $a_a$ on $A$  (itself induced by the dependence of~$S$ on~$\rho$), all conspire to produce a significant coefficient of quadratic expansion for the isoscalar density around a given location~${\pmb r}$.  Further evidence, for a~reduced range of validity of the implications of the second-order $E_a$-expansion, for a stronger  $\rho$-variation of $S$, is the fact that the enhanced differences between the isovector and isoscalar densities should lead to an earlier breakdown of the condition \eqref{eq:break}, whether when staying at a given location in the matter and increasing $\eta_V$ or when departing from nuclear surface at a fixed $\eta_V$.

While the $\rho_0$-parameter $L$, Eq.~\eqref{eq:Sexp}, is commonly used to quantify the $\rho$-dependence of $S$, obviously variation of $S$ over a wider range of subnormal densities must impact the properties of matter associated with the asymmetry.  In the local expressions \eqref{eq:rhoaS} and~\eqref{eq:AaaA}, it is seen that the inverse $S$ is weighted with density.  In the practical calculations of half-infinite nuclear matter of the next section, we find that local values of $S$ impact the isovector density and $a_a(A)$ within the isoscalar density range of $\rho_0/4 \lesssim \rho \lesssim \rho_0$.  For positions in the self-bound matter where $\rho \lesssim \rho_0/4$, the operator ${\mathcal S}$ becomes delocalized and, thus, any potential impact of the specific values of $S(\rho)$ gets smeared out.  Since any realistic $S(\rho)$ eventually tends to zero as $\rho \rightarrow 0$, a negative value of $L$ at $\rho_0$, see Fig.~\ref{fig:syme}, leads in practice to an $S$ with a weaker variation with density over the pertinent region, than a~similar in magnitude but positive $L$.

Selected results, of the local approximation pursued here, have been arrived at in the past following the local Thomas-Fermi approximation~\cite{PhysRevC.67.034305,Danielewicz:2003dd,Steiner:2004fi}.  Specifically, Bodmer and Usmani~\cite{PhysRevC.67.034305} (see also~\cite{Danielewicz:2003dd}) have demonstrated, in the absence of Coulomb effects, an inverse proportionality of $\rho_{np}$ to the symmetry energy at a local net density.  In \cite{Danielewicz:2003dd} (see also \cite{Steiner:2004fi} by Steiner {\em et al.}), the ratio $A/a_a(A)$, i.e.\ up to a factor the capacitance of a nuclear system for asymmetry, was found to be given by the integral on the r.h.s.\ of Eq.~\eqref{eq:AaaA}.

\subsection{Functional Continuity and Kohn-Sham Methodology}

In employing the Hohenberg-Kohn functional, we found it important to invoke a smoothing and interpolation procedure making the resulting functional analytic in nucleonic densities.  If one were to find the Hohenberg-Kohn functional in reality, though, e.g.~progressing over certain directions in the density space, the functional would be defined for discrete particle numbers only and would, otherwise, exhibit definite discontinuities across the density-ranges of interest.  The latter are implied by the abrupt changes in experimental ground-state energies, superimposed on top of a smooth variation of the energies with nucleon numbers.  Those abrupt changes can be, in particular, attributed to filling of single-particle energy shells.  In Kohn-Sham methodology \cite{PhysRev.140.A1133}, the difference between the actual (in most practice hypothetical) Hohenberg-Kohn functional, on one hand, and the kinetic energy of such form as for a noninteracting system, on the other, is represented by a continuous functional of density.  The advantage of the Kohn-Sham methodology is that the assumed functional yields the exact theory in the limit of no interactions.  Otherwise, discontinuities in the approximate functional, due to its kinetic part, on top of what would follow from using discrete particle numbers, may bear similarities to those for the exact Hohenberg-Kohn functional.

The Skyrme-Hartree-Fock model (SHFM), discussed in the next subsection and well-tried in nuclear physics, see e.g.~\cite{Reinhard:1991} and~\cite{Bennaceur:2005mx}, expands the Kohn-Sham methodology, in that the kinetic-energy terms for nucleons in the approximated functional get multiplied by factors dependent on density.  The latter allows for some control over the discontinuities resulting from the functional.  The remaining part of the functional is kept only minimally nonlocal through the use of lowest-order terms in density gradients.  SHFM stems from modeling internucleon forces; a strategy relying from the outset on a Kohn-Sham type functional has been pioneered in nuclear physics by Fayans {\em et al.} \cite{Smirnov88}.

In an application of the SHFM, later in the paper, to the domain of semi-infinite matter at changing asymmetry, it might seem that no smoothing of the approximate energy functional is required to make our preceding considerations, based on a functional expandable in asymmetry  (cf.~Eq.~\eqref{eq:EaS}), directly applicable.  Such a conclusion might be inferred from the fact that the half-infinite matter is characterized by infinite nucleon numbers and by a continuum energy spectrum.  However, it will turn out that, on account of the so-called Friedel oscillations extending from nuclear surface into the matter, the $\eta$-derivatives of nucleon density from functional minimization, of order third and higher, steadily increase with distance away from the surface.  The~implication of this behavior is that the expansion in asymmetry, which we have considered for the functional, must be just asymptotic in the specific domain.  A direct application of our Hohenberg-Kohn considerations to SHFM for half-infinite nuclear matter, without any smoothing, could be then regarded particularly interesting as an extrapolation of our considerations.  On the other hand, regarding the formal issue of smoothing and of functional analyticity, any smoothing that would lead to a smoothing of the resulting densities over a~small range of Fermi wavevectors, would tame the mentioned derivatives with respect to asymmetry.  In~contrast to half-infinite matter, the SHFM functional applied to uniform matter renders fully analytic results for energy without any smoothing.

\subsection{Skyrme-Hartree-Fock Model}

The Skyrme-Hartree-Fock model
gets exploited in this paper in different ways.  On~one hand, the model has been already utilized to illustrate properties of uniform matter.
On the other hand, the model will be utilized in the half-infinite matter calculations.  Results of those calculations will be exploited to verify assertions
concerning the Hohenberg-Kohn energy functional.  Since the SHFM energy functional is expressed in terms of single-particle
wavefunctions~$\lbrace \phi_\alpha \rbrace$, rather than in terms of densities $\rho_n$ and $\rho_p$, the inferences regarding the HK functional
can be nontrivial.

In SHFM, the energy is
\beq
E = E_\text{Skyrme} + E_C + E_\text{pair} \, ,
\label{eq:ESkyrme}
\eeq
where the two last terms $E_C$ and $E_\text{pair}$ are, respectively, the Coulomb and pairing energies.  The Skyrme functional is
\beq
\label{eq:Skyrme}
\begin{split}
E_\text{Skyrme}  \equiv  &  \int \text{d} {\pmb r}  \,  e_\text{Skyrme}({\pmb r}) \\
=  & \int \text{d} {\pmb r}  \Bigg\lbrace  \frac{\hbar^2}{2 m} \, \tau + \frac{t_0}{2} \,\Big[ \Big( 1 + \frac{x_0}{2} \Big)
\, \rho^2 -  \Big( \frac{1}{2} + x_0 \Big) \sum_q \rho_q^2 \Big]
\\
& + \frac{t_3}{12} \, \Big[ \Big( 1 + \frac{x_3}{2} \Big) \, \rho^{\alpha +2} -  \Big(
 \frac{1}{2} + x_3 \Big) \, \rho^\alpha \, \sum_q \rho_q^2 \Big] \\
& + \frac{t_1}{4} \, \Big[ \Big( 1 + \frac{x_1}{2} \Big) \, \rho   \tau - \Big( \frac{1}{2} + x_1 \Big) \, \sum_q \rho_q \, \tau_q \Big] \\
& + \frac{t_2}{4} \, \Big[ \Big( 1 + \frac{x_2}{2} \Big) \, \rho   \tau + \Big( \frac{1}{2} + x_2 \Big)  \, \sum_q \rho_q \, \tau_q \Big]  \\
& + \frac{3 t_1}{16} \, \Big[ \Big( 1 + \frac{x_1}{2} \Big) \, \big( \nabla \rho \big)^2 - \Big( \frac{1}{2} + x_1 \Big) \,
\sum_q \big( \nabla \rho_q \big)^2 \Big] \\
& - \frac{t_2}{16} \, \Big[ \Big( 1 + \frac{x_2}{2} \Big)  \, \big( \nabla \rho \big)^2 +  \Big(\frac{1}{2} + x_2 \Big)  \,
\sum_q \big( \nabla \rho_q \big)^2 \Big]  \\
 & - b_4 \, \rho \, {\pmb \nabla J} - b_4' \, \sum_q \rho_q \, {\pmb \nabla J}_q \Bigg\rbrace \, ,
\end{split}
\eeq
where the $q$-index summations are over neutrons and protons, $\tau = \tau_n + \tau_p$ and ${\pmb J} = {\pmb J}_n + {\pmb J}_p$.  Particle densities
and scaled kinetic energy and spin densities are obtained from summations over the single-particle wavefunctions:
\begin{align}
\label{eq:rhoq=}
\rho_q ({\pmb r}) & = \sum_\alpha  n_\alpha^q \, \phi_\alpha^\dagger ({\pmb r}) \, \phi_\alpha ({\pmb r}) \\
\label{eq:tauq=}
\tau_q ({\pmb r}) &  = \sum_\alpha n_\alpha^q \, {\pmb \nabla} \phi_\alpha^\dagger ({\pmb r}) \, {\pmb \nabla} \phi_\alpha ({\pmb r})  \\
\label{eq:Jq=}
{\pmb J}_q ({\pmb r}) & = \frac{1}{2i} \sum_\alpha n_\alpha^q \, \phi_\alpha^\dagger ({\pmb r}) \, \Big[
\big( \overrightarrow{\pmb \nabla} - \overleftarrow{\pmb \nabla} \big) \times {\pmb \sigma} \Big] \,
\phi_\alpha ({\pmb r}) \,  .
\end{align}
Here, $\lbrace n_\alpha^q \rbrace$ are occupation numbers of the single-particle states.  In less elaborate Skyrme parameterizations, the two spin-parameters are combined into one:
\beq
\label{eq:b_4p}
b_4 = b_4'= t_4/2 \, .
\eeq
The gradient terms in the interaction part of the energy functional (\ref{eq:Skyrme}) (multiplied by adjustable constants), represent a simplified form of interaction nonlocalities.

Minimization
of the SHFM functional (\ref{eq:ESkyrme}), with respect to the wavefunctions,
produces self-consistent equations for the latter
\beq
\label{eq:hq}
(h_q + U_C) ({\pmb r})  \, \phi_\alpha ({\pmb r}) = \epsilon_\alpha  \, \phi_\alpha ({\pmb r}) \, .
\eeq
In the equations, $U_C$ is Coulomb potential and the nuclear single-particle hamiltonian $h_q$ is of the form
\beq
\label{eq:hq=}
h_q ({\pmb r}) = - {\pmb \nabla} \, B_q ({\pmb r}) \, {\pmb \nabla} + U_q ({\pmb r}) - i {\pmb W}_q ({\pmb r}) \left( {\pmb \nabla} \times {\pmb \sigma} \right) \, .
\eeq
The mass potential in $h_q$ is given by
\beq
\begin{split}
B_q = \frac{\hbar^2}{2 m_q^* } = \frac{\hbar^2}{2 m} & +  \frac{t_1}{4} \, \Big[ \Big( 1 + \frac{x_1}{2} \Big) \, \rho
- \Big( \frac{1}{2} + x_1 \Big) \, \rho_q \Big] \\
& +  \frac{t_2}{4} \, \Big[\Big( 1 + \frac{x_2}{2} \Big)  \, \rho + \Big( \frac{1}{2} + x_2 \Big) \, \rho_q  \Big]   \, .
\end{split}
\eeq
and $m_q^*$ is position-dependent effective mass.
The potential $U_q$ in (\ref{eq:hq=}) is
\beq
\label{eq:Uq=}
\begin{split}
U_q  = & {t_0} \,\Big[ \Big( 1 + \frac{x_0}{2} \Big) \, \rho
- \Big( \frac{1}{2} + x_0 \Big) \, \rho_q \Big] \\
& + \frac{t_3}{12} \,   \Big[ \big( 2 + \alpha \big) \Big( 1 + \frac{x_3}{2} \Big) \, \rho^{2}
- \Big(  \frac{1}{2} + x_3 \Big)  \Big( \alpha  \sum_{q'} \rho_{q'}^2 + 2 \rho \, \rho_q \Big) \Big] \, \rho^{\alpha - 1} \\
& + \frac{t_1}{4} \, \Big[  \Big( 1 + \frac{x_1}{2} \Big) \, \tau - \Big( \frac{1}{2} + x_1 \Big) \, \tau_q  \Big] +
\frac{t_2}{4} \, \Big[ \Big( 1 + \frac{x_2}{2} \Big)  \, \tau  + \Big( \frac{1}{2} + x_2 \Big) \, \tau_q \Big] \\
 & - b_4 \, {\pmb \nabla J} - b_4' \,  {\pmb \nabla J}_q \, .
\end{split}
\eeq
Finally, the form factor ${\pmb W}_q$ in (\ref{eq:hq=}) is
\beq
\label{eq:W=}
{\pmb W}_q = b_4 \, {\pmb \nabla} \rho + b_4' \, {\pmb \nabla} \rho_q \, .
\eeq
Above, we suppress the terms in the SHFM functional and single-particle hamiltonian that are normally small, such as those proportional to the square of spin density.

In the context of investigating the effects of neutron-proton asymmetry, we shall introduce average and deviation-from-average quantities, for the terms in the single-particle hamiltonian and for the hamiltonian itself.  These will be denoted, respectively, with either the lack of species index or with that index replaced by 'a'.  (Unlike in the case of density, no special normalization will be adopted here for the quantities with the index 'a'.) Thus, e.g.\ for the mass potential, we will have
\beq
B = \frac{B_n + B_p}{2} =  \frac{\hbar^2}{2 m} + \frac{\rho}{16} \, \big[ 3 t_1 + (5+4 x_2) t_2 \big] \, ,
\eeq
and
\beq
B_a = \frac{B_n - B_p}{2} = \frac{\rho_{np}}{16} \, \big[ (1 + 2 x_2)t_2 - (1 + 2 x_1)t_1 \big] \, .
\eeq
For the effective mass, we will have
\beq
m^* = \frac{1}{4}  \left( \frac{1}{B_n} + \frac{1}{B_p} \right) = \frac{B}{2(B^2 - B_a^2)}
\simeq \frac{1}{2B} \, ,
\eeq
where the approximation holds for low asymmetries,
and
\beq
m_a^* = \frac{1}{4}  \left( \frac{1}{B_n} - \frac{1}{B_p} \right) = - \frac{B_a}{2(B^2 - B_a^2)}
= - \frac{m^* \, B_a}{B} \, .
\eeq
For the optical potential, we will have, one hand, $U=(U_n+U_p)/2$ and, on the other,
\beq
\label{eq:Ua=}
\begin{split}
U_a = & \frac{U_n - U_p}{2} = - \frac{t_0}{4} \, \big(1 + 2 x_0 \big) \rho_{np}
- \frac{t_3}{24} \, \big( 1 + 2 x_3 \big) \rho^\alpha \, \rho_{np} \\
& + \frac{1}{16} \big[ (1+ 2x_2) t_2 - (1+2 x_1) t_1 \big] \tau_{np} - \frac{b_4'}{2} \,
{\pmb \nabla J}_{np} \, ,
\end{split}
\eeq
where $\tau_{np} = \tau_n - \tau_p$ and ${\pmb J}_{np} = {\pmb J}_n -{\pmb J}_p$.

In uniform matter, the kinetic energy densities are
\beq
\label{eq:tauq}
\tau_q = \frac{3}{5} \, \rho_q \, k_{Fq}^2  \, ,
\eeq
with the Fermi wavenumbers given by
\beq
\label{eq:kFq}
k_{Fq} = \left( 3 \pi^2 \rho_q \right)^{1/3} \, .
\eeq
In symmetric matter, the partial densities are equal and, thus,
\beq
\rho_q = \rho/2 \hspace*{2em} \text{and}  \hspace*{2em} \tau_q = \tau/2 \, .
\eeq
With $\tau_q \propto \rho_q^{5/3}$, the partial kinetic energies may be expanded in asymmetric matter around their symmetric value,
\beq
\label{eq:tauqexp}
\tau_q = \frac{3}{10} \rho \, k_F^2 \left[ 1 \pm \frac{5}{3} \frac{\rho_{np}}{2 \rho} + \frac{10}{9} \left( \frac{\rho_{np}}{2 \rho} \right)^2 \pm \ldots \right] \, ,
\eeq
where $k_F$ is wavevector for the symmetric matter at density~$\rho$.
Upon inserting the symmetric values into~(\ref{eq:Skyrme}), the energy per nucleon of symmetric uniform matter becomes
\beq
\frac{E_0}{A} (\rho) =
\frac{3 \hbar^2}{10 m} \, k_F^2 + \frac{3 t_0}{8} \, \rho
+ \frac{t_3}{16} \, \rho^{\alpha+1} + \frac{3}{80} \big( 3 t_1 + 5 t_2 + 4 x_2 \, t_2 \big)
\rho \, k_F^2 \, .
\eeq
For a future use, from \eqref{eq:tauqexp}, the difference in partial kinetic energies may be represented at small asymmetries as
\beq
\tau_{np} \simeq \rho_{np} \, k_F^2 \, .
\eeq
Finally, upon exploiting \eqref{eq:tauqexp} in the energy \eqref{eq:Skyrme} for uniform matter, the density-dependent expansion coefficient in asymmetry becomes
\beq
\label{eq:Srho=}
\begin{split}
S(\rho) =
\frac{\hbar^2}{6 m} \, k_F^2 & - \frac{t_0}{8} \big( 1 + 2 x_0 \big) \rho
- \frac{t_3}{48} \big( 1 + 2 x_3 \big) \rho^{\alpha+1} \\ & - \frac{1}{24}
\big[ 3 t_1 \, x_1 - ( 4 +5 x_2) t_2 \big] \rho \, k_F^2 \, .
\end{split}
\eeq

As is apparent, for uniform matter the same powers of density appear in the symmetry energy $S$ and the energy per nucleon $E_0/A$ of symmetric matter.  From the interaction constants, the dimensionless constants $x_0$, $x_1$ and $x_3$ affect the strength of the three interaction-terms in $S$, with different density powers, without affecting $E_0/A$.   From among the different SHFM quantities, the mass potential has a relatively simple linear density dependence.  As should be obvious from Table \ref{tab:skypro}, the number of Skyrme interaction constants in use in the literature is large and it is, obviously, growing.  For the values of constants, we refer the reader to the references indicated in the table.  Otherwise, one larger compilation of constant values can be found in Ref.~\cite{PhysRevC.68.034324}.

A combination of SHFM with the semiclassical expansion for kinetic energy~\cite{Kirzhnits67} in terms of density,
\beq
\label{eq:kirz}
\tau_q  \approx \frac{3}{5} \left( 3 \pi^2 \right)^{2/3} \,
\rho_q^{5/3} + \frac{1}{36 \rho_q} \left( {\pmb \nabla} \rho_q \right)^2 + \frac{1}{3} \Delta \rho_q \, ,
\eeq
valid for slowly varying densities,
allows one to gain an insight into the symmetry operator ${\mathcal S}$ and its inverse.  For simplicity, we assume the spin densities ${\pmb J}_q$ to be zero.  Upon inserting~(\ref{eq:kirz}) into (\ref{eq:Skyrme}) and employing the identity
\beq
{\mathcal S} (\rho, {\pmb r}_1, {\pmb r}_2) = 2 \left. \frac{\delta^2 E_\text{nucl}}{\delta \rho_{np}({\pmb r}_1) \, \delta \rho_{np}({\pmb r}_2)} \right|_{\rho_{np}=0} \, ,
\eeq
we find for the operator
\beq
\label{eq:SSkyrme}
\begin{split}
{\mathcal S} (\rho, {\pmb r}_1, {\pmb r}_2) \approx & \, {\mathcal S}_0 (\rho, {\pmb r}_1) \, \delta({\pmb r}_1 - {\pmb r}_2) \\ & + \int \text{d}{\pmb r} \, {\mathcal S}_1 (\rho, {\pmb r}) \, \frac{{\pmb \nabla} \rho}{\rho} \cdot \big( \delta({\pmb r} - {\pmb r}_1) \, {\pmb \nabla} \delta({\pmb r} - {\pmb r}_2) + \delta({\pmb r} - {\pmb r}_2) \, {\pmb \nabla} \delta({\pmb r} - {\pmb r}_1) \big) \\
& + \int \text{d}{\pmb r} \, {\mathcal S}_2 (\rho, {\pmb r}) \, {\pmb \nabla} \delta({\pmb r} - {\pmb r}_1)
\cdot {\pmb \nabla} \delta({\pmb r} - {\pmb r}_2) \, ,
\end{split}
\eeq
where
\beq
\label{eq:SiSkyrme}
\begin{split}
{\mathcal S}_0 & =  \frac{S}{\rho} + \frac{1}{18 \rho}
\left( \frac{{\pmb \nabla} \rho }{\rho} \right)^2
\bigg\lbrace \frac{\hbar^2}{m} + \frac{1}{4}  \big[ (2 + x_1) \, t_1  + (2 + x_2) \, t_2  \big] \, \rho \bigg\rbrace \, , \\[.5ex]
{\mathcal S}_1 & = - \frac{1}{72 \rho}  \bigg\lbrace \frac{\hbar^2}{m} + \frac{1}{4}  \big[ (2 + x_1) \, t_1  + (2 + x_2) \, t_2  \big] \, \rho \bigg\rbrace \, , \\[.5ex]
{\mathcal S}_2 & = \frac{1}{72 \rho}  \left\lbrace \frac{\hbar^2}{m} - \frac{1}{16}
 \big[ 35 \, (2 + x_1) \, t_1  + 21 \, (2 + x_2) \, t_2  \big] \, \rho   \right\rbrace \, .
\end{split}
\eeq
Those nonlocal gradient terms in the combination of results (\ref{eq:SSkyrme}) and (\ref{eq:SiSkyrme}), which are multiplied by force constants, are associated with the nonlocality of interactions.  On the other hand, those gradient terms, which are multiplied by $\hbar^2$, are associated with quantum nonlocality.  It may be noted that, for an approximately uniform $\rho$, the operator ${\mathcal S}$ given by~(\ref{eq:SSkyrme}) becomes a~function of $({\pmb r}_1 - {\pmb r}_2)$ only.  In addition, when both $\rho$ and $\rho_{np}$ onto which~${\mathcal S}$ acts approach uniformity, then ${\mathcal S}$ given by~(\ref{eq:SSkyrme}) approaches the local result (\ref{eq:Slocal}).

\section{Hartree-Fock Calculations for Half-Infinite Matter}
\label{sec:nucal}

\subsection{Skyrme-Hartree-Fock Model for Half-Infinite Matter}

\begin{figure}
\centerline{\includegraphics[width=.50\linewidth]{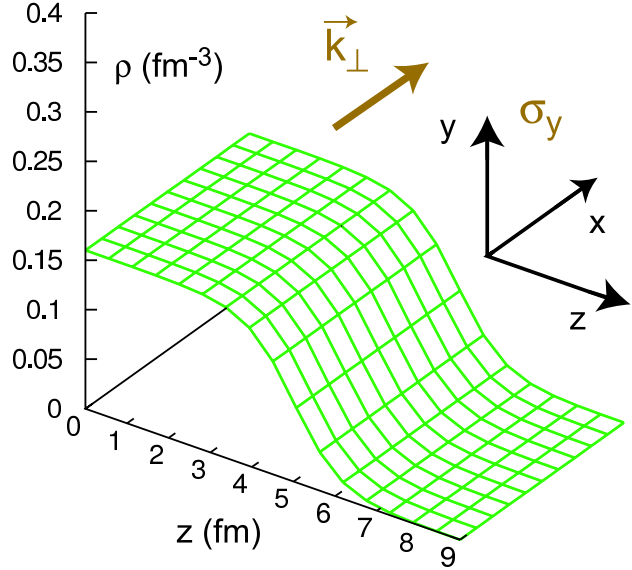}}
\caption{
Within our half-infinite matter SHFM-calculations, the $z$-axis is aligned with the direction of nonuniformity of matter, while the $x$-axis is aligned with the single-particle transverse momentum.  The single-particle particle states are chosen to have a definite projection of spin onto the remaining $y$-axis.
}
\label{fig:xyz}
\end{figure}

With the interest in low-density features of symmetry energy, we now turn to an examination of half-infinite nuclear matter \cite{Kohler:1969,Cote:1978} within SHFM.  The half-inifinite matter is uniform in two cartesian directions and generally nonuniform in the third that we choose to be~$z$.  Given the transverse uniformity, the solutions to the SHFM equations~(\ref{eq:hq}) may be looked for as the eigenstates of transverse momentum, $\phi ({\pmb r}) = \Psi (z) \, \text{exp}(i {\pmb k}_\perp \, {\pmb r}_\perp)$.  Here, ${\pmb k}_\perp$ is the transverse wavevector and $\Psi$ is a spinor depending on~$z$.  In the single-particle hamiltonian~(\ref{eq:hq=}), only the last-r.h.s.\ spin-orbit term has a vector character.  With the form factor~${\pmb W}_q$ of Eq.~(\ref{eq:W=}) pointing in the $z$-direction, that spin-orbit term, acting on a~wavefunction, becomes
\beq
- i {\pmb W}_q ({\pmb r}) \left( {\pmb \nabla} \times {\pmb \sigma} \right) \, \phi ({\pmb r})
= - i {W}_q (z) \left( {\pmb \nabla} \times {\pmb \sigma} \right)_z \, \phi ({\pmb r})
= {W}_q (z) \left( {\pmb k}_\perp \times {\pmb \sigma} \right)_z \, \phi ({\pmb r}) \, ,
\eeq
where
\beq
{W}_q (z) = b_4 \, \frac{\text{d} \rho}{\text{d} z} +  b_4' \, \frac{\text{d} \rho_q}{\text{d} z} \, .
\eeq
If we next orient the $x$-axis along the ${\pmb k}_\perp$-direction, see~Fig.~\ref{fig:xyz}, we find that the spin-orbit term in the hamiltonian becomes proportional to the Pauli matrix along the $y$-direction:
\beq
\label{eq:hphi}
h_q ({\pmb r}) \,  \phi ({\pmb r}) \,
= \bigg(
- \frac{\text{d}}{\text{d}z} \,  \frac{\hbar^2}{2 m_q^* (z)} \, \frac{\text{d}}{\text{d}z}
+ \frac{\hbar^2 \, k_\perp^2}{2 m_q^* (z)}
+ U_q (z) +
{W}_q (z) \, k_\perp \, \sigma_y \bigg) \,  \phi ({\pmb r}) \, .
\eeq
With this, it turns out that the eigenstates of the hamiltonian may be searched for as the eigenstates of operator of spin projection onto the $y$-direction, recognized as the direction of orbital angular momentum calculated relative to the far-away interior of matter:
\beq
\label{eq:phipsichi}
\phi_{{\pmb k}  \lambda q} ({\pmb r}) = \psi_{ k_\perp  k_z  \lambda q} (z) \, \chi_\lambda \, \text{e}^{i {\pmb k}_\perp \, {\pmb r}_\perp}
\, .
\eeq
Here, $\chi_\lambda$ is an eigenstate of $\sigma_y$,
\beq
\label{eq:sigchi}
\sigma_y \, \chi_\lambda = \lambda \, \chi_\lambda \, ,
\eeq
the eigenvalues are $\lambda = \pm 1$, and $\psi (z)$ is a scalar wavefunction.

From (\ref{eq:hq}) and \eqref{eq:hphi}-\eqref{eq:sigchi}, the wavefuctions $\psi$ satisfy
the differential equation
\beq
\label{eq:epsi}
\epsilon_{k q} \, \psi_{ k_\perp  k_z^\infty  \lambda q} (z) =
\bigg(
- \frac{\text{d}}{\text{d}z} \,  \frac{\hbar^2}{2 m_q^* (z)} \, \frac{\text{d}}{\text{d}z}
+ \frac{\hbar^2 \, k_\perp^2}{2 m_q^* (z)}
+ U_q (z) + \lambda \,
{W}_q (z) \, k_\perp  \bigg) \, \psi_{ k_\perp  k_z^\infty  \lambda q} (z) \, .
\eeq
The single-particle energy above can be represented as
\beq
\epsilon_{k q} = \frac{\hbar^2 \, k^2}{2 m_{ q \, \infty}^*} + U_{q \, \infty} \, ,
\eeq
where the quantities with the infinity index refer to a uniform interior of the matter far away from the surface, and where the wavevector squared is
\beq
k^2 = k_\perp^2 + (k_z^{\infty})^{2} \, .
\eeq
The last two equations principally define $k_z^\infty \ge 0$ in terms of $\epsilon$.  The $z$-component of the wavevector must be defined in the asymptotic region in the matter, as the $z$-component of single-particle momentum is generally not conserved, due to matter nonuniformity.  Otherwise, in the asymptotic region, the eigenstates of single-particle hamiltonian combine waves moving towards and away from the boundary of matter, following a reflection from that boundary,
\beq
\label{eq:psi=}
\psi_{ k_\perp  k_z^\infty  \lambda q} (z) \simeq 2 \sin { ( k_z^\infty \, z + \delta_{k_\perp  k_z^\infty  \lambda q}) } \, .
\eeq
Here, the individual plane waves in the combination are normalized to unity and $\delta$ is the phase of the incident relative to the reflected wave.  The phase shift depends on the choice of origin for the $z$-axis.  However, changes in the phase shift with species or with momentum, such as in~${\pmb \nabla_k} \, \delta$, do not depend on the choice of origin.  The arbitrary overall phase-factor multiplying the wavefunction in (\ref{eq:psi=}) is chosen so as to make the wavefunction~$\psi$ real asymptotically.  With the wavefunction and its derivative being real, and with the coefficients in the wave-equation~(\ref{eq:epsi}) being real, the wavefunction $\psi$, after setting the phase-factor, becomes real everywhere.  Far outside of the nuclear surface, in the classically forbidden region, the wavefunction must decrease exponentially with distance away from the surface,
\beq
\label{eq:psi-}
\psi_{{\pmb k} \, q} (z) \propto \text{e}^{-\kappa_{{\pmb k} \, q} \, z} \, ,
\eeq
where we assume that the matter is positioned towards the negative direction of the $z$-axis, and
\beq
\kappa_{{\pmb k} \, q} = \sqrt{ k_\perp^2 - \frac{ 2 m \, \epsilon_{k q}}{\hbar^2} } \, .
\eeq

Provided self-consistent solutions for the half-infinite matter may be found, different observables can be obtained
by carrying out integrations over wavefunction continuum, with the summations over discrete states, such as in (\ref{eq:rhoq=}), specifically replaced by
\beq
\label{eq:igral}
\sum_\alpha (\cdot ) \rightarrow \sum_{\lambda} \frac{1}{(2 \pi)^3} \int \text{d} {\pmb k}_\perp \int_0^\infty
\text{d} k_z^\infty \, (\cdot ) = \sum_{\lambda} \frac{1}{(2 \pi)^2} \int_0^1 \text{d} \cos{\theta} \int_0^\infty
\text{d} k \, k^2 \, (\cdot ) \, ,
\eeq
in accordance with the normalization of the wavefunctions in (\ref{eq:psi=}).  On the r.h.s., the wavevector components are $k_z = k \cos{\theta}$ and $k_\perp = k \sin{\theta}$.  In particular, with the Hartree-Fock occupations of $n_k^q = \theta (k_F^q - k)$, the nucleon and kinetic-energy densities take, respectively, the forms
\beq
\label{eq:rhoqz}
\rho_{q} (z) = \sum_\lambda \frac{1}{(2 \pi)^2} \int_0^1 \text{d} \cos{\theta} \int_0^{k_{F}^q}
\text{d} k \, k^2 \, \left| \psi_{ k_\perp  k_z^\infty  \lambda q} (z) \right|^2 \, ,
\eeq
and
\beq
\label{eq:tauqz}
\tau_{q} (z) = \sum_\lambda \frac{1}{(2 \pi)^2} \int_0^1 \text{d} \cos{\theta} \int_0^{k_{F}^q}
\text{d} k \, k^2 \, \left( k_\perp^2 \, \left| \psi_{ k_\perp  k_z^\infty  \lambda q} (z) \right|^2
+ \left| \frac{\text{d} \psi_{ k_\perp  k_z^\infty  \lambda q} }{\text{d}z} \right|^2
 \right) \, .
\eeq
Given the symmetry of half-infinite matter, only the $z$-component of spin density can be finite.
With~\eqref{eq:Jq=}, \eqref{eq:phipsichi} and~\eqref{eq:sigchi}, cf.\ Fig.~\ref{fig:xyz}, that component takes the form
\beq
\label{eq:Jzz}
J_{z}^{q} (z) = \sum_{\lambda}  \frac{\lambda}{(2 \pi)^2} \int_0^1 \text{d} \cos{\theta} \int_0^{k_{F}^q}
\text{d} k \, k^2 \, k_\perp \, \left| \psi_{ k_\perp  k_z^\infty  \lambda q} (z) \right|^2 \, .
\eeq

\subsection{Numerical Solution of the SHFM Equations}

As the first step, we solve the SHFM problem for nuclear matter for a certain assumed asymmetry $\eta_V$ in the interior of the matter.  Specifically, under the constraint of fixed $\eta_V$, with $\rho_{n,p}^V = \rho_V \, (1 \pm \eta_V)/2$ and employing \eqref{eq:tauq} and \eqref{eq:kFq}, we seek $\rho_V$ that minimizes $E/A$ of Eq.~\eqref{eq:Skyrme} for uniform matter.  That solution provides features of the semi-infinite matter in the asymptotic region.

Given the densities $\rho_{n,p}^V$ and the corresponding Fermi momenta $k_F^{n,p}$, we next discretize the wavevector space.  Specifically, aiming at a calculation of the integrals \eqref{eq:rhoqz}-\eqref{eq:Jzz} for observables, we discretize the space according to the abscissas $\lbrace u_j^M \rbrace_{j=1}^M $ for Gaussian integration in angle,
\beq
\label{eq:cotj}
\cos{\theta_j} = \frac{1}{2} \big( 1 + u_j \big) \, , \hspace*{2em} j=1, \cdots , M \, ,
\eeq
and according to the abscissas $\lbrace t_i^N \rbrace_{i=1}^N$ for Gaussian integration in wavenumber,
\beq
\label{eq:ki}
( k_{iq})^3 = \frac{1}{2} \big( 1 + t_i \big) \, (k_{Fq})^3 \, , \hspace*{2em} i=1, \cdots , N \, .
\eeq
The choices \eqref{eq:cotj} and \eqref{eq:ki} for the discretization are made to provide a uniform coverage of wavevector space, when integrations such as in \eqref{eq:rhoqz}-\eqref{eq:Jzz} are approximated numerically by sums, following
\beq
\label{eq:disc}
\int_0^1 \text{d} \cos{\theta} \int_0^{k_{F}^q}
\text{d} k \, k^2 \, (\cdot ) = \frac{(k_{Fq})^3}{12} \int_{-1}^{1} \text{d}u \, \text{d}t
\, (\cdot )
\simeq \frac{(k_{Fq})^3}{12} \sum_{i,j} w_i^N \, w_j^M \, (\cdot ) \, .
\eeq
Here, $w$ are the respective Gaussian weights.
The wavevector components for the expressions summed over on the r.h.s.\ of~\eqref{eq:disc} are then
\beq
k_{zq}^{ij} = k_{iq} \, \cos{\theta_j} \, ,
\eeq
and
\beq
k_{\perp q }^{ij} = k_{iq} \, \sin{\theta_j} \, .
\eeq
The wavefunction equations \eqref{eq:epsi} are solved self-consistently for the chosen discrete wavevector values, at two spin orientations.

We integrate the differential equations (\ref{eq:epsi}), for $\psi$ at individual wavevectors, by starting out far out from the surface.  Over there, we relate the derivative to wavefunction in accordance with the asymptotic relation (\ref{eq:psi-}), i.e.\ impose the condition
\beq
\label{eq:psip}
\psi'(z) = - \kappa \, \psi(z) \, .
\eeq
 and we move with our solution towards the matter and into its interior.  Formal solutions to the differential equations (\ref{eq:epsi}) generally consist of two components:
\beq
\psi (z) = \psi^- (z) + \psi^+ (z) \, .
\eeq
In the above, the $\psi^-$ component drops in magnitude when moving away from the nuclear surface, according to the asymptotics of Eq.~(\ref{eq:psi-}).  That component represents, up to a~factor, the solution we are looking for.  On the other hand, the $\psi^+$ component grows when moving away from the surface, behaving as
\beq
\psi^+ (z) \propto \text{e}^{+\kappa \, z} \, .
\eeq
Have we moved away from the surface in integration, any inaccuracy in (\ref{eq:psip}) compared to an exact relation for $\psi^-$, or any accuracies in the integration, would admix the $\psi^+$ component to our numerical solution.  The $\psi^+$ component would eventually begin to dominate over $\psi^-$ due to its exponential growth.  However, if we move towards the matter, the opposite takes place, i.e.\ any $\psi^+$ admixture fades away due to its exponential decrease.  For such reasons, in fact, it is not very important to insist on (\ref{eq:psip}) in the starting conditions, as long as any results very close to the starting point are not utilized.  E.g.\ one could start integration with $\psi'=0$.  Due to the linearity of the wavefunction equation, further, the starting magnitude of the wavefunction is not even important, i.e.\ e.g.\ $\psi=1$ could be principally used.  This is because the wavefunction normalization can be reset after the integration arrives into the interior of matter, by imposing the condition
\beq
\label{eq:psinorm}
|\psi_{{\pmb k}} (z) |^2 + \left| \frac{1}{k_z^\infty} \, \frac{\text{d} \psi_{{\pmb k}}}{\text{d}z} \right|^2 = 4 \, ,
\eeq
following (\ref{eq:psi=}).  In practice, because of the Friedel oscillations that will be discussed, which die off slowly in~$z$, we prefer to impose the condition \eqref{eq:psinorm} with the l.h.s.\ in the condition averaged over a distance in $z$ of $\sim \pi/k_F$.

An attractive high-accuracy method for solving second-order differential equations, such as the Schr\"{o}dinger equation, of fifth order in the employed step, is the Cowell-Numerov (CN) method~\cite{Cowell:1910,Numerov:1923,Numerov:1924}.  High order of a method generally allows for large steps in integration preventing accumulation of rounding errors, when numerous small steps need to be taken. The CN method applies directly, though, to the equations where the second derivative is expressed exclusively in terms of the independent variable and the integrated function.  For those Skyrme interactions in Table \ref{tab:skypro}, for which $m^*/m \ne 1$, the position-dependent effective mass in the wavefunction equation (\ref{eq:epsi}) introduces, however, the first-order derivative of wavefunction into the single-particle equation, preventing a direct application of the CN method.  With $B_q(z)= \hbar^/2 m_q(z)$, we specifically get for the derivative term in~(\ref{eq:epsi}):
\beq
\label{eq:difpsi}
\frac{\text{d}}{\text{d}z} \,  B \, \frac{\text{d}}{\text{d}z} \, \psi
=  B \,  \frac{\text{d}^2 \, \psi}{\text{d}z^2} +
 \frac{\text{d} B}{\text{d}z} \,  \frac{\text{d} \psi}{\text{d}z} \, .
\eeq
A straightforward work-around has been put forward by Dobaczewski {\em et al.} \cite{Dobaczewski:1984}.  Thus, on introducing
\beq
\label{eq:doba}
\upsilon_{ k_\perp  k_z^\infty  \lambda q} (z) = B_q^{1/2} (z) \, \psi_{ k_\perp  k_z^\infty  \lambda q}(z) \, ,
\eeq
the r.h.s.\ of (\ref{eq:difpsi}) can be rewritten as
\beq
\label{eq:Bupsilon}
B^{1/2} \, \frac{\text{d}^2 \, \upsilon}{\text{d}z^2} + \frac{1}{2 B^{3/2}} \,
\bigg[ \frac{1}{2} \bigg(  \frac{\text{d} B }{\text{d}z} \bigg)^2 - B \, \frac{\text{d}^2 \, B}{\text{d}z^2} \bigg] \, \upsilon = B^{-1/2} \bigg( \frac{\text{d}^2 \, \upsilon}{\text{d}z^2} - \Delta U \, \upsilon
\bigg) \, ,
\eeq
where
\beq
\Delta U_q (z) = \frac{1}{2} \, \frac{\text{d}^2 \, B_q}{\text{d}z^2} - \frac{1}{4 B_q(z)}
\, \bigg(  \frac{\text{d} B_q }{\text{d}z} \bigg)^2 \, .
\eeq
Upon rewriting of the derivative term in \eqref{eq:Bupsilon}, the first-order derivative disappears from the expression at the cost of the appearance of a term linear in the function, but the latter is not a problem for the CN method.  Upon inserting the derivative term into \eqref{eq:epsi}, the~differential equation for $\upsilon$ takes the form
\beq
\label{eq:eusi}
\epsilon_{k q} \, \upsilon_{ k_\perp  k_z  \lambda q} (z) =
\bigg(
- \frac{\hbar^2}{2 m_q^* (z)} \, \frac{\text{d}^2}{\text{d}z^2} \,
+ \frac{\hbar^2 \, k_\perp^2}{2 m_q^* (z)}
+ U_q^\text{eff} (z) + \lambda \,
{W}_q (z) \, k_\perp  \bigg) \, \upsilon_{ k_\perp  k_z  \lambda q} (z) \, ,
\eeq
where
\beq
U_q^\text{eff} (z) = U_q  (z) + \Delta U_q (z) \, .
\eeq
The equation for $\upsilon$ above can now be solved with the CN method and $\psi$ can be obtain from~$\upsilon$ following \eqref{eq:doba}.  The asymptotic $z \rightarrow \pm \infty$ behaviors are of the same type for $\upsilon$ as for $\psi$, as $B$ approaches a constant either inside or outside of the matter.

In the discretization of wavevector space, we typically utilize abscissa numbers of $M \sim 14$ and~$N \sim 28$.  The~typical spatial step that we employ, when integrating the wavefunction equations with the CN method, is~$\Delta z = 0.02 \, \text{fm}$.  We start from a distance~$z_\text{max}$ displaced from the matter surface position of~$z_0$ by~$z_\text{max} - z_0 \simeq 14 \, \text{fm}$ and we continue the integration of the wavefunction equations down to a position $z_\text{min}$ in the matter, away from the surface by $z_0 - z_\text{min} \simeq 16 \, \text{fm}$.  After each solution pass, the local particle and kinetic energy densities, of Eqs.~\eqref{eq:rhoqz}~and~\eqref{eq:tauqz}, are updated, with the integrals~\eqref{eq:igral} in the densities represented in terms of the summations~\eqref{eq:disc}.

For initialization of the solution iterations, starting densities must be assumed and we employ the combination of Fermi shapes:
\beq
\rho_{n,p}(z) = \frac{1}{2} \big[ \rho^F(z) + \overline{\rho_{np}^F}(z) \big] \, ,
\eeq
where
\beq
\rho^F (z) = \frac{\rho_V}{1 + \text{exp}\big[(z-z_0)/d\big]} \, ,
\eeq
and
\beq
\label{eq:orho}
\overline{\rho_{np}^F}(z) = \text{sgn}{(\rho_{np}^V)} \, \text{min}{(|\rho_{np}^F(z)|,\rho(z))} \, ,
\eeq
with
\beq
\rho_{np}^F (z) = \frac{\rho_n^V - \rho_p^V}{1 + \text{exp}\big[(z-z_0-\Delta R_a)/d_a\big]} \, .
\eeq
In the above,
$z_0$ is the starting position of nuclear surface, $d$ is the diffuseness of isoscalar density, $\Delta R_a$ is the displacement of isovector density relative to the isoscalar density, and $d_a$ is the diffuseness of isovector density.  The role of Eq.\ \eqref{eq:orho} is to ensure that the magnitude of asymmetry does not exceed 1 locally. The initial kinetic energy densities for the wave equations are estimated from the densities using Eq.~\eqref{eq:kirz}.  The spin densities are at first assumed to vanish~\cite{Cote:1978} and are thereafter found to rise to finite values during iterations.  The surface diffusness is first taken in the initialization equal to $d \sim 0.55 \, \text{fm}$.  After self-consistency is first reached for symmetric matter, the~value of $d$ is read off from the obtained $\rho(z)$ and that value is then used to initialize a~restarted set of iterations for the given interaction.  For calculations of asymmetric matter, the initial values used are $\Delta R_a = 0.6 \, \text{fm}$ (cf.\ Table \ref{tab:skypro}) and $d_a = d$.  After the self-consistency is first reached for a small asymmetry $\eta_V$, the values of $\Delta R_a$ and $d_a$ are read off from the small-$\eta_V$ solution and used subsequently in the initialization of all other calculations of asymmetric matter for the specific interaction.

Within iterations for self-consistency, an~attempt to use directly output densities from an $I$'th solution iteration, $\big[\rho \big]_\text{out}^I$ and $\big[\tau \big]_\text{out}^I$, as input densities to the ($I$+1)'th iteration, generally introduces problems of two types.  One problem is an undesired gradual shift of the solution in the direction outside of matter, towards a boundary of the computational region.  Another problem are instabilities.  To remedy both problems while optimizing convergence, as an input to the subsequent iteration, we employ, see also \cite{Cote:1978}, a linear combination of the preceding input and of the shifted output,
\beq
\big[ \rho_q (z) \big]_\text{in}^{I+1} = \beta (z) \, \big[ \rho_q (z + u_I) \big]_\text{out}^{I}
+ \big[ 1 -  \beta (z)\big] \, \big[ \rho_q (z) \big]_\text{in}^{I} \, ,
\eeq
with the same $\beta(z)$ and $u_I$ for $\rho_q$ and $\tau_q$.  In the practice of setting the value of $u_I$, we found the requirements of either an unchanged nucleon number within the computational region, or of a fixed position for the net density reaching half of its asymptotic value, to be comparably effective in speeding up the iteration convergence and in preventing the undesired drift.  The~factor $\beta$ has been made to vary smoothly across the computational region, and to reach larger values, up to 0.4, outside of the matter and smaller inside the matter.  For two of the Skyrme parameterizations, Skz0 and Skzm1 \cite{PhysRevC.66.014303}, we could not arrive at consistency at finite asymmetry, no matter what $\beta(z)$ we employed.  For any of the Skyrme parametrizations, the~use of an excessive $z_0 - z_\text{min}$ might induce long-wavelength instabilities in the solution, inside the matter, evidenced by a convergence of the solution iterations, but to slightly different density shapes $\rho_{n,p}(z)$ in the matter, depending on the step $\Delta z$ and on the starting position~$z_\text{min}$.  Except for the immediate vicinity of the boundaries of the computational region, the solutions for interactions represented in Table \ref{tab:skypro}, for our choices of $z_\text{min,max}$ and~$\Delta z$, are stable with respect to variations in the boundaries and in step magnitude.  When comparing to each other our solutions for different interactions and different asymmetries in the next section, we reposition the solutions, so that the location where the net density drops to half of its asymptotic value does not change from one solution to another, $\rho(z_0) = \rho_V$/2.

\subsection{Role of the Spin-Orbit Coupling}

\begin{figure}
\centerline{\includegraphics[width=.85\linewidth]{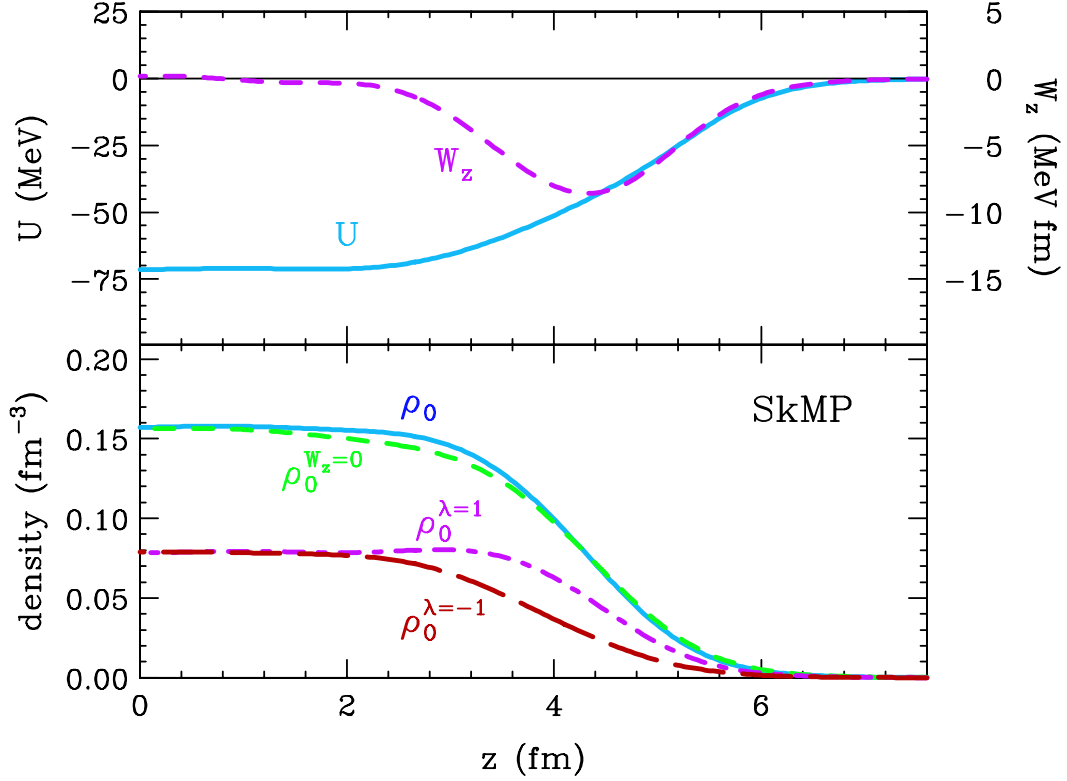}}
\caption{Results of reaching self-consistency in symmetric half-infinite matter for the SHFM equations with the SkMP interaction, shown as a function of position along the axis perpendicular to the surface.  The top panel displays the central potential $U$ and the spin-orbit form factor~$W_z$.  The~bottom panel displays the net nucleon density $\rho_0$ and partial densities of nucleons with spin parallel $\rho_0^{\lambda=1}$ and antiparallel $\rho_0^{\lambda=-1}$ to orbital momentum.  Also displayed is the net density $\rho_0^{W_z=0}$ obtained when the spin-orbit potential is put to zero, $W_z=0$.
}
\label{fig:uwan}
\end{figure}

Figure \ref{fig:uwan} shows the typical central potential $U$, spin-orbit form factor $W_z$ and net density~$\rho$, arrived at in the course of solving the SHFM equations until consistency, in symmetric half-infinite nuclear matter.  As seen in the figure, the density $\rho$ and potential $U$ exhibit different diffusenesses, with $\rho$ being steeper.  Otherwise, consistently with \eqref{eq:W=} and with general physical requirements, the~spin-orbit form factor $W_z$ is finite only in the surface region.  With this, it can affect macroscopic features of the nuclear surface.

Quantitatively, with $k_\perp \lesssim k_F \approx 1.33 \, \text{fm}^{-1}$, the spin-orbit term $k_\perp \, W_z$ is small in the wavefunction equation \eqref{eq:epsi}, lower by as much as a factor of $\sim 4$ or more at the maximum of its magnitude compared to the central potential $U$, see Fig.~\ref{fig:uwan}.  The sign of $W_z$ makes the spin-orbit potential attractive for spins parallel to the orbital angular momentum, $\lambda=+1$, and repulsive for antiparallel, $\lambda = - 1$.  On account of the potential, nucleons with $\lambda=+1$ get, on the average, displaced outward relative to the nucleons with $\lambda=-1$, as illustrated in terms of the respective densities in Fig.~\ref{fig:uwan}.  This displacement, in turn, produces a positive spin density~$J_z$ in the surface region, cf.~\eqref{eq:Jzz}.  A finite spin density modifies the central potential which involves a divergence of that density, see Eq.~\eqref{eq:Uq=}.  With $b_4 + b_4'/2 > 0$ for a Skyrme parameterization, the correction to the central potential is attractive in the more inner part of the surface and repulsive in the outer.  This steepens the central potential and, in consequence, makes the net density steeper in the surface region compared to the case when the spin-orbit term is ignored, see Fig.~\ref{fig:uwan} where the density arrived at without the spin-orbit term is also shown.

A steeper profile of net density due to the spin-orbit term has been first predicted by Stocker, within a variational consideration \cite{Stock70}.  Details behind his prediction, however, differ from our results.  Thus, he has anticipated a steeper density-profile for $\lambda=-1$ than $\lambda=1$ nucleons, while we find the opposite to hold, see Fig.~\ref{fig:uwan}.  We next turn to the systematics of density with changing asymmetry.  We will come back to the role of spin-orbit coupling in the surface region in the context of coefficients characterizing surface energy.

\subsection{Densities for Different Asymmetries}
\label{ssec:Skys}

\begin{figure}
\centerline{\includegraphics[width=.76\linewidth]{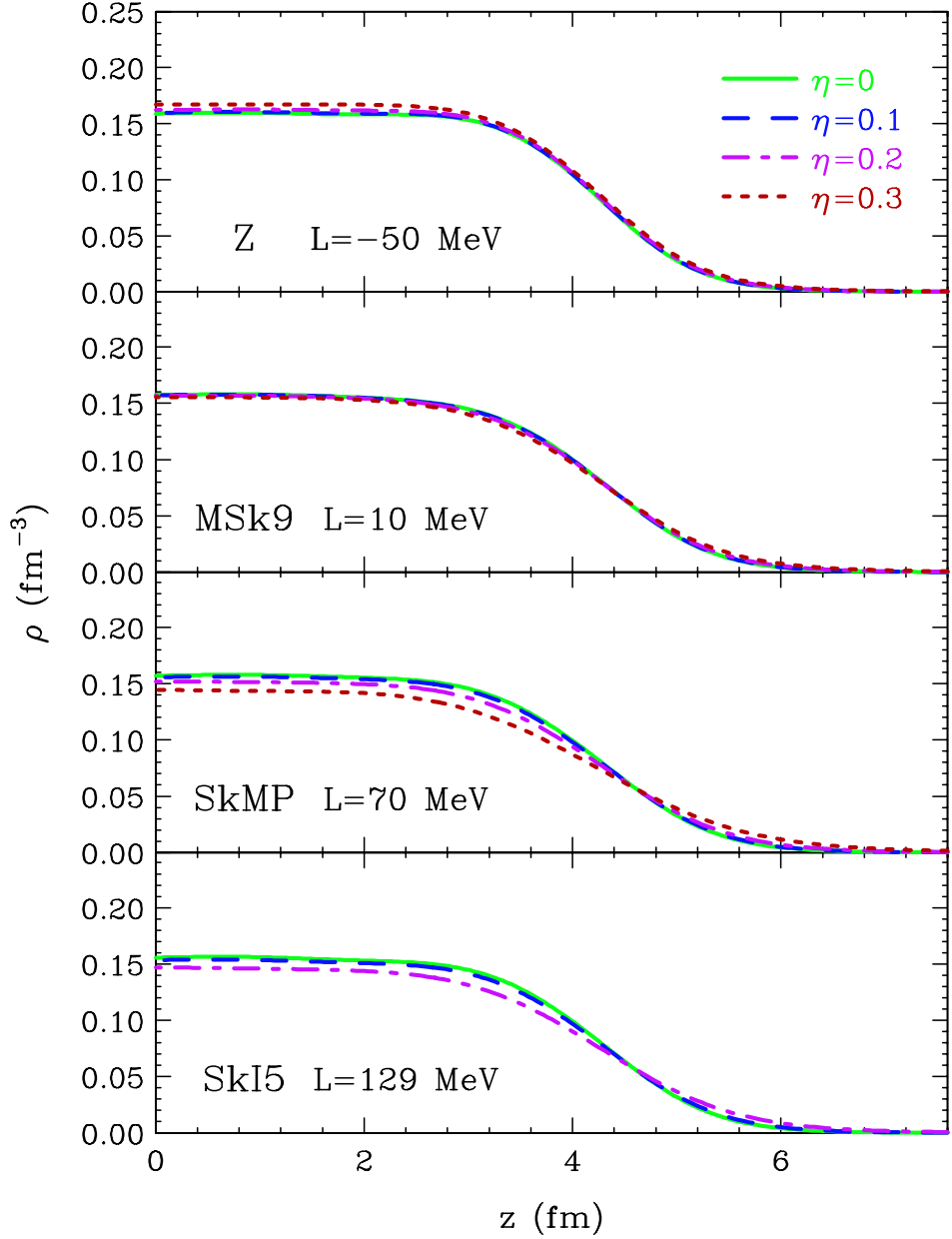}}
\caption{Net density profiles at different asymmetries in half-infinite nuclear matter within SHFM, for sample interactions.  For the SkI5 interaction, the asymmetry of 0.3 is already past the drip value of $\eta_V^d = 0.288$ and the corresponding profile is not shown.  The $\eta=0$ SkMP density is the same as the $\rho_0$-density in Fig.~\ref{fig:uwan}.
}
\label{fig:density}
\end{figure}

\begin{figure}
\centerline{\includegraphics[width=.76\linewidth]{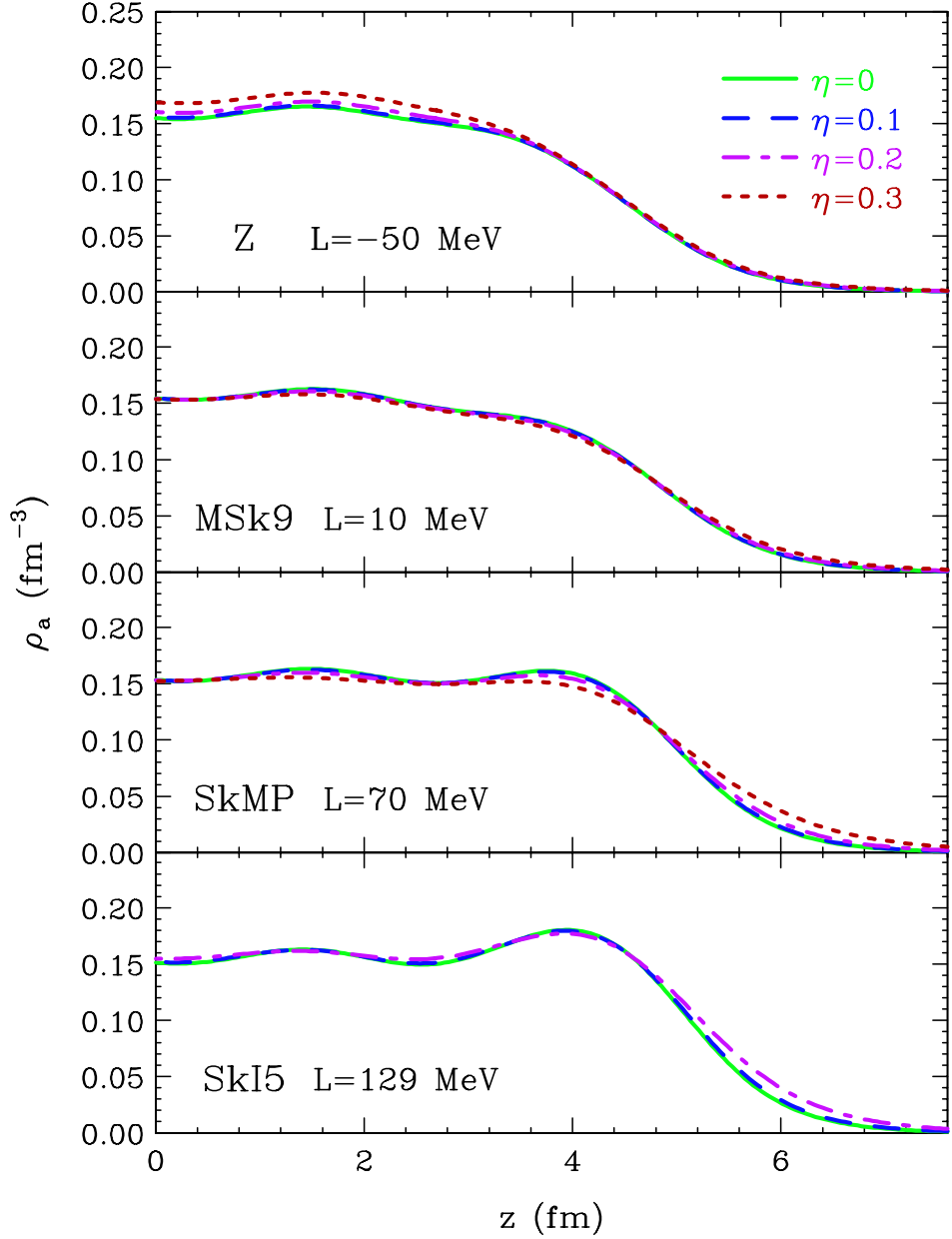}}
\caption{Profiles of the isovector density, Eq.~\eqref{eq:rhoa=}, at different asymmetries in half-infinite nuclear matter within SHFM, for sample interactions.  For the SkI5 interaction, the asymmetry of 0.3 is already past the drip value of $\eta_V^d = 0.288$ and the corresponding profile is not shown.  The abscissa scale is numerically identical to that in Fig.~\ref{fig:density}.
}
\label{fig:densvec}
\end{figure}

In Section
\ref{sec:enfu}, we have indicated that, in consequence of the quadratic dependence of $E_a$ on $\rho_{np}({\pmb r})$, the normalized difference of nucleon densities for a given mass number, $\rho_{a(A)}({\pmb r})$, as well as the net density, $\rho ({\pmb r})$, should both weakly depend on asymmetry.  Figures~\ref{fig:density} and~\ref{fig:densvec} show, respectively, the isoscalar and isovector densities from solving the SHFM equations for half-infinite matter at several $\eta$, for sample Skyrme parameterizations.  It is apparent in the figures that the anticipated scaling is generally very well satisfied.
The~scaling behavior holds even at $\eta =0.3$ for the MSk9 and SkMP interactions, when the drip asymmetries for those interactions are close-by, at $\eta_V^d = 0.339$ and $0.336$, respectively.  For the SkI5 interaction, the asymmetry $\eta = 0.3$ is already above the drip asymmetry of 0.288 and the corresponding densities are not displayed in the figures.
A closer examination of the densities in Figs.~\ref{fig:density} and~\ref{fig:densvec} reveals that, in the interior of matter, the scaling worsens a bit with an increase in the {\em magnitude} of~$L$ and, in the density tails, the scaling worsens with an increase in the {\em value} of $L$.  These findings are consistent with the expectations developed in Subsection~\ref{ssec:local}.  Regarding implications of Eq.~\eqref{eq:rhofmu}, the overall symmetry coefficient for half-infinite matter is identical to that for the volume, $a_a \equiv a_a^V$.  Consistently either then with \eqref{eq:rhofmu} or with \eqref{eq:rhomin}, in the interior of the matter in Fig.~\ref{fig:density}, the net density goes up or down depending on the sign of $L$, to the order of $\eta^2$.  Further, consistently with the expectations developed in the context of the local approximation in Subsection~\ref{ssec:local}, in the density tails, the scaling is seen to be a bit worse satisfied for the isovector densities in Fig.~\ref{fig:densvec}, than for the isoscalar densities in Fig.~\ref{fig:density}.  Nonetheless, one can claim that, in the end, the scaling for the isovector densities is more impressive than for isoscalar.  This is because the isoscalar densities are rather bland across the interactions and they are forced to take on a value of $\rho_V/2$ at $z_0$.  In contrast to the isoscalar,
the isovector densities exhibit much variation across the interactions and, for each interaction, the scaling needs to be exclusively due to an approximately invariant interdependence between the isovector and isoscalar densities and cannot be attributed to any imposed auxiliary condition.

\begin{figure}
\centerline{\includegraphics[width=.76\linewidth]{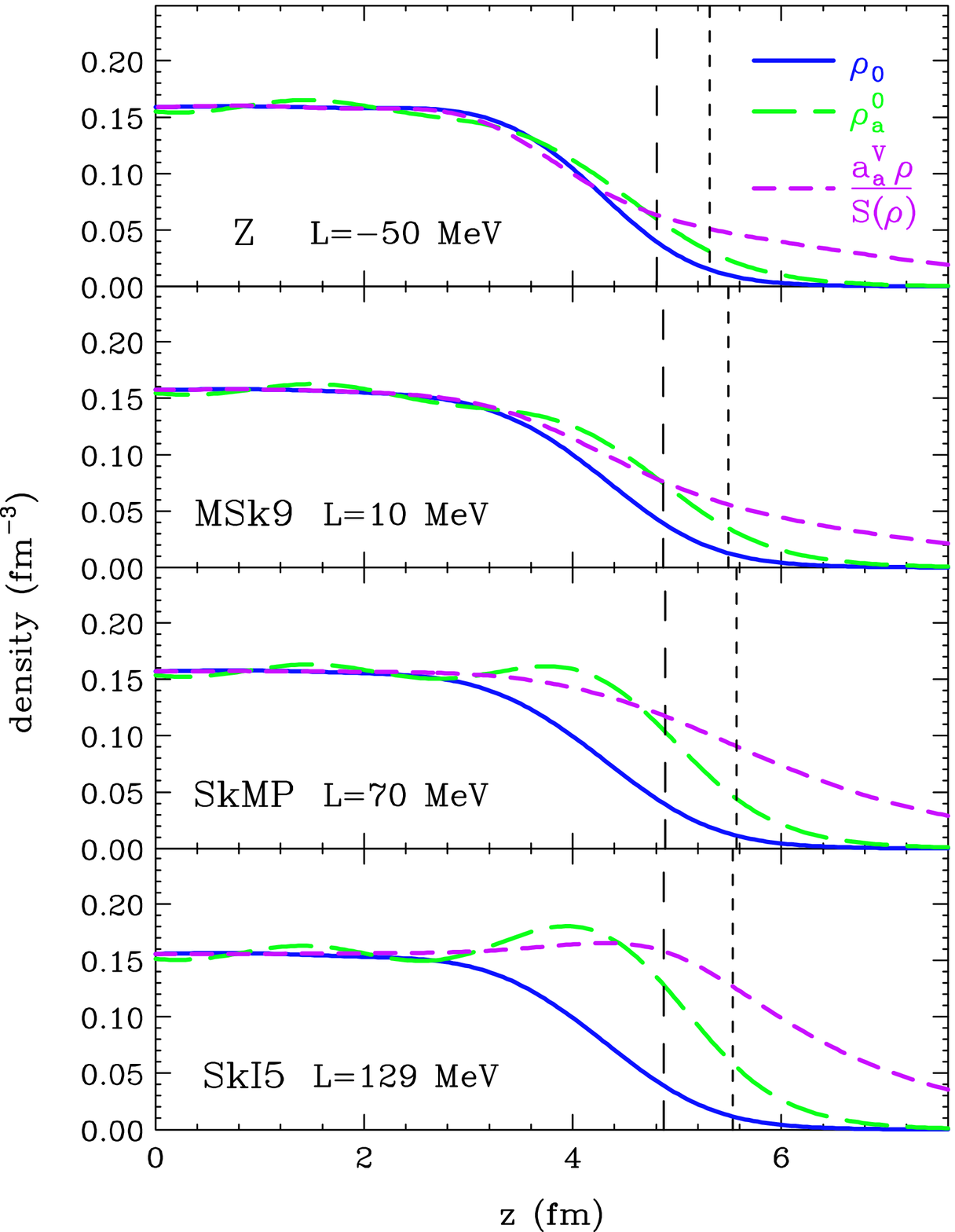}}
\caption{Comparison of isoscalar and isovector densities, and of the local approximation to the isovector density, in symmetric half-infinite nuclear matter within SHFM, for sample interactions.  The~longer- and shorter-dashed vertical lines for the specific interaction indicate, respectively, the~location where the net density is equal to the quarter of normal density and the location of a classical return point for the Fermi wavevector directed along the $z$-axis, when the spin-orbit potential is disregarded.
}
\label{fig:denscomp}
\end{figure}

Figure \ref{fig:denscomp} next compares the isoscalar and isovector densities in symmetric half-infinite matter, for the same sample interactions.  For each interaction, the isovector density is generally pushed out relative to the isoscalar density.  Amongst the interactions, for the Z interaction with the most negative $L$, though,  the two densities are rather close to each other.  As~$L$ increases, the densities separate more and more, with the most pronounced differences between these densities appearing for the SkI5 interaction in the figure.  The~growing difference with growing~$L$ again conforms with the expectations developed in Subsection \ref{ssec:local}.

Given the qualitative utility of the local approximation, it can be of interest to test that approximation quantitatively.  For that purpose, in addition to the already displayed $\eta=0$ isoscalar and isovector densities, we plot in Fig.~\ref{fig:denscomp} the local approximation to the isovector density, of Eq.~\eqref{eq:rhoaS}.  It can be seen that the actual isovector density $\rho_a^0(z)$ oscillates around the local approximation at the higher isoscalar densities, down to the net density reaching about a quarter of the normal density, $\rho_0/4$, for the interactions in the figure.  At~lower net densities in the matter, the actual density and the approximation separate from each other, with the local approximation thereafter strongly overestimating $\rho_a^0(z)$.  It follows that, for arguments ${\pmb r}$ of the SHFM-operator ${\mathcal S}$ such that $\rho({\pmb r}) < \rho_0/4$, this operator becomes significantly nonlocal.

Increasingly nonlocal character of the operator ${\mathcal S}$, with a drop of density in the surface, is~expected on the basis of the result \eqref{eq:SSkyrme}.  In that result, see \eqref{eq:SiSkyrme}, the gradient corrections proportional to $t_1$ and $t_2$ represent effects of the range of interaction, while the corrections proportional to~$\hbar^2$ represent quantal nonlocalities.  It is seen in the expressions \eqref{eq:SiSkyrme} that the $\hbar^2$ correction-terms will dominate over the finite-range terms at sufficiently low densities and eventually also overwhelm the zeroth-order term.  Overall, it appears that the nonlocalities enhance the operator ${\mathcal S}$ in the far-out surface region making its inverse reduced and producing physically expected exponential fall-off for the isovector density $\rho_a$.

The close proximity of the isovector density $\rho_a$ to the local approximation, for the wide range of higher densities, opens up a chance of determining details in the density dependence of symmetry energy, from the systematics of proton densities alone.  Correspondingly, it~becomes of interest to understand better both the agreement of isovector density with the local expectation and the deviations.  For this, we turn, in the following, to the Wentzel-Kramers-Brillouin-Jeffreys (WKBJ) approximation for the single-particle wavefunctions.

\subsection{WKBJ Analysis}

The wavelength for the oscillations of isovector density in Fig.~\ref{fig:denscomp}, around the local approximation, can be easily recognized as that expected for the Friedel oscillations \cite{Friedel54}, $\lambda = \pi/ k_F \simeq \pi/(1.33 \, \text{fm}^{-1}) = 2.36 \, \text{fm} $.  Corresponding oscillations can also be detected in the isoscalar density, but they are there of a considerably lesser amplitude than in the isovector density.  Friedel oscillations generally arise when a disturbance of the system, forcing nonuniformity, such as the surface, acts to synchronize the density oscillations for individual component wavefunctions.  Other than generating the oscillations, the specific synchronization of components due to the surface acts also to slow down the rise of density with distance away from the surface, when moving into the interior of matter.  The~Friedel oscillations are much stronger for the isovector density, due to the  dominant contribution of the states near the Fermi surface, to the difference of neutron and proton densities.

To understand the oscillations mathematically, we invoke approximate WKBJ solutions to the SHFM equations \eqref{eq:epsi}.
In the classically allowed region, where the equation
\beq
\label{eq:wkben}
\epsilon_{k q} = \frac{\hbar^2 \, k_{ k_{\perp} \lambda q}^{z2} (z) + \hbar^2 \, k_{\perp}^2 }{2 m_q^*(z)} + U_q(z) + \lambda \, W_q(z) \, k_\perp \, ,
\eeq
has real solutions $k^z (z)$, the WKBJ wavefunctions are of the form
\beq
\psi_{ k_\perp  k_z^\infty  \lambda q} (z) \simeq 2 \sqrt{ \frac{v_{z}^{ \infty}}{v_z(z)} } \,
\sin{ \left( \int_{z}^{z_{k_\perp  k_z^\infty  \lambda q}} \text{d}z' \, k_{ k_{\perp}  k_z^\infty  \lambda q}^{z} (z') + \frac{\pi}{4} \right)  } \, .
\eeq
In the above, $z_{k_\perp  k_z^\infty  \lambda q}$ is the classical return point ending the classically allowed region that is assumed to extend here up to infinity in the negative $z$-direction, and $v_z$ is the $z$-component of the velocity, $v_z = \hbar^{-1} \partial \epsilon/\partial k_z = \hbar k_z/m^*$.  In the classically forbidden region, where \eqref{eq:wkben} has purely imaginary solutions, the WKBJ wavefunctions are of the form
\beq
\label{eq:psiforb}
\psi_{ k_\perp  k_z^\infty  \lambda q} (z) \simeq \sqrt{ \frac{v_{z}^{ \infty}}{|v_z(z)|} } \,
\text{exp} \left( - \int_{z_{k_\perp  k_z^\infty  \lambda q}}^z \text{d}z' \, |k_{ k_{\perp}  k_z^\infty  \lambda q}^{z} (z')| \right) \, .
\eeq
Those wavefunction forms actually apply some distance away from the return point.  Deep into the forbidden region, the wavefunctions are exponentially depressed and, in the considerations below, we will just put them to zero beyond the return point. With this, the density of nucleons~$q$, at position~$z$, becomes
\beq
\label{eq:rhoqwkb0}
\begin{split}
\rho_q(z) & \simeq  \frac{4}{(2 \pi)^3} \sum_{\lambda} \int \text{d}k_{z}^{ \infty} \, \text{d} {\pmb k}_\perp \, \frac{v_{z}^{ \infty}}{v_z(z)} \, \sin^2{\left( \cdots \right)} \\
& =  \frac{4}{(2 \pi)^3} \sum_{\lambda} \int \text{d}k_{z} \, \text{d} {\pmb k}_\perp \,  \sin^2{\left( \cdots \right)}  \\
& = \frac{2}{(2 \pi)^3} \sum_{\lambda} \int \text{d}k_{z} \, \text{d} {\pmb k}_\perp \,
\left[ 1 + \sin{ \left( 2 \int_{z}^{z_{k_\perp  k_z^\infty  \lambda q}} \text{d}z' \, k_{ k_{\perp}  k_z^\infty  \lambda q}^{z} (z') \right) }  \right] \, .
\end{split}
\eeq
In obtaining the second to the last expression, we have converted the wavevector integration from one over the  asymptotic to one over the {\em local} wavevector-components.  It should be noted that those wavevector integrations are confined to a half of the Fermi sphere.  In~arriving at the last expression, we have used the identity $\sin^2{(\alpha + \frac{\pi}{4})}= \frac{1}{2} + \frac{1}{2} \sin{(2 \alpha)}$.
Effects of the spin-orbit term on nucleon densities cancel out to the lowest order.  For that reason and also to make the local Fermi sphere isotropic, those effects will be further disregarded.  In consequence, the density of nucleons~$q$ becomes
\beq
\label{eq:rhowkb}
\begin{split}
\rho_q(z) & \simeq  \frac{4}{(2 \pi)^3} \int_{k < k_{F q}(z)} \text{d} {\pmb k} \, \left[ 1 + \sin{ \left( 2 \int_{z}^{z_{k_\perp  k_z^\infty  q}} \text{d}z' \, k_{ k_{\perp}  k_z^\infty   q}^{z} (z') \right) }  \right] \\
& = \frac{k_{Fq}^3(z)}{3 \pi^2} + \frac{1}{2 \pi^3} \int_{k < k_{Fq}(z)} \text{d}{\pmb k} \,
\sin{ \left( 2 \int_{z}^{z_{k_\perp  k_z^\infty  q}} \text{d}z' \, k_{ k_{\perp}  k_z^\infty   q}^{z} (z') \right) } \, .
\end{split}
\eeq
With the disregard of the spin-orbit term, the local Fermi wavevectors in \eqref{eq:rhowkb} follow from
\beq
\label{eq:muq}
\mu_q = \frac{\hbar^2 \, k_{Fq}^2(z)}{2m_q^*(z)} + U_q(z) \, .
\eeq
If we were to consider the general case of mass potential $B$, we would need to decompose the local kinetic energy $\tau_q$ in a similar manner to~\eqref{eq:rhowkb}.  However, since our primary goal is to gain an insight, we shall confine ourselves to the case of $m^* \equiv m$, avoiding the need for specific decomposition.

Mathematically, the Friedel oscillations can be tied to the sine term on the r.h.s.\ of \eqref{eq:rhowkb} or \eqref{eq:rhoqwkb0}.  Integration over wavevector components for that term leads to an averaging of the sine over argument values.  While, in general, wide-range variations of the argument will make that average approach zero, any residue of averaging is likely to reflect possible nonuniformities in the integration, as far as values of the sine argument are concerned.  An~obvious discontinuity is the termination of integration at $k_{Fq}$, producing oscillations of the form $\sin{(2 k_{Fq} \, z + \delta )}$.  Otherwise, the local amplitude of oscillations depends on the distance from the surface.  The larger the distance, the faster will be the variation of the argument of sine in~\eqref{eq:rhoqwkb0}, with the variation of wavevector, and the more suppressed will be the contribution of the sine to a~density.  Furthermore, different amplitudes will emerge for different types of densities.  Thus, isoscalar density involves 3-dimensional integrals over parameters affecting the sine argument, cf.~\eqref{eq:rhowkb}.  However, isovector density involves differentiation of nucleonic densities with respect to asymmetry, cf.~\eqref{eq:rhoadef}, which produces a~sine-function integral reduced in the dimension to~2. In effect of the reduced averaging of the sine function, the~Friedel oscillations turn out to be significantly larger, and they fall off slower with distance, in the isovector than in the isoscalar density.

Considering the vicinity of zero asymmetry for the functions, we now represent nucleon densities as
\beq
\label{eq:rhoqg}
\rho_q(z) = \frac{k_{Fq}^3(z)}{3 \pi^2} \big(1+ {\mathcal G}_q(z)\big) \, ,
\eeq
where
\beq
{\mathcal G}_q(z) \simeq \frac{3}{2 \pi \, k_{Fq}^3(z)} \int_{k < k_{Fq}(z)} \text{d}{\pmb k} \,
\sin{ \left( 2 \int_{z}^{z_{k_\perp  k_z^\infty  q}} \text{d}z' \, k_{ k_{\perp}  k_z^\infty   q}^{z} (z') \right) } \, .
\eeq
The net nucleon density at $\eta=0$ is then
\beq
\label{eq:rhoG}
\rho_0(z) =  \frac{2k_{F}^3(z)}{3 \pi^2} \big(1+ {\mathcal G}(z)\big) \, .
\eeq
We write, moreover, the $\mu_a$-derivative of $\rho_{np}$ at $\eta=0$, for use in the isovector density~\eqref{eq:rhoadef}, as
\beq
\label{eq:rhonpF}
\frac{\partial \rho_{np}(z)}{\partial \mu_a} = 2 \frac{\partial \rho_{n}(z)}{\partial \mu_a}
= \frac{2 k_F^2(z)}{\pi^2} \big(1 + {\mathcal F}(z) \big) \, ,
\eeq
where, with \eqref{eq:rhoqg},
\beq
{\mathcal F}(z)= {\mathcal G}(z) + \frac{k_F(z)}{3\, (\frac{\partial k_{Fn}}{\partial \mu_a})} \,
\frac{\partial }{\partial \mu_a } \, {\mathcal G}_n(z) \, .
\eeq
On evaluating the $\mu_a$-derivative of both sides of Eq.~\eqref{eq:muq} for neutrons, and on using Eqs.~\eqref{eq:rhoadef}, \eqref{eq:munp}, \eqref{eq:Ua=}, \eqref{eq:Srho=}, \eqref{eq:rhoG} and \eqref{eq:rhonpF}, we arrive at the following expression for the isovector density
\beq
\label{eq:rhoaFG}
\rho_a^0 = \frac{a_a^V \, \rho}{S(\rho)} \, \frac{1}{1 + \frac{\hbar^2 \, k_F^2}{6m\, S} \, \frac{{\mathcal G} - {\mathcal F}}{1 + {\mathcal F}}} \,
\approx \frac{a_a^V \, \rho}{S(\rho)} \, \left( 1 +  \frac{\hbar^2 \, k_F^2}{6m\, S} \, {\mathcal F} \right) \, .
\eeq
The last approximation is valid at large distances from the surface, in the classically allowed region, where we expect $1 \gg {\mathcal F} \gg {\mathcal G}$.

Equation \eqref{eq:rhoaFG} represents the isovector density in terms of the local approximation to that density combined with an oscillatory correction.  Coefficient for the correction, whether examined in the denominator of the middle expression or on the r.h.s.~of~\eqref{eq:rhoaFG}, involves the ratio of the kinetic contribution to symmetry energy to the net symmetry energy.  Correspondingly, the Friedel oscillations are expected to be stronger for interactions with lower symmetry-energy values around the normal density, which indeed appears to be the case for the examples in Fig.~\ref{fig:denscomp}.  For~the simplistic model of a step-like barrier at $z_0$, representing a~rapidly rising isoscalar optical potential~$U$, the functions ${\mathcal G}$ and ${\mathcal F}$ may be arrived at in the following analytic forms:
\bea
{\mathcal G}(z) & = & \frac{3}{4 k_F^2 \, (z_0 - z)^2} \left[ \cos{\left(2 k_F \, (z_0 -z) \right)} - \frac{\sin{\left(2 k_F \, (z_0 -z) \right)}}{2 k_F \, (z_0 -z) } \right] \, , \\
{\mathcal F} (z) & = & - \frac{1}{2 k_F \, (z_0 -z)} \sin{\left(2 k_F \, (z_0 -z) \right)} \, .
\eea
Following those results, oscillations in the isovector density, governed by ${\mathcal F}$, should die out rather slowly with the distance $(z_0 -z)$, as distance inverse, which appears to be borne out by the results in Fig.~\ref{fig:denscomp}.  On the other hand, oscillations in the isoscalar density should die out rather quickly, as distance inverse squared, consistently with the previous qualitative expectations and with practical findings.

A couple of other observations can be made in connection with the considerations above. Thus, it is apparent that the region of potential validity of the local approximation in the $\eta \rightarrow 0$ matter is necessarily limited by the farthest lying classical return point for symmetric matter.  At the general level, this criterion is consistent with the expectation that the range effects must be ignorable for the approximation to apply.  An interesting aspect of this criterion is, however, that it is independent of the symmetry energy and tied only to the features of symmetric matter.  Indeed, in spite of the widespread variation of the symmetry energy in Fig.~\ref{fig:denscomp}, the position where local approximation breaks down, in terms of net density, remains approximately the same.  The farthest return point is of course that for a Fermi wavevector directed along the $z$-axis and, for reference, we indicate the location of those points in Fig.~\ref{fig:denscomp}.  The return points for lower wavenumbers or for wavevectors not pointing along the $z$-axis are closer to the interior of the matter; the coarse end-location for the validity of the local approximation, of isoscalar density being about $\rho_0/4$, represents the return point for a Fermi wavevector directed at about 35$^\circ$ to the $z$-axis.

\begin{figure}
\centerline{\includegraphics[width=.85\linewidth]{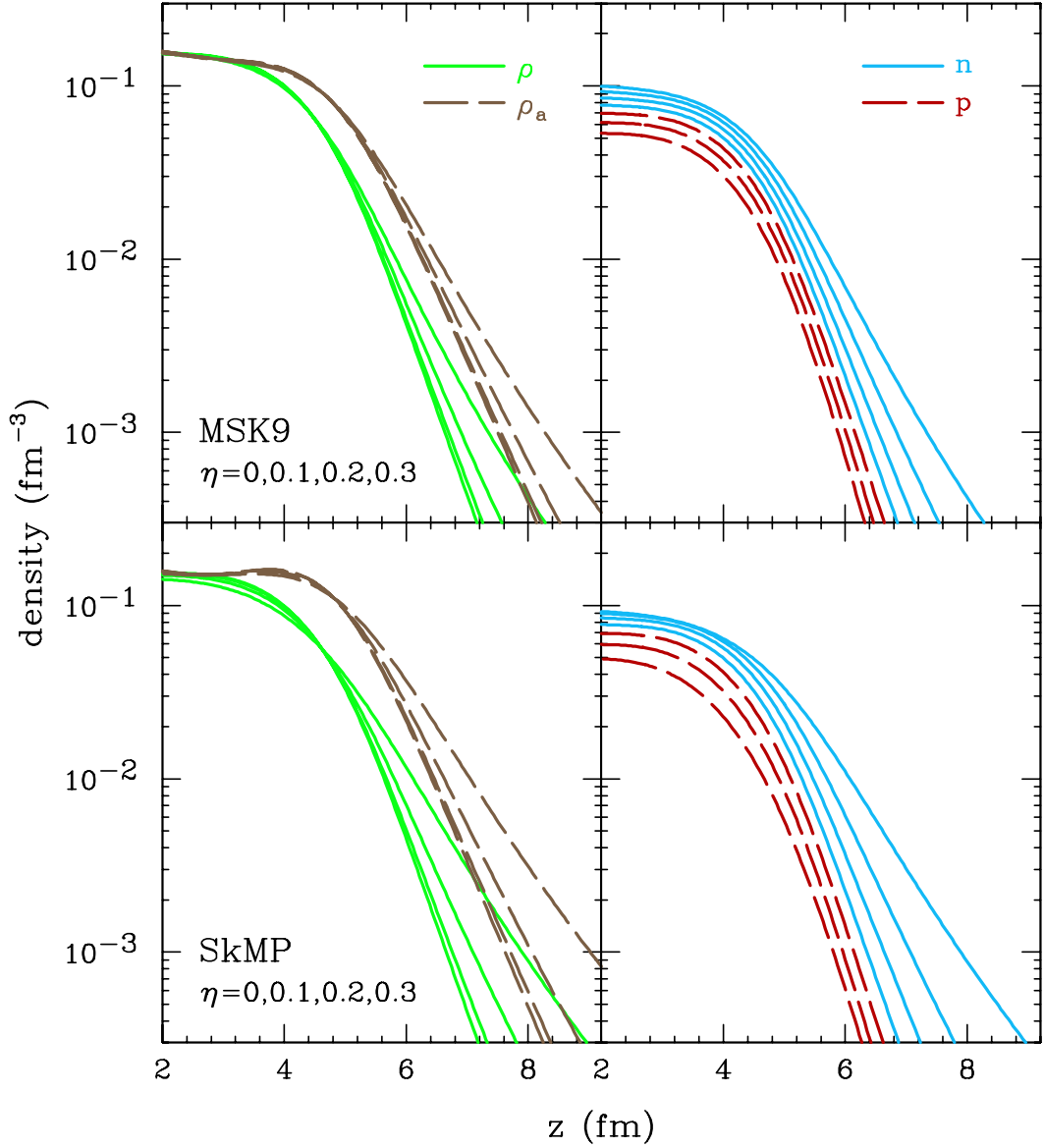}}
\caption{Densities on a logarithmic scale, in the surface region of half-infinite nuclear matter, for sample MSk9 and SkMP interactions in the top and bottom panels, respectively.  The abscissa scale is numerically identical to that in Figs.~\ref{fig:density}, \ref{fig:densvec} and~\ref{fig:denscomp}.  The left panels show the approximately invariant densities, isoscalar (solid lines) and isovector (dashed), with the order from left to right in the tails corresponding to $\eta = 0$, 0.1, 0.2 and 0.3, respectively.  The right panels show proton densities (dashed lines) for, respectively, $\eta = 0.3$, 0.2 and 0.1, from left to right in the tail, and neutron densities (solid) for, respectively, $\eta =0$, 0.1, 0.2 and 0.3, from left to right in the density tail.
}
\label{fig:denslog}
\end{figure}

Another observation, related to the preceding considerations, pertains to the $\eta$-systematics of the tails of isoscalar and isovector densities.  The {\em relative} variation of density in those tails appears to be significant with higher~$\eta$ in Figs.~\ref{fig:density} and \ref{fig:densvec}, for at least two of the sample interactions, SkMP~and~SkI5.  Features of the tails combine the effects of symmetry energy close to the classical return points and the effects of chemical potentials on the forbidden region.  Figure~\ref{fig:denslog}, which displays details of different densities on a logarithmic scale, at different $\eta$ for two of the sample interactions, MSk9~and~SkMP, can serve as an illustration in the discussion of understanding of the tails.  At $\rho \gtrsim \rho_0/4$, larger deviations of symmetry energy from $a_a^V$, for SkMP than for MSk9 interaction, and larger~$\text{d}S/\text{d}\rho$, yield both stronger changes in the isoscalar~$\rho({\pmb r})$ with~$\eta^2$ for~SkMP, as~expected from Eqs.~\eqref{eq:mufrho} and~\eqref{eq:rhofmu}.  Changes within the second order in one of the densities, such as for~SkMP, generally impact the other density within that order.  Regarding the forbidden region, insights come from the behavior of the WKBJ wavefunctions far into the forbidden region, described in~Eq.~\eqref{eq:psiforb}.  With integrations in the wavevector space over squared wavefunctions producing densities, it is apparent that the neutron and proton densities need to fall off primarily exponentially in the far-out forbidden region, as $\text{exp}(-2\kappa_{Fq} \, z) \equiv \text{exp}(-z/d_q)$, where $\kappa_{Fq} = \sqrt{-2m \, \mu_q} /\hbar$ and, thus, $d_q \sim 0.57 \, \text{fm}$ for $\mu_q \sim 16 \, \text{MeV}$.  For $\eta \ge 0$ then, on account of the neutron contribution, with neutron chemical potential being greater or equal to the proton potential, the isoscalar and isovector densities should fall off as $\text{exp}(-2\sqrt{-2m \, \mu_n} \, z /\hbar)$.  Given a~milder neutron than proton slope for $\eta > 0$, in fact, sufficiently far out, the neutron density needs to eventually strongly dominate over the proton density.  Looking at the tails of proton and neutron densities in Fig.~\ref{fig:denslog}, far from the surface, it is seen that these densities indeed eventually begin to fall off exponentially, by about $e$ per $0.6 \, \text{fm}$ at $\eta=0$.  As has been anticipated, the fall-off slopes are always about the same for $\rho$ and $\rho_a$ at a given~$\eta$, and, at $\eta \ge 0$, about the same for these two densities and~$\rho_n$.  As neutron chemical potential increases with an increase in~$\eta$, the slope of fall-off of the densities becomes milder.  When one moves then farther and farther out of the matter, for $\eta >0$, the accumulated effects of changed density slope, while possibly small on absolute scale, can become arbitrarily large on the scale of density scaled with the density for $\eta=0$.  However, besides slopes, the behavior of densities in the allowed region matters as well for the forbidden region, in setting of the initial conditions for the fall-off.  Incidentally, as critical asymmetries nearly coincide for MSk9 and SkMP, see Table~\ref{tab:skypro}, there is a similarity in evolution of $\mu_n$ with changes in~$\eta$, for the two interactions.  That similarity implies a~degree of similarity in the slopes alone, that can be checked in~Fig.~\ref{fig:denslog}.

\subsection{Coefficients Characterizing Nuclear Surface Energy, $a_S$ and $a_a^S$}

For a self-sustained system in the absence of long-range interactions, the energy associated with a boundary can be simply defined as the difference between the actual energy and the energy expected if the boundary were not there,
\beq
\label{eq:E_SSky}
E_S = E - \left(\frac{E}{A}\right)_V \, A = \int \text{d} {\pmb r} \, \left[ e_\text{Skyrme}({\pmb r}) -
\left(\frac{E}{A}\right)_V  \, \rho ({\pmb r}) \right] \, .
\eeq
The last expression has been arranged to make the integration over depth within the matter finite in the limit of $A \rightarrow \infty$, when the net energy can be represented in Skyrme form.  In the case of half-infinite matter, Eq.~\eqref{eq:E_SSky} yields surface energy per unit area of the form
\beq
\label{eq:ESS}
\frac{E_S}{\Sigma} = \int \text{d} z \, \left[ e_\text{Skyrme}(z) -
 \left(\frac{E}{A}\right)_V  \, \rho (z)  \right] \, .
\eeq
From the energy per unit area, principally, the coefficients $a_S$ and $a_a^S$ in an energy formula may be inferred.  Specifically, with the area of $\Sigma \simeq 4 \pi \, r_0^2 \, A^{2/3}$ for a large spherical nucleus at $\eta \rightarrow 0$, we find, from \eqref{eq:es=}, that the coefficients are related to the energy of half-infinite matter with
\beq
\label{eq:ESSeta}
\begin{split}
4 \pi \, r_0^2 \, \frac{E_S}{\Sigma} & = a_S +  \frac{ \mu_a^2}{4 \, a_a^S} + {\mathcal O}(\eta^4) \equiv
 a_S + \frac{(a_a^V)^2}{a_a^S} \, \left( \eta_V' \right)^2 + {\mathcal O}(\eta^4) \\[.3ex]
& = a_S + \frac{(a_a^V)^2}{a_a^S} \, \eta_V^2 + {\mathcal O}(\eta^4) \, .
\end{split}
\eeq
In practice, some problems emerge when trying to utilize Eqs.~\eqref{eq:ESS} and \eqref{eq:ESSeta} directly for determination of the coefficients characterizing nuclear surface.

Thus, while the subintegral function in \eqref{eq:ESS} nominally approaches zero with increasing depth into the matter, i.e.~$z \rightarrow - \infty$ in our convention, in practice it is not possible to employ~\eqref{eq:ESS} down to any depth in matter.  This is because errors on the values of~$e(z)$ and~$\rho(z)$, from numerical integration of the SHF equations, do not vanish with increasing depth.  If the use of~\eqref{eq:ESS} were insisted upon down to any depth in the matter, the accumulated effects of errors across the volume would eventually overpower any surface contribution to the sought integral.  Outside of the matter, errors from the integration of the SHF equations surge, at least relatively, close to the start of integration.  In consequence, in integrations across surface of half-infinite matter, when seeking $E_S/\Sigma$ or other surface quantities, we employ a~profile function $P(z)$ to control errors accumulating with distance away from the surface, by~choosing to approximate integration such as in \eqref{eq:ESS} with
\beq
\label{eq:ESP}
\frac{E_S}{\Sigma} \simeq \int \text{d} z \, P(z) \, \left[ e_\text{Skyrme}(z) -
\left(\frac{E}{A}\right)_V  \, \rho (z)  \right] \, .
\eeq
The profile function takes on the values
\beq
P(z) = \left\lbrace \begin{aligned} &1 \, , \, \, \text{at} \, \, |z-z_0| < z_P \\
&0  \,  ,  \, \, \text{at} \, |z-z_0| > z_P + \Delta z_P \end{aligned}\right. \, .
\eeq
where $z_0$ represents surface location.  We further make $P(z)$ vary smoothly between $z_P$ and $z_P + \Delta z_P$ aiming at a reduction of the error in integration associated with an interplay between the termination of integration and the Friedel oscillations in nucleon and energy densities inside the matter.  We typically employ $z_P \sim 6 \, \text{fm}$ and $\Delta z_P \sim 6 \, \text{fm}$.  Results that will be presented, obtained using~$P$, are stable with respect to variations in $z_P$ and $\Delta z_P$ around our choices.

When trying to extract the $a_a^S$ coefficient from Eq.~\eqref{eq:ESP}, an additional to the above problem arises, particularly serious for interactions with negative~$L$.  Thus, when extracting~$a_a^S$, the energy values at small $\eta$ must be considered.  This makes the contribution of surface-symmetry energy to the net energy to be of net third-order in smallness, with one order due to the surface-to-volume ratio and with the two remaining orders due to the asymmetry.  For a~negative~$L$, the surface contribution to the net symmetry energy is particularly small.  The~problem can be circumvented by employing Eq.~\eqref{eq:aaaa} for calculating $a_a^S$, rather than~\eqref{eq:ESS}.  Namely, upon adding and subtracting $A/a_a^V$ from the right-hand side of~\eqref{eq:aaaa}, for large but finite spherical system of mass number~$A$, we find
\beq
\frac{A}{a_a(A)} = \frac{A}{a_a^V} + \frac{1}{a_a^V} \int \text{d}{\pmb r} \, \left( \rho_a^0({\pmb r}) - \rho^0({\pmb r}) \right) \simeq  \frac{A}{a_a^V} + \frac{A^{2/3}}{a_a^S} \, .
\eeq
From the above, we find that the surface symmetry coefficient may be expressed as
\beq
\label{eq:aaSrho}
\frac{1}{a_a^S} = \frac{4 \pi r_0^2}{a_a^V} \int \text{d} z \, \left( \rho_a^0(z) - \rho^0(z) \right) \, ,
\eeq
where the r.h.s.\ integration is over the direction perpendicular to the surface.  In practical calculations with Eq.~\eqref{eq:aaSrho}, we also employ a profile function such as in~\eqref{eq:ESP}.  Reaching a~desired accuracy for $a_a^S$ with \eqref{eq:aaSrho} requires operating within two orders of smallness, rather than three, due to surface-to-volume ratio and due to difference $\rho_n - \rho_p$, of the order of $\eta$,  in calculating $\rho_a^0$.  With the typical values of asymmetry that can be afforded in our extractions of $a_a^S$, $\eta \simeq \text{(0.03--0.15)}$, the error on $a_a^S$ turns out actually to be (1--2) orders of magnitude smaller when using \eqref{eq:aaSrho} than when using \eqref{eq:ESS}.  In~the following, we will extract the coefficients $a_S$ and $a_a^S$, relying on the formulas above, and we will compare the results to those of preceding Hartree-Fock calculations for half-infinite matter by Kohler~\cite{Kohler:1976} and by Farine, Pearson {\em et al.}~\cite{Farine:1980,PhysRevC.24.303,Farine97}.  Also, we shall compare the extracted coefficients to those deduced from spherical Hartree-Fock calculations~\cite{reinhard:014309} for large, up to $A \sim (10^5 - 10^6)$, systems.

Interestingly, our Eq.~\eqref{eq:aaSrho} is equivalent to a droplet-model expression considered in the context of their calculations by Farine, Cote and Pearson \cite{Farine:1980,PhysRevC.24.303} (see \cite{Fari85} for a functional derivation by Farine), and ultimately exploited \cite{samyn:044316} in the context of Farine's code, where
\beq
\label{eq:aaSDM}
\frac{1}{a_a^S} = \frac{3}{2 a_a^V \, r_0} \lim_{\eta_V \rightarrow 0} \frac{1}{\eta_V}
\int \text{d}z \, \left\lbrace \frac{\rho_n(z)}{\rho_n^V} - \frac{\rho_p(z)}{\rho_p^V} \right\rbrace \, .
\eeq
The equivalence of the results \eqref{eq:aaSDM} and \eqref{eq:aaSrho} can be seen by noting that at low volume asymmetries, $\eta_V \rightarrow 0$, the asymptotic nucleon densities become $\rho_q^V \simeq 3(1 \pm \eta_V/2)/(8 \pi r_0^3)$, allowing to rewrite the r.h.s.\ of \eqref{eq:aaSDM} as
\beq
\begin{split}
\frac{1}{a_a^S} & = \frac{4 \pi \, r_0^2}{a_a^V} \lim_{\eta_V \rightarrow 0} \frac{1}{\eta_V}
\int \text{d}z \, \left\lbrace \frac{\rho_n(z)}{1+\eta_V} - \frac{\rho_p(z)}{1-\eta_V} \right\rbrace \\
&
=  \frac{4 \pi \, r_0^2}{a_a^V} \lim_{\eta_V \rightarrow 0} \int \text{d}z \, \left\lbrace
\frac{\rho_n(z)- \rho_p(z)}{\eta_V} - \rho(z)
\right\rbrace
 \, ,
 \end{split}
\eeq
where the r.h.s.\ yields next the r.h.s.\ of \eqref{eq:aaSrho}, cf.~Eq.~\eqref{eq:rhoadef}.  Notably, because of the Friedel oscillations, Farine {\em et al.}~ended up never employing Eq.~\eqref{eq:aaSDM} directly in their analyses.  Instead, they have examined the separation between neutron and proton surfaces at finite~$\eta_V$, with which separation they have replaced the integral that appears on the r.h.s.\ of Eq.~\eqref{eq:aaSDM}.  With this, they proceeded along the lines of the droplet model \cite{Myers:1969}, see also~\cite{Kohler:1976}.

\begin{figure}
\centerline{\includegraphics[width=.72\linewidth]{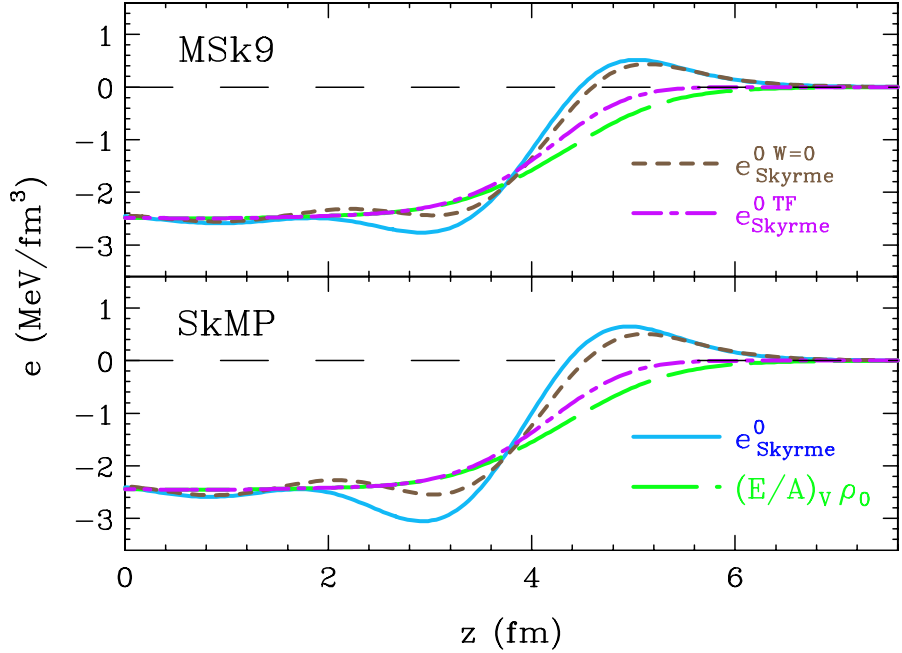}}
\caption{Different energy densities in the surface region of symmetric half-infinite nuclear-matter, for sample Skyrme interactions.  Shown are the Skyrme energy density in the standard SHFM calculations, $e_\text{Skyrme}^0$ from Eq.~\eqref{eq:Skyrme}, the energy density from the calculations with spin-orbit term ignored, $e_\text{Skyrme}^{0 \, \, W=0}$, the energy density in the Thomas-Fermi approximation with gradient terms ignored, $e_\text{Skyrme}^{0 \, \, \text{TF}}$, and, finally, the energy density expected if the energy per nucleon were constant, $(E/A)_V \,  \rho_0$.  The~abscissa scale is numerically identical to that in Figs.~\ref{fig:density}, \ref{fig:densvec} and~\ref{fig:denscomp}.
}
\label{fig:edens}
\end{figure}

Figure \ref{fig:edens} displays different energy densities in the surface region of symmetric half-infinite nuclear matter, for two sample Skyrme interactions, MSK9 and SkMP.  Those densities include the two energy densities, $e_\text{Skyrme}^0$ and $ (E/A)_V \, \rho_0$, of which the integrated difference yields the surface energy~\eqref{eq:ESS}, and they further include the energy density $e_\text{Skyrme}^{0 \, W=0}$ calculated with the spin-orbit term suppressed and, finally, the energy density calculated in the Thomas-Fermi approximation, $e_\text{Skyrme}^{0 \, \text{TF}}$.  In obtaining the last energy density, the density of kinetic energy is taken in the local approximation of Eq.~\eqref{eq:tauq} and the gradient terms in any {\em density} are ignored.

For moderately subnormal nucleon densities in Fig.~\ref{fig:edens}, the energy densities $(E/A)_V  \, \rho_0 (z)$ and $e_\text{Skyrme}^{0 \, \text{TF}}(z)$ are not much different from each other.  This just reflects the fact that the energy per nucleon minimizes in uniform matter at normal density and, in consequence, it changes little with moderate deviations of density from normal.  Across nucleon densities, the energy density $e_\text{Skyrme}^{0}(z)$ oscillates in the figure in the vicinity of $e_\text{Skyrme}^{0 \, \text{TF}}(z)$.  Dominant and positive contribution to the difference of energy densities $e_\text{Skyrme}^0 - (E/A)_V  \, \rho_0$, in the expression~\eqref{eq:ESS} for the surface energy, is seen to come from the far-out tail of the density distribution, from essentially the classically forbidden region.
In fact, in the outmost tail, the potential energy approaches zero faster than does the kinetic energy, making the net energy density even positive.
The positive contribution to the difference in~\eqref{eq:ESS} is next, in the matter direction, much reduced by the negative contribution from just moderately subnormal nucleon densities in the density tail.  When moving further in position into interior of the matter, the energy densities in the difference in~\eqref{eq:ESS} and, correspondingly, the difference itself oscillate in a~Friedel fashion, with a~wavelength of $\sim \pi/k_F \simeq 2.4 \, \text{fm}$.  The~precise location of maxima and minima in $e_\text{Skyrme}$, and in the energy difference, depends on the choice that was made for the kinetic energy density. Thus, when expressing the kinetic energy density in terms of a square of wavefunction gradient in~\eqref{eq:tauq=}, the kinetic energy oscillates approximately out of phase with respect to the density.  If the kinetic energy were expressed, though, in terms of a laplacian, rather than gradient squared, the oscillations of kinetic-energy density would have been approximately in phase with the nucleon density.

Surface energy coefficients, calculated by integrating the $\eta=0$ difference of energy densities in \eqref{eq:ESP}, to obtain the energy for elementary surface area,
\beq
\label{eq:aSES}
a_S = 4 \pi \, r_0^2 \, \frac{E_S^0}{\Sigma} \, ,
\eeq
are provided for various Skyrme interactions in Table \ref{tab:skypro}.  The value distribution, across interactions, is rather narrow, with the average coefficient value of $17.6 \, \text{MeV}$ combined with the standard deviation of $1.1 \, \text{MeV}$.

An interesting issue is how the surface coefficient may be affected by the effects of a spin-orbit coupling acting in the surface region.  Looking at the Skyrme energy functional~\eqref{eq:Skyrme}, where nucleon densities multiply spin-density divergencies in the coupling, we see that this coupling allows a nuclear system to lower its energy by developing spin densities.  Divergence of a spin density obviously integrates to zero.  However, energy of a system can change when positive and negative divergence values are correlated with different magnitudes of nucleon densities.  For positive $b_4$-coefficient values in the functional~\eqref{eq:Skyrme}, the energy gets lowered when a spin density directed outward develops in the surface region.  In that case, the divergence is positive in the inner part of the surface and negative in the outer allowing the energy density to drop more in the inner than in the outer part of the surface.  The~polarization in the system is moderated, in~particular, by the kinetic energy density that grows, but only quadratically, with spin density.  For reference, in addition to other energy densities in Fig.~\ref{fig:edens}, we show there also the Skyrme energy density $e_\text{Skyrme}^{0 \, W=0}$ obtained when suppressing the spin-orbit coupling.  It~is seen that, indeed, inclusion of the coupling lowers energy density more in the inner part of the surface than in the outer. Lowering of the system energy leads to lowering of the surface coefficient.  Notably, inclusion of the coupling affects also $(E/A)_V \, \rho$ in the difference of energy densities for the surface energy~\eqref{eq:ESP}, but not as strongly as $e_\text{Skyrme}$ there, cf.~Fig.~\ref{fig:uwan}.  Quantitatively, for~the surface coefficients of the interactions illustrated in Fig.~\ref{fig:edens}, we find $a_S^{W=0} = 18.7 \, \text{MeV}$ for both the MSk9 and SkMP, when the spin-orbit coupling is suppressed.  With the coupling, we find $a_S = 17.1$ and $16.6 \, \text{MeV}$, respectively, for the the MSk9 and SkMP interactions, see Table \ref{tab:skypro}.  A $\sim 10\%$ magnitude in reduction of $a_S$ is typical for the Skyrme parameterizations, when switching on the spin-orbit coupling, see also the forthcoming Table~\ref{tab:comparison}.  To our knowledge, Stocker~\cite{Stock70} was the first to assess the sign and magnitude of the change in the surface symmetry coefficient when including the spin-orbit coupling in the surface description.

\begin{figure}
\centerline{\includegraphics[width=.72\linewidth]{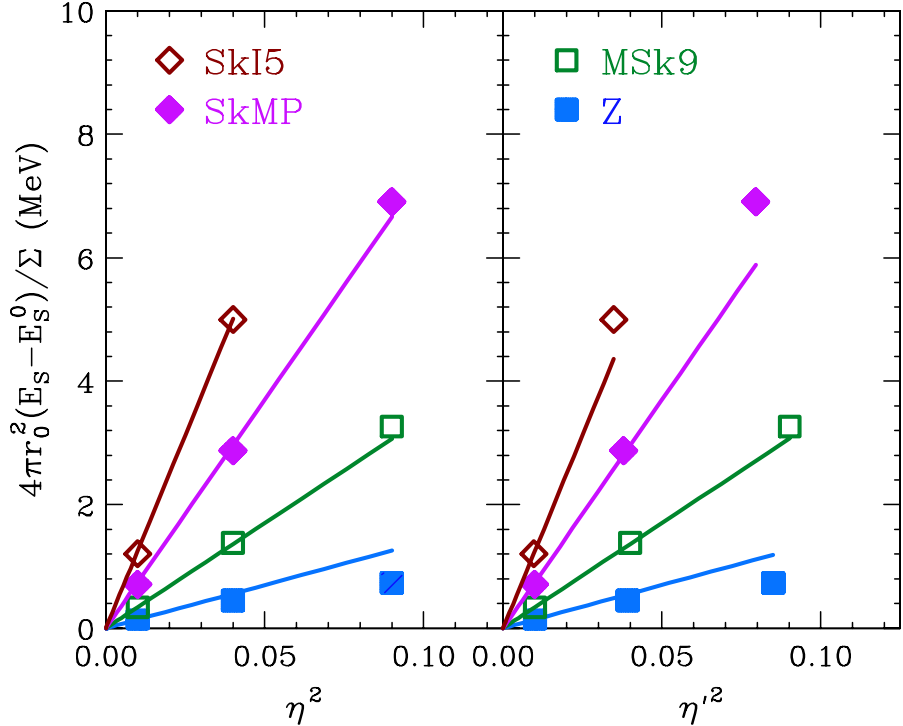}}
\caption{Symbols represent change in the surface energy of an elementary area, from \eqref{eq:ESP}, for different Skyrme parameterizations, plotted vs square of asymmetry: volume asymmetry in the left panel and effective volume asymmetry, $\eta_V' \equiv \mu_a/(2 a_a^V)$, in the right panel.  Lines represent expectations based on Eq.~\eqref{eq:ESSeta}, with surface symmetry coefficients taken from Eq.~\eqref{eq:aaSrho}.
}
\label{fig:aset}
\end{figure}

Besides the values of $a_S$, Table \ref{tab:skypro} gives, further, the surface symmetry coefficients $a_a^S$ obtained by integrating the $\eta=0$ difference of isovector and isoscalar densities in Eq.~\eqref{eq:aaSrho}.  As might be expected from  Fig.~\ref{fig:denscomp}, where the relation between isoscalar and isovector densities varies widely from one interaction to another, the values of $a_a^S$ vary widely between individual interactions, from 10 to 60\,MeV.  Figure \ref{fig:aset} illustrates next the changes in surface energy~\eqref{eq:ESP} of an elementary area $4\pi r_0^2$, with changing asymmetry, for the sample Skyrme interactions.  Magnitude of the changes in energy should be compared to the magnitude of the surface coefficient \eqref{eq:aSES}, of the order of 17\,MeV for the Skyrme interactions.  Note that the rise of surface energy with asymmetry in a finite nucleus would be accompanied by a~drop in the volume symmetry energy, compared to a system with no surface, due to asymmetry moving out to the surface region.  As evident in Fig.~\ref{fig:denscomp}, the surface energy generally changes quadratically with asymmetry.  At small asymmetries, barely visible deviations from the lines, representing expectations from Eqs.~\eqref{eq:ESSeta} and~\eqref{eq:aaSrho}, stem from numerical inaccuracies, primarily arising from the use of Eq.~\eqref{eq:ESP} to calculate the surface energy.  At~higher asymmetries generally true fourth-order terms in asymmetry come into play.  In~particular, some differences between the regularities in the left and right panels become apparent at higher $\eta$ and $L$; these are due to the differences that, depending on interaction, develop between $\eta_V$ and $\eta_V'$ at higher asymmetries, as a consequence of fourth-order terms in the energy of uniform matter.

For the Skyrme parameterizations with $b_4 = b_4'$, cf.~Eq.~\eqref{eq:b_4p}, the spin-orbit coupling has even less effect on $a_a^S$ than on $a_S$.  Unlike $a_S$, the coefficient $a_a^S$ increases rather than decreases with the switching on of the spin-orbit coupling, usually by less than 10\% and often by much less.  Notably, though, at constant volume asymmetry, the increase in $a_a^S$ implies dropping of the surface energy, cf.~\eqref{eq:ESSeta}, just as a~decrease in~$a_S$ does.  Without the coupling, we find e.g.~$a_a^{S \, W=0} = 22.5$ and $11.1 \, \text{MeV}$, respectively, for the MSk9 and SkMP parameterizations, and $a_a^S = 23.0$ and $12.1 \, \text{MeV}$, respectively, for these parameterizations, with the coupling.  Additional exemplary results from $b_4 = b_4'$ Skyrme parameterizations can be found in the forthcoming Table~\ref{tab:comparison}.  For parameterizations with different $b_4$ and $b_4'$, the spin-orbit coupling may have a~stronger effect on the $a_a^S$ coefficient, than for $b_4=b_4'$, seemingly due to a~strengthening of the spin-isospin correlations in the surface region.  E.g.~for the SkO parameterization, we find $a_a^{S \, W=0} = 13.2 \, \text{MeV}$ and $a_a^{S} = 14.9 \, \text{MeV}$, without and with the coupling, respectively.  For SkI4, we find $a_a^{S \, W=0} = 17.5 \, \text{MeV}$ and $a_a^{S} = 20.8 \, \text{MeV}$.

\subsection{Comparison to Coefficient Values in the Literature}

Our results for the coefficients of surface energy are next compared in Table \ref{tab:comparison} to the results obtained by others in the literature within the SHFM calculations of asymmetric systems.
Specifically, our coefficients are compared to those obtained within the calculations of semi-infinite matter, done by Kohler \cite{Kohler:1976} and by Pearson {\em et al.} \cite{Cote:1978,PhysRevC.24.303,Farine97,Samyn03,PhysRevC.68.054325,samyn:044309,samyn:044316,Gor05,goriely:031301}. The~latter results stem predominantly from the code by Farine \cite{Farine81}. The dependence of the surface energy on asymmetry has been expressed in the past calculations of semi-infinite matter in terms of the coefficient $Q$ from the droplet model, cf.~Subsection~\ref{ssec:droplet}.  That coefficient differs from $a_a^S$ by simple factor, cf.~\eqref{eq:aaSQ}, allowing for a simple transcription of the results obtained for semi-infinite matter.  The calculations by Kohler were done without spin-orbit coupling and, in comparisons to his results, we have suppressed that coupling as well.  In addition, in Table~\ref{tab:comparison}, our coefficients for the surface are compared to those deduced in the calculations of spherical nuclei with large mass, up to $A \sim (10^5 - 10^6)$, by Reinhard {\em et al.}~\cite{reinhard:014309}.  Those authors have assessed the  coefficients of expansion of energy in powers of~$A^{-1/3}$ and~$\eta^2$, for different interactions.  If we expand the symmetry coefficient~\eqref{eq:aa} in~$A^{-1/3}$, we find that we can compute $a_a^S$ from the coefficient $a_\text{ssym}$ in~\cite{reinhard:014309}, with
\beq
a_a^S = - \frac{\left(a_a^V \right)^2} {a_\text{ssym}} \, .
\eeq
As will be mentioned later, in addressing the likely coefficient values, the expansion of the symmetry coefficient~\eqref{eq:aa} lacks justification for nuclear masses encountered in nature.

\begin{center}
\setlength{\LTcapwidth}{5in}
\begin{longtable}
                {l|c|c|c|c|c}
\caption{Comparison of coefficients characterizing nuclear surface energy, from SHFM calculations of nuclear systems by different authors, with sources for outside results provided in the superscripts.  The surface symmetry coefficient $a_a^S$ is related to the coefficient $Q$, stemming from the droplet model, with $a_a^S = 4Q/9$, and to coefficient $a_\text{ssym}$ in the large-$A$ expansion of nuclear energy~\cite{reinhard:014309}, with $a_a^S = - (a_a^V)^2/a_\text{ssym}$.}
\label{tab:comparison}\\

\hline
      &     & \multicolumn{4}{c}{Value (MeV)} \\ \cline{3-6}
Name  & $a$ & Kohler & Pearson {\em et al.} & Reinhard {\em et al.} & Our \\
\hline
\endfirsthead

\multicolumn{6}{l}{{\hspace*{6em}\tablename} \thetable{} -- Continued}\\
\hline
      &     & \multicolumn{4}{c}{Value (MeV)} \\   \cline{3-6}
Name  & $a$ & Kohler & Pearson {\em et al.} & Reinhard {\em et al.} & Our \\
\hline
\endhead

\multicolumn{6}{l}{\hspace*{6em}Table Continued on Next Page\ldots}
\endfoot

\hline
\endlastfoot

SII        & $a_S^{W=0}$      & 19.8$^\text{\cite{Kohler:1976}}$ &  &    & 20.20 \\
           & $a_a^{S \, W=0}$ & 11.6$^\text{\cite{Kohler:1976}}$ &  &    & 17.27 \\
           & $a_S$            &                    & 19.8$^\text{\cite{PhysRevC.24.303}}$ &  & 19.38  \\
           & $a_a^S$          &                    & 17.8$^\text{\cite{PhysRevC.24.303}}$ &  & 17.54  \\ \hline
SIII       & $a_S^{W=0}$      & 19.6$^\text{\cite{Kohler:1976}}$ & 19.9$^\text{\cite{Cote:1978}}$ &  & 19.85 \\
           & $a_a^{S \, W=0}$ & 21.3$^\text{\cite{Kohler:1976,PhysRevC.24.303}}$ &  &  & 21.79 \\
           & $a_S$            &   & 18.8$^\text{\cite{PhysRevC.24.303}}$ &   & 18.54 \\
           & $a_a^S$          &   & 23.1$^\text{\cite{PhysRevC.24.303}}$ &   & 21.77  \\
           \hline
SIV        & $a_S$            &   & 19.6$^\text{\cite{PhysRevC.24.303}}$ &   & 18.87  \\
           & $a_a^S$          &   & 15.1$^\text{\cite{PhysRevC.24.303}}$ &   & 13.39  \\
           \hline
SV         & $a_S^{W=0}$      & 19.6$^\text{\cite{Kohler:1976}}$ &   &  & 20.44  \\
           & $a_a^{S \, W=0}$ & 4$^\text{\cite{Kohler:1976}}$    &   &  &  9.95 \\
           & $a_S$            &   & 19.8$^\text{\cite{PhysRevC.24.303}}$ &   & 19.16 \\
           & $a_a^S$          &   & 11.1$^\text{\cite{PhysRevC.24.303}}$ &   & 10.27  \\
           \hline
SVI        & $a_S$            &   & 18.3$^\text{\cite{PhysRevC.24.303}}$ &   & 18.10  \\
           & $a_a^S$          &   & 26.7$^\text{\cite{PhysRevC.24.303}}$ &   & 26.27  \\
           \hline
SkMs       & $a_S$            &   & 18.0$^\text{\cite{Farine97}}$ & 17.6$^\text{\cite{reinhard:014309}}$  & 17.46  \\
           & $a_a^S$          &   & 26.7$^\text{\cite{Farine97}}$ & 17.4$^\text{\cite{reinhard:014309}}$  & 14.48  \\
           \hline
SKa        & $a_S^{W=0}$      & 19.8$^\text{\cite{Kohler:1976}}$ &   &  & 19.97 \\
           & $a_a^{S \, W=0}$ & 9.3$^\text{\cite{Kohler:1976}}$  &   &  &  12.06 \\
           \hline
SKb        & $a_S^{W=0}$      & 19.8$^\text{\cite{Kohler:1976}}$ &   &  & 19.97 \\
           & $a_a^{S \, W=0}$ & 7.6$^\text{\cite{Kohler:1976}}$  &   &  &  10.03 \\
           \hline
SkP        & $a_S$            &   &   & 18.2$^\text{\cite{reinhard:014309}}$ & 18.18 \\
           & $a_a^S$          &   &   & 20.0$^\text{\cite{reinhard:014309}}$ & 18.02 \\
           \hline
SkI3       & $a_S$            &   &   & 18.0$^\text{\cite{reinhard:014309}}$ & 17.77 \\
           & $a_a^S$          &   &   & 16.2$^\text{\cite{reinhard:014309}}$ & 12.77 \\
           \hline
SkI4       & $a_S$            &   &   & 17.7$^\text{\cite{reinhard:014309}}$ & 17.48 \\
           & $a_a^S$          &   &   & 25.6$^\text{\cite{reinhard:014309}}$ & 20.83 \\
           \hline
SLy4       & $a_S$            &   &   & 18.4$^\text{\cite{reinhard:014309}}$ & 18.24 \\
           & $a_a^S$          &   &   & 19.0$^\text{\cite{reinhard:014309}}$ & 16.60 \\
           \hline
SLy6       & $a_S$            &   & 17.74$^\text{\cite{samyn:044316}}$ & 17.7$^\text{\cite{reinhard:014309}}$  & 17.53  \\
           & $a_a^S$          &   & 14.8$^\text{\cite{samyn:044316}}$  & 20.0$^\text{\cite{reinhard:014309}}$  & 17.12  \\
           \hline
SkO        & $a_S$            &   &   & 17.3$^\text{\cite{reinhard:014309}}$ & 17.14 \\
           & $a_a^S$          &   &   & 17.6$^\text{\cite{reinhard:014309}}$ & 14.94 \\
           \hline
BSk1       & $a_S$            &   & 17.54$^\text{\cite{samyn:044316}}$ & 17.5$^\text{\cite{reinhard:014309}}$  & 17.22  \\
           & $a_a^S$          &   & 20.3$^\text{\cite{samyn:044316}}$  & 21.5$^\text{\cite{reinhard:014309}}$  & 23.72  \\
           \hline
BSk2       & $a_S$    &   & 17.54$^\text{\cite{PhysRevC.68.054325,Samyn03,samyn:044316}}$       &  & 17.14  \\
& $a_a^S$ &&20.4$^\text{\cite{samyn:044316}}$, 30.2$^\text{\cite{PhysRevC.68.054325,Samyn03}}$  &  & 23.15 \\
           \hline
BSk3       & $a_S$            &   & 17.5$^\text{\cite{PhysRevC.68.054325,Samyn03}}$ &   & 17.16  \\
           & $a_a^S$          &   & 29.8$^\text{\cite{PhysRevC.68.054325,Samyn03}}$ &   & 23.37  \\
           \hline
BSk4       & $a_S$            &   & 17.3$^\text{\cite{PhysRevC.68.054325}}$ &   & 16.91  \\
           & $a_a^S$          &   & 33.8$^\text{\cite{PhysRevC.68.054325}}$ &   & 23.01  \\
           \hline
BSk5       & $a_S$            &   & 17.5$^\text{\cite{PhysRevC.68.054325}}$ &   & 17.04  \\
           & $a_a^S$          &   & 23.1$^\text{\cite{PhysRevC.68.054325}}$ &   & 20.65  \\
           \hline
BSk6       & $a_S$ & & 17.18$^\text{\cite{PhysRevC.68.054325,samyn:044309,samyn:044316}}$& 17.3$^\text{\cite{reinhard:014309}}$ & 16.74  \\
& $a_a^S$ && 19.9$^\text{\cite{samyn:044316}}$, 23.7$^\text{\cite{samyn:044309}}$, 36.9$^\text{\cite{PhysRevC.68.054325}}$& 23.8$^\text{\cite{reinhard:014309}}$  & 22.73  \\
           \hline
BSk7       & $a_S$            &   & 17.3$^\text{\cite{PhysRevC.68.054325,samyn:044316}}$    &  & 16.70  \\
& $a_a^S$  &   & 20.1$^\text{\cite{samyn:044316}}$, 35.6$^\text{\cite{PhysRevC.68.054325}}$ &  & 22.81  \\
           \hline
BSk8       & $a_S$            &   & 17.64$^\text{\cite{samyn:044309,samyn:044316,Gor05}}$   &  & 17.12  \\
& $a_a^S$  &   & 20.2$^\text{\cite{samyn:044316,Gor05}}$, 24.0$^\text{\cite{samyn:044309}}$ &  & 23.13  \\
           \hline
BSk9       & $a_S$            &   & 17.92$^\text{\cite{samyn:044316,Gor05}}$  &   & 17.42  \\
& $a_a^S$  &   & 15.8$^\text{\cite{samyn:044316,Gor05}}$                      &   & 17.83  \\
           \hline
BSk10      & $a_S$            &   & 18.0$^\text{\cite{goriely:2006}}$  &   & 17.51 \\
& $a_a^S$  &   & 15.6$^\text{\cite{goriely:2006}}$                     &   & 17.48 \\
           \hline
BSk11      & $a_S$            &   & 17.7$^\text{\cite{goriely:2006}}$  &   & 17.32 \\
& $a_a^S$  &   & 15.6$^\text{\cite{goriely:2006}}$                     &   & 17.42 \\
           \hline
BSk12      & $a_S$            &   & 17.7$^\text{\cite{goriely:2006}}$  &   & 17.30 \\
& $a_a^S$  &   & 16.0$^\text{\cite{goriely:2006}}$                     &   & 17.51 \\
           \hline
BSk13      & $a_S$            &   & 17.7$^\text{\cite{goriely:2006}}$  &   & 17.34 \\
& $a_a^S$  &   & 15.6$^\text{\cite{goriely:2006}}$                     &   & 17.34 \\
           \hline
BSk14      & $a_S$            &   & 17.6$^\text{\cite{goriely:031301,goriely:064312}}$ &   & 17.17  \\
& $a_a^S$  &   & 15.6$^\text{\cite{goriely:031301,goriely:064312}}$                    &   & 17.22  \\
           \hline

\end{longtable}
\setlength{\LTcapwidth}{4in}
\end{center}

\begin{figure}
\centerline{\includegraphics[width=.55\linewidth]{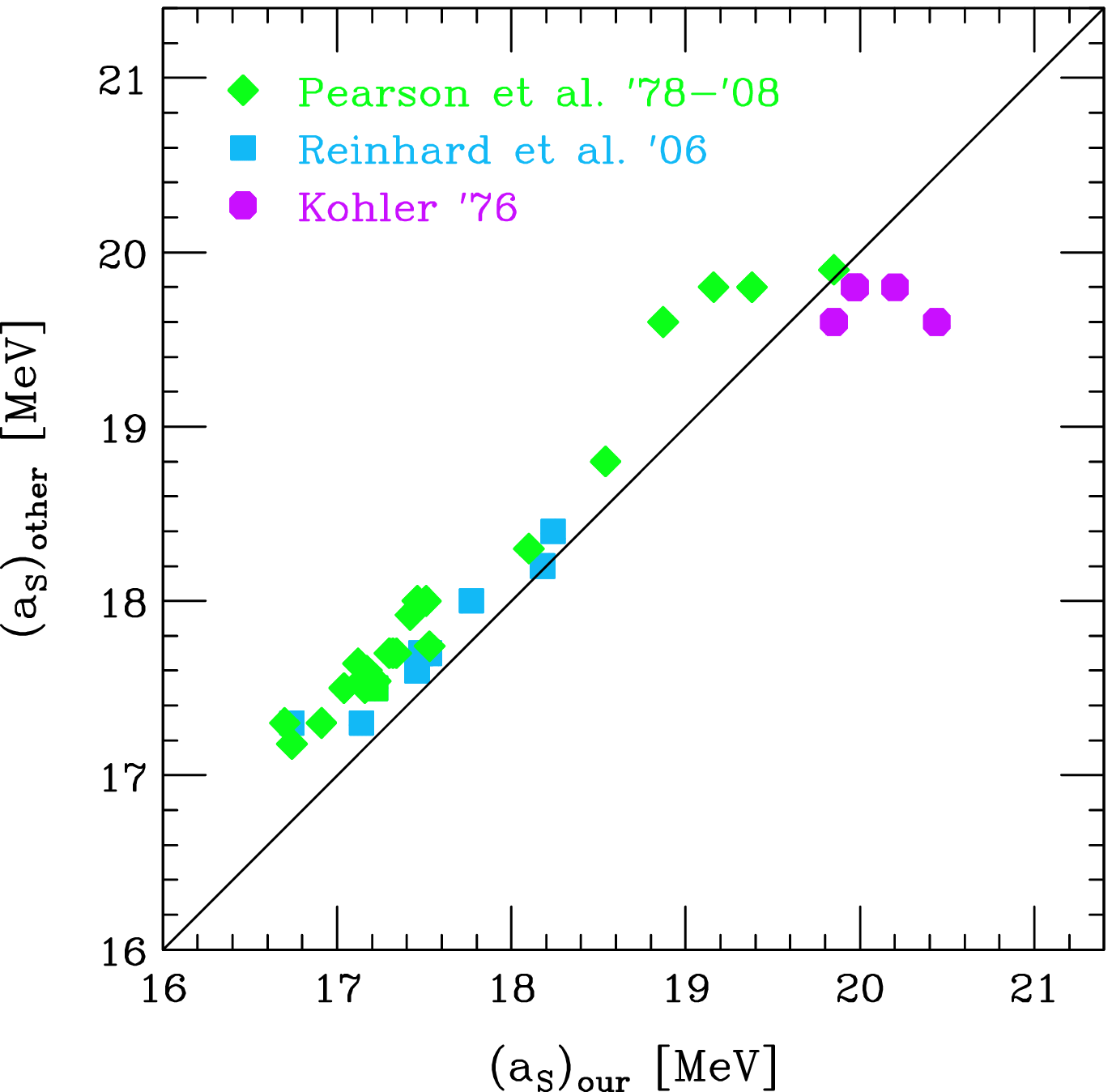}} \vspace*{.18in}
\centerline{\includegraphics[width=.54\linewidth]{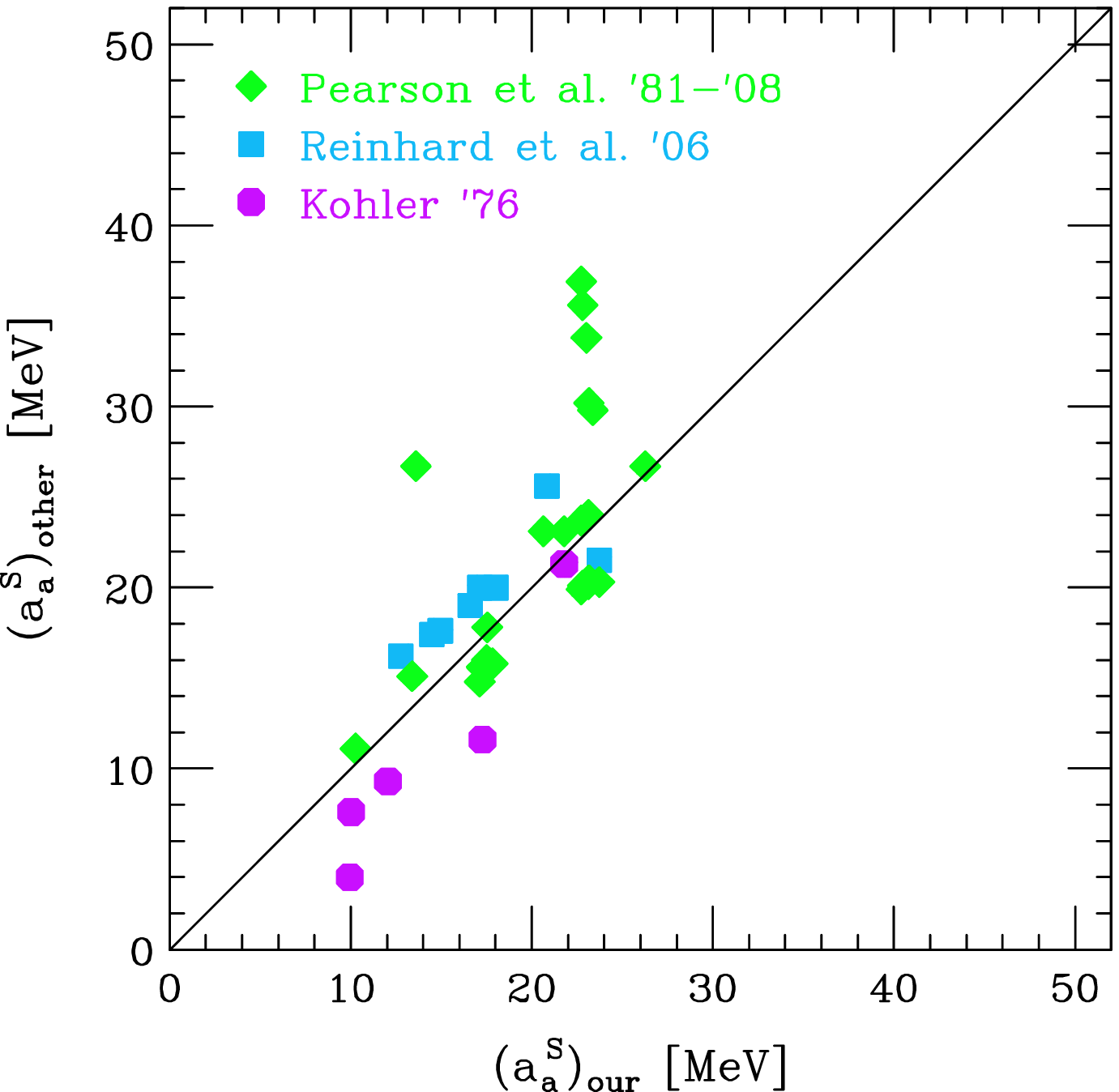}}
\caption{Comparison of the values of coefficients characterizing nuclear surface energy in our calculations and elsewhere in the literature \cite{Kohler:1976,Cote:1978,PhysRevC.24.303,Farine97,Samyn03,PhysRevC.68.054325,samyn:044309,samyn:044316,Gor05,goriely:031301,goriely:2006,goriely:064312,reinhard:014309}.  The first and second panels show, respectively, the comparison of surface coefficients~$a_S$ and of surface symmetry coefficients~$a_a^S$.  Respective coefficient values are given in Table \ref{tab:comparison} and the comparisons of coefficients with and without spin-orbit coupling are lumped together.  In the case of an ideal agreement, the results, as plotted, should line up with the diagonals in the panels.
}
\label{fig:ascomp}
\end{figure}

When carrying a case-by-case examination of the entries in Table~\ref{tab:comparison}, it becomes apparent that results for the surface coefficient $a_S$ agree fairly well between different authors.  As~to results for the surface symmetry coefficient $a_a^S$, a degree of agreement is found in many cases but in a number of other cases there is a disagreement by factors even in excess of~1.5.  The status of the two coefficients in the literature is further illustrated with two panels in Fig.~\ref{fig:ascomp}.  For different Skyrme parameterizations with or without spin-orbit coupling, the~values obtained in the literature are plotted in those panels against our values.  In~case of an ideal agreement, the results should line up with the diagonal lines in the panels.  In~assessing the level of agreement between the results in literature, on the basis of the figure, it should be noted that the range of values displayed in the panel for $a_S$ is lower by one order of magnitude than in the panel for $a_a^S$.

Regarding the three author groups, we appear to have the best overall agreement with Reinhard {\em et al.}~\cite{reinhard:014309}. Their results, though, are also later than most of other results that we compare against.  On the average, their results for $a_S$ are higher by just $\sim 0.25 \, \text{MeV}$ than ours for the coefficient.  The results by Pearson {\em et al.}, for the surface coefficient, appear systematically shifted by a bit more, $\sim 0.4 \, \text{MeV}$, which still represents just 2\% of coefficient value.  By contrast, in the overall assessment, the agreement between the results for $a_a^S$ turns out to be somewhat unsatisfactory, given the discrepancies encountered that are of the same order as coefficient values.  In practice, this demonstrates the difficulty in working with contributions to system properties that are small due to their simultaneous association with surface and with asymmetry.  That difficulty is further underscored by the fact that Pearson {\em et al.} have quoted significantly different values characterizing surface symmetry energy, for the same interaction, in different publication years, cf.\ Table~\ref{tab:comparison}.

Realizing the fragility of their results for the coefficient characterizing surface symmetry-energy, Pearson {\em et al.}\ have eventually reassessed the procedures for extracting the coefficient values, specifically in Appendix A of Ref.~\cite{samyn:044316}.  They have examined there the stability of extracted coefficient values, when changing asymmetry in the extraction.  From the three tested procedures tested, they have found the one based on Eq.~\eqref{eq:aaSDM} to be the most stable and, thus, preferred.  Note that before \cite{Farine:1980}, for the sake of result stability, Pearson {\em et al.}\ have found it beneficial to approximate the integral on the r.h.s.~of~\eqref{eq:aaSDM}, with a displacement of densities.  While there is a significant improvement in the agreement of our results with those of Pearson {\em et al.}\ following the reassessment \cite{samyn:044316,samyn:044316,goriely:031301,goriely:064312}, disagreements between the two sets still can reach the order of~20\%, cf.~Table~\ref{tab:comparison}.  Reinhard {\em et al.}~\cite{reinhard:014309} have employed a strategy close to that in Eq.~\eqref{eq:ESSeta} and disagreements between their and our $a_a^S$-coefficient values reach $\sim 25 \%$.

Our own tests of the stability of our results for $a_a^S$ from Eq.~\eqref{eq:aaSrho}, with respect to different technical aspects of calculations, indicate that errors of those coefficient values do not exceed~0.5\%.  The aspects of calculations that have been varied include the already mentioned mesh in wavevector space, step of integration in space, and the profile function in space.  In the context of the investigations in \cite{samyn:044316}, it may be still worthwhile to compare the results on~$a_a^S$ obtained with different methods.  We have mentioned that errors for a~method involving dividing out two powers of $\eta$, can exceed errors for a method involving dividing out just one power, by (1--2) orders of magnitude, for the small values of $\eta$ we employ.  Partly, this can be compensated by employing a higher $\eta$ in extracting $a_a^S$ with the inferior method: when one is resigned to a larger error, effects of any anharmonicity of the symmetry energy in $\eta$ matter less.  Overall, though, when the quality of calculations increases, errors decrease, whether for the more favored or the more inferior method of $a_a^S$ extraction.  Thus, differences between the results obtained with different methods can be exploited in general assessment of calculations.

As to the intrinsic comparisons,
Fig.~\ref{fig:aset}, for one, can be used for testing visually the consistency between Eqs.~\eqref{eq:ESSeta} and~\eqref{eq:aaSrho} in our calculations, for the several interactions in that figure.  Otherwise, Table~\ref{tab:aasEqs} shows the values of~$a_a^S$ extracted with three different methods from our half-infinite matter calculations, when employing Eq.~\eqref{eq:aaSrho}, Eq.~\eqref{eq:ESSeta} and also~Eq.~\eqref{eq:es=}.
The last two methods both involve dividing out factors of the second order in asymmetry and, thus, are both expected to provide inferior results to the first method that is our standard.  Our calculations for the table have been done for those Skyrme parameterizations for which Pearson {\em et al.}\ have also done calculations with different methods.  Their results are also shown in Table~\ref{tab:aasEqs}, for comparison.  The largest deviation between the preferred and an inferior method in Table~\ref{tab:aasEqs} is 7\% for our calculations and 34\% for Pearson {\em et al.}

\begin{table}
\caption{Values of surface symmetry coefficient, extracted following different indicated equations from the SHFM calculations of half-infinite nuclear matter, done for the different indicated Skyrme parameterizations.  The results that were ultimately decided to be superior by the respective authors are represented in the roman font and other results are represented in italic.\\[-.2ex]}
\begin{minipage}{\textwidth}
\begin{center}
\begin{tabular}{|l|c c c | c c  c|}
\hline
       & \multicolumn{5}{c}{$a_a^S \, \text{[MeV]}$} & \\
       \cline{2-7}
Name   & \multicolumn{3}{c|}{Our} & \multicolumn{3}{c|}{Pearson {\em et al.}}  \\
        & Eq.~\eqref{eq:es=} & Eq.~\eqref{eq:aaSrho} & Eq.~\eqref{eq:ESSeta}
       & Eq.~\eqref{eq:aaSDM}\footnote{An additional approximation is employed \cite{Farine:1980}.} & Eq.~\eqref{eq:ESSeta} & Ref.\
      \\
\hline
SII    & {\em 16.63} & 17.54 & {\em 18.26} & 15.1 & {\em 17.8} &  \cite{PhysRevC.24.303} \\
\hline
SIII   & {\em 20.83} & 21.77 & {\em 22.75} & 19.1 & {\em 23.1} & \cite{PhysRevC.24.303} \\
\hline
SIV    & {\em 12.88} & 13.39 & {\em 14.27} & 12.9 & {\em 15.1} & \cite{PhysRevC.24.303} \\
\hline
SV     & {\em 10.40} & 10.27 & {\em 10.80} & 8.9 & {\em 11.1} & \cite{PhysRevC.24.303} \\
\hline
SVI    & {\em 26.13} & 26.27 & {\em 26.66} & 32.4 & {\em 26.7} & \cite{PhysRevC.24.303} \\
\hline
BSk8   & {\em 22.80}  & 23.13 & {\em 23.32}& 20.0 & {\em 13.3} & \cite{samyn:044316} \\
\hline
\end{tabular}
\end{center}
\end{minipage}
\label{tab:aasEqs}
\end{table}

\subsection{Further Discussion of the Isovector Density $\rho_a^0$ and of the Coefficient $a_a^S$}

As we have shown, in the classically allowed region of the surface, $\rho \gtrsim \rho_0/4$, the isovector density $\rho_a$ follows closely the expectation of Eq.~\eqref{eq:rhoaS} from uniform matter, with $\rho_a$ inversely proportional to the symmetry energy $S(\rho)$ at a given net density~$\rho$.   The form of the adherence to the expectation, and the partial consequences of that adherence, are the following for the specific limits of~$S(\rho)$.
When the symmetry energy remains significant at subnormal densities, for low positive or even negative values of the dimensionless parameter-ratio $L/a_a^V$, the~isovector density $\rho_a$ stays close to the isoscalar density $\rho$ over a significant portion of the classically allowed region of nuclear surface, as expected from Eq.~\eqref{eq:rhoaS}.  Outside of the classically allowed region, both densities drop rapidly to zero, in an approximately exponential fashion.  On the other hand, when, for high $L/a_a^V$-ratio, the symmetry energy drops rapidly with density at subnormal densities, the isovector density $\rho_a$ remains significantly higher than $\rho$, across the classically allowed region, cf.\ Fig.~\ref{fig:denscomp}.  For the exponential fall-off of the densities in the classically forbidden region, the starting value for $\rho_a$ is then further significantly higher than for $\rho$, extending the region where $\rho_a$ dominates over $\rho$, cf.~Fig.~\ref{fig:denslog}.  Note that, while $L/a_a^V$ ratio is suitable for characterizing the {\em shape} of~$S(\rho)$, due to small variations of $a_a^V$ relative to variations of $L$ between different interactions, the relative magnitudes of $L$ and the relative magnitudes of $L/a_a^V$ are usually interchangeable when comparing different interactions.

In Table \ref{tab:skypro} different parameters can be found, quantifying characteristics of the densities $\rho_0$ and~$\rho_a^0$, including respective diffuseness parameters, $d_0$ and $d_a^0$.  The isoscalar diffuseness $d_0$ is of obvious importance as quantifying the pace, independent of the symmetry energy, at~which the isoscalar density approaches zero.  We define each diffuseness parameter in terms of the derivative of respective density at half of asymptotic value:
\beq
\frac{\rho_0}{4 d_0} = \left. \frac{\text{d}\rho_0}{\text{d} z} \right|_{\rho_0(z) = \rho_0/2} \, \hspace*{1em} \text{and} \hspace*{1em} \frac{\rho_0}{4 d_a^0} = \left. \frac{\text{d}\rho_a^0}{\text{d} z} \right|_{\rho_a^0(z) = \rho_0/2} \, .
\eeq
We elect such a definition over the more common relation of diffuseness to the surface thickness as distance over which the density drops from 90\% to 10\% of asymptotic value.  This is because of our intention to have consistent definitions for $\rho$ and $\rho_a$ and because of the potentially complicated features of~$\rho_a$ around normal density.  The latter features are due to the relation of $\rho_a$ to symmetry energy and due to amplified Friedel oscillations in~$\rho_a$, as compared to~$\rho$, cf.~Fig.~\ref{fig:denscomp}.  For the isoscalar density, the definitions in terms of derivative and surface thickness yield nearly identical results.

\begin{figure}
\centerline{\includegraphics[width=.48\linewidth]{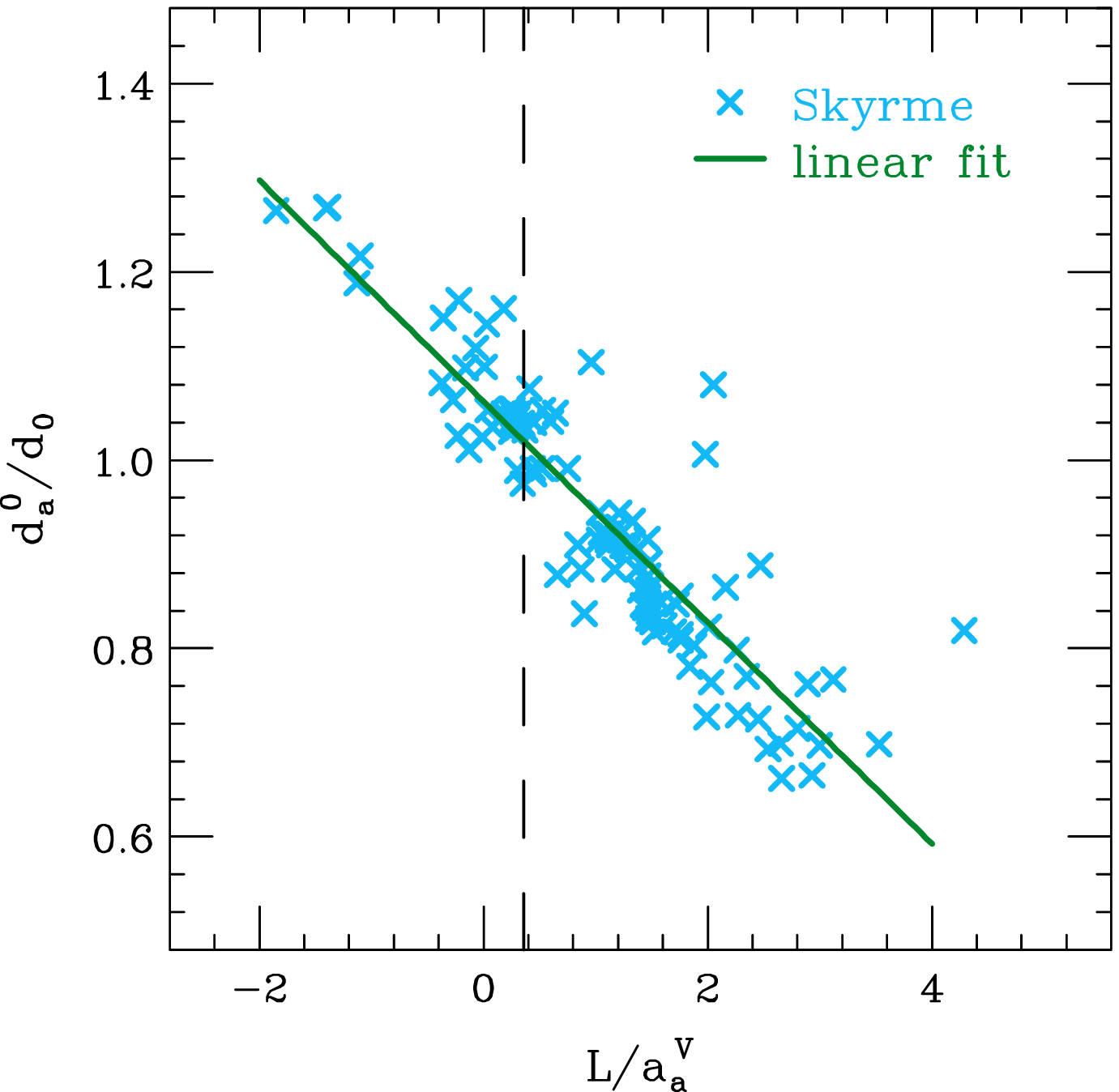} \hspace*{1em}
\includegraphics[width=.48\linewidth]{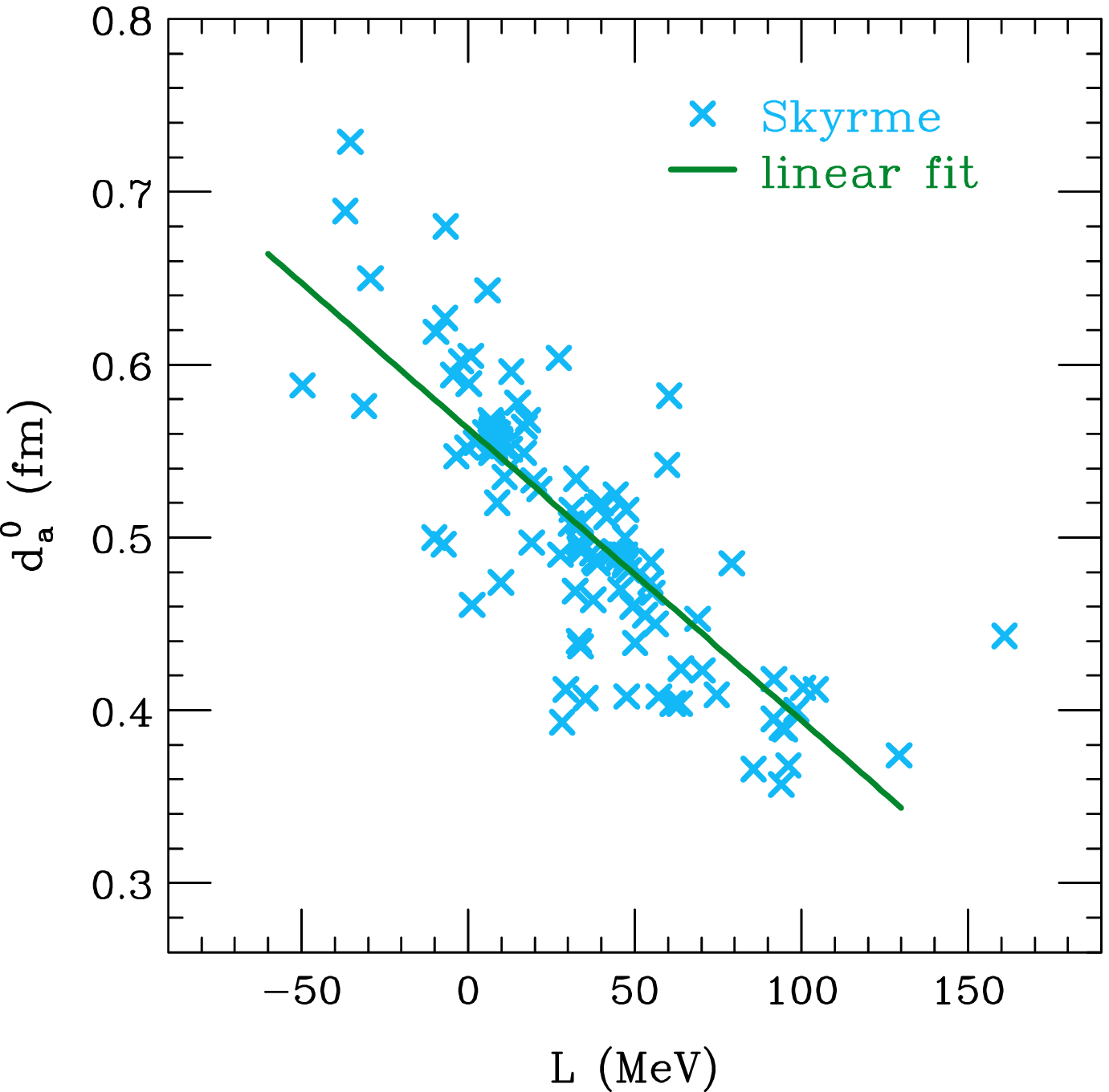}}
\caption{Correlation between the diffuseness $d_a^0$ of isovector density in symmetric half-infinite matter and the slope parameter $L$ of symmetry energy in uniform matter.  In the first panel, the~isovector diffuseness $d_a^0$ and the slope parameter $L$ are scaled, respectively, with the diffuseness $d_0$ of isoscalar density and with the value $a_a^V$ of symmetry energy at $\rho_0$.  No scalings are applied in the second panel.  Symbols in the panels represent results for Skyrme interactions in Table \ref{tab:skypro}.  Solid lines represent linear fits to the results for Skyrme interactions.  The dashed vertical line in the first panel separates coarsely those interactions, to the left of the line, for which the $\rho_0/2$-point in isovector density appears in the classically allowed region, from those interactions, to the right of the line, for which the $\rho_0/2$-point appears in the classically forbidden region.
}
\label{fig:ddla}
\end{figure}

The isoscalar diffuseness parameters tend to be fairly consistent between different Skyrme interactions, averaging at 0.541~fm in Table~\ref{tab:skypro}, with an rms deviation of 0.031~fm.  When a~larger deviation from the average is found for an interaction in Table~\ref{tab:skypro}, a correlated change, compared to other interactions, is typically found to occur in other distance parameters for the surface.  For the lowest $L$-values, for which the $\rho_0/2$-position of the isovector density lies within the classically allowed region, the~isovector diffuseness needs obviously to depend on the shape of the dependence of symmetry energy on density, cf.~Eq.~\eqref{eq:rhoaS}.  At any fixed shape for the symmetry energy, such as characterized by the ratio $L/a_a^V$, though, the isovector diffuseness should further scale in proportion to the isoscalar diffuseness that sets the scale for variation of isoscalar density in space.  Figure \ref{fig:ddla} shows the correlation between, on one hand, the isovector diffuseness from the SHFM calculations, scaled with the isoscalar diffuseness, and, on the other, the slope parameter $L$ scaled with~$a_a^V$.
A quite tight correlations is observed between the scaled parameters, with a linear fit to the results yielding
\beq
d_a^0 \simeq d_0 \, (1.062 - 0.117 \, L/a_a^V) \, .
\eeq
Interestingly, the tight correlation extends well into the region of the isovector $\rho_0/2$-point lying beyond the $\rho_0/4$-point in isoscalar density, representing the start of a classically forbidden region.  Figure \ref{fig:ddla} shows also the correlation between the isovector diffuseness $d_a^0$ and slope~$L$, when the parameter scaling is removed.  The lack of scaling results in a deterioration of the correlation; analogous deteriorations may be found for other tight correlations in the paper, when scalings implied by the physics are removed.

\begin{figure}
\centerline{\includegraphics[width=.55\linewidth]{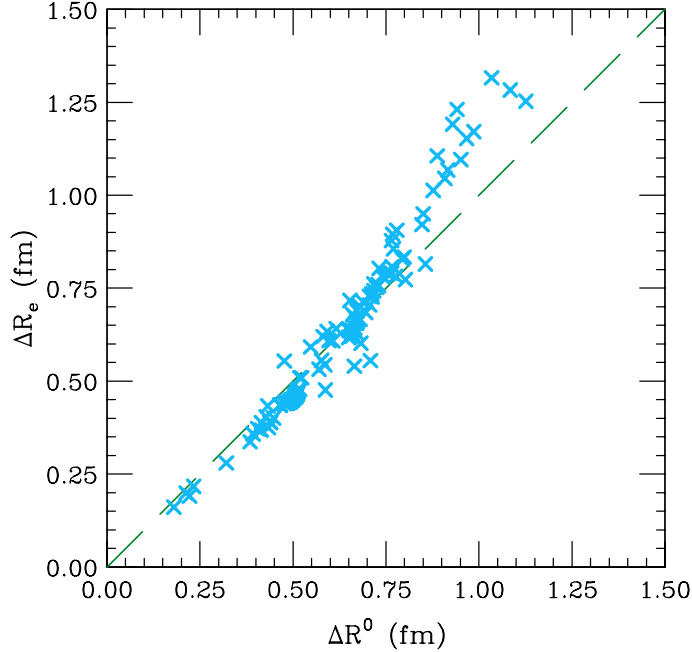}}
\caption{Correlation between, on one hand, the relative displacement $\Delta R^0$ of isovector and isoscalar densities in half-infinite symmetric matter, from the $\rho_0/2$-points, and, on the other hand, the effective displacement $\Delta_e R = (a_a^V/a_a^S)  \, (r_0/3)$.  Symbols represent the variety of Skyrme interactions from Table~\ref{tab:skypro}.  The diagonal dashed line serves to guide the eye.
}
\label{fig:drdr}
\end{figure}

More significant than the difference in diffusenesses, between the isovector and isoscalar densities, is the overall displacement of those densities relative to each other.  This displacement can be quantified in different ways.  One possible quantification is in terms of the displacement $\Delta R^0$ of the $\rho_0/2$-density points for isovector and isoscalar densities, at $\eta = 0$, see Table~\ref{tab:skypro}.  Another possible quantification is in terms of the ratio of symmetry coefficients~$a_a^V/a_a^S$, proportional to the integral over the difference of those densities at $\eta=0$, in~the direction perpendicular to nuclear surface, cf.~Eq.~\eqref{eq:aaSrho}. To the extent that the densities could be both described in terms of Fermi functions or, otherwise, were simply translations one of another, the ratio of coefficients could, following~\eqref{eq:aaSrho}, be approximated with
\beq
\label{eq:aaVSDR}
\frac{a_a^V}{a_a^S} \simeq \frac{3 \, \Delta R^0}{r_0} \, .
\eeq
Following \eqref{eq:aaVSDR}, we define an effective displacement in terms of the coefficient ratio as
\beq
\label{eq:DeR}
\Delta_e R = \frac{r_0}{3} \ \frac{a_a^V}{a_a^S} \, ,
\eeq
which is provided in Table~\ref{tab:skypro} in addition to~$\Delta R^0$.  Surface displacement $\Delta R^0$ is found to range in the table from 0.21 to 1.13~fm.  Figure \ref{fig:drdr} shows the correlation between the two displacements, for the Skyrme interactions.
Over much of the range of variation, the~displacements are rather close to each other.  Only for the highest displacement values, $\Delta_e R$ tends to prevail over $\Delta R^0$.

\begin{figure}
\centerline{\includegraphics[width=.57\linewidth]{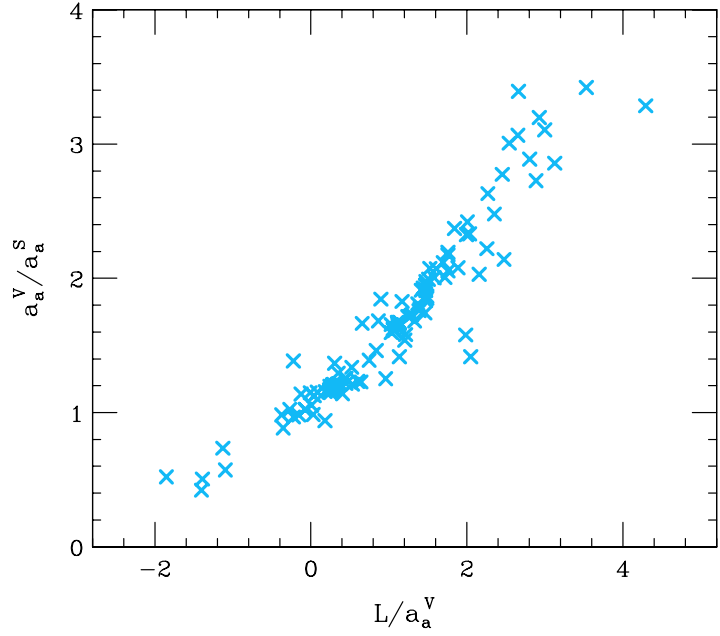}}
\caption{Ratio of symmetry energy coefficients $a_a^V/a_a^S$, for different Skyrme interactions, plotted vs the slope parameter $L$ of the symmetry energy, scaled with the value of symmetry energy $a_a^V$.
}
\label{fig:lavas}
\end{figure}

\begin{figure}
\centerline{\includegraphics[width=.57\linewidth]{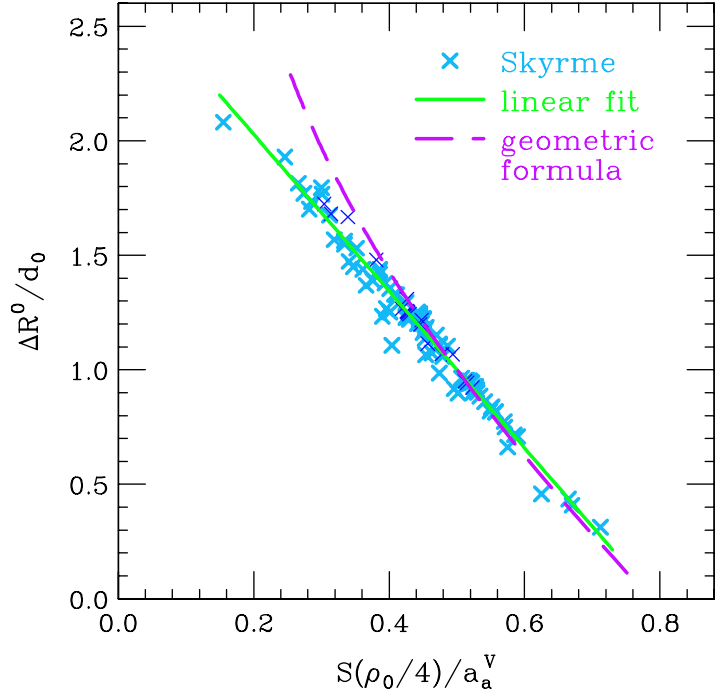}}
\caption{Displacement of isovector- relative to isoscalar-density in symmetric half-infinite matter, plotted vs value of symmetry energy at $\rho_0/4$.  The displacement and the energy value at $\rho_0/4$ are scaled, respectively, with the isoscalar diffuseness and the symmetry-energy value at $\rho_0$.  Symbols represent results for Skyrme interactions from Table~\ref{tab:skypro}. Solid line represents a linear fit to the Skyrme results.  Dashed line represents predictions of formula \eqref{eq:drdsge} based on simplified geometric considerations.
}
\label{fig:drds}
\end{figure}

\begin{figure}
\centerline{\includegraphics[width=.63\linewidth]{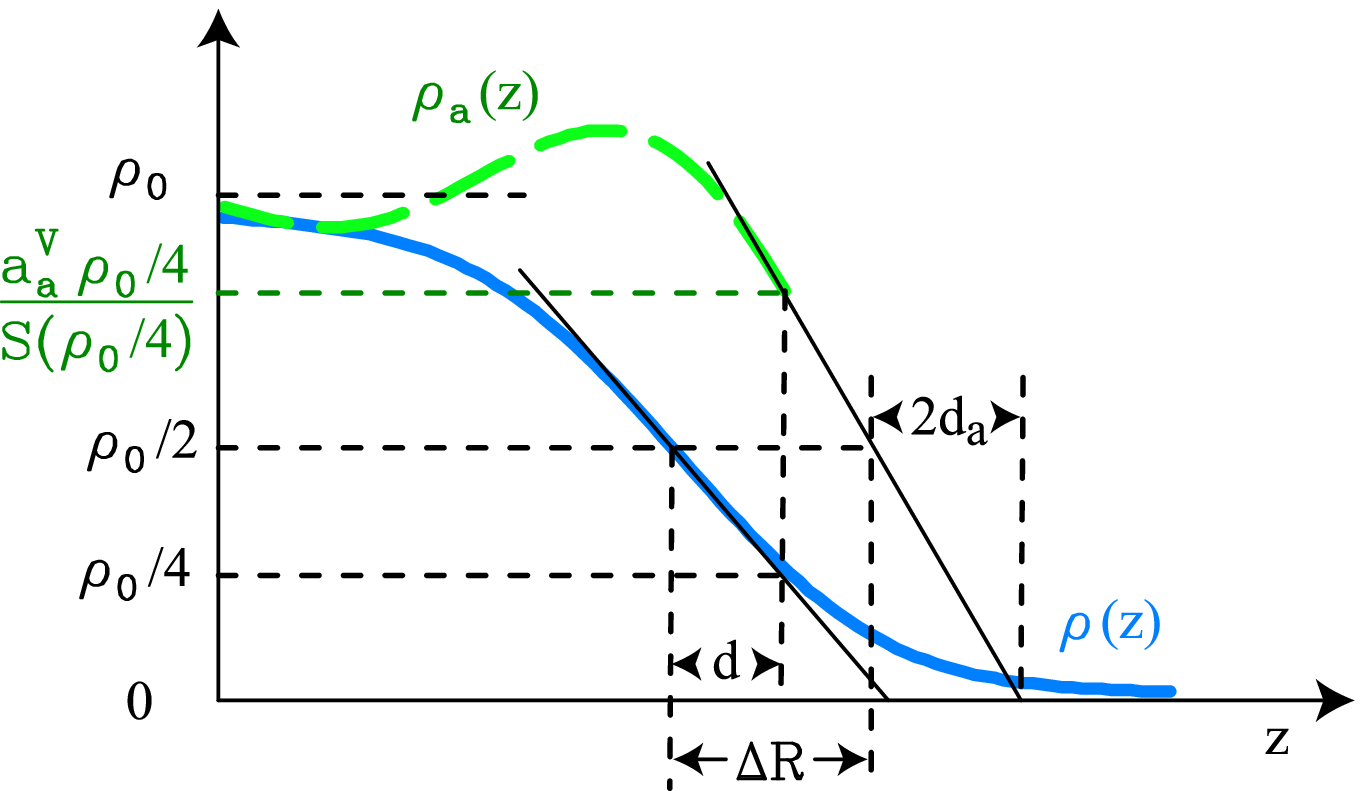}}
\caption{Geometric construction behind Eq.~\eqref{eq:drdsge}.  In the surface region, the dependence of isoscalar density and of the isovector density on position, beyond the breakdown of local approximation for $\rho_a^0$, is approximated in a linear form, with the slope expressed in terms of the respective diffuseness.
}
\label{fig:sketch}
\end{figure}

Judging from the local approximation, and from Fig.~\ref{fig:denscomp}, the most significant differences between the isovector and isoscalar densities should be associated with the highest $L$-values.  Indeed, high $L$-values imply a rapid drop of symmetry energy at subnormal densities and, correspondingly, high isovector compared to isoscalar density values.  Correlation between, on one hand, the symmetry parameter ratio $a_a^V/a_a^S$, proportional to the integral over density difference, cf.~Eq.~\eqref{eq:aaSrho}, and, on the other hand, the scaled slope parameter $L/a_a^V$, is shown in~Fig.~\ref{fig:lavas}.
The lowest values of the symmetry parameter ratio, of the order of~0.4, are indeed found for the lowest values of scaled $L$, and the highest values of the ratio, of the order of~3.4, are found for the highest values of~$L$.  In more detail, the isovector density is expected to stay elevated, following the symmetry energy in the local approximation of Eq.~\eqref{eq:rhoaS}, down to the isoscalar classical-return density of~$\sim \rho_0/4$.  With this, a correlation can be expected between the displacement $\Delta R^0$ of the isovector from isoscalar surfaces and the value of symmetry energy at~$\rho_0/4$.  Figure \ref{fig:drds} shows the correlation between $\Delta R^0$ scaled with isoscalar diffuseness and the symmetry energy at $\rho_0/4$, scaled with~$a_a^V$.
The correlation is the tightest from among those explored here for surface properties.  Linear fit to the correlation, represented by a solid line in Fig.~\ref{fig:drds}, produces
\beq
\label{eq:drds}
\Delta R^0 \simeq d_0 \, (2.71 - 3.42 \, S(\rho_0/4)/a_a^V) \, .
\eeq
The dashed line in Fig.~\ref{fig:drds} represents prediction based on the simplified geometric consideration illustrated in Fig.~\ref{fig:sketch}, where the densities are taken to vary linearly with position.
That consideration yields
\beq
\label{eq:drdsge}
\Delta R^0 \simeq d_0 + d_a^0 \, \left(a_a^V/S(\rho_0/4) - 2 \right) \, ,
\eeq
and, for Fig.~\ref{fig:drds}, we use $d_a^0 \simeq d_0 \, (0.336 - 1.336 \, S(\rho_0/4)/a_a^V)$ from fitting the correlation of $d_a^0$ with symmetry energy.  While the formula from the simple consideration begins to overestimate the shift for the symmetry energies with a strongest density-dependence, it is apparent that the simple consideration grasps the essence of impact of the low-density symmetry-energy on the shift.

\begin{figure}
\centerline{\includegraphics[width=.63\linewidth]{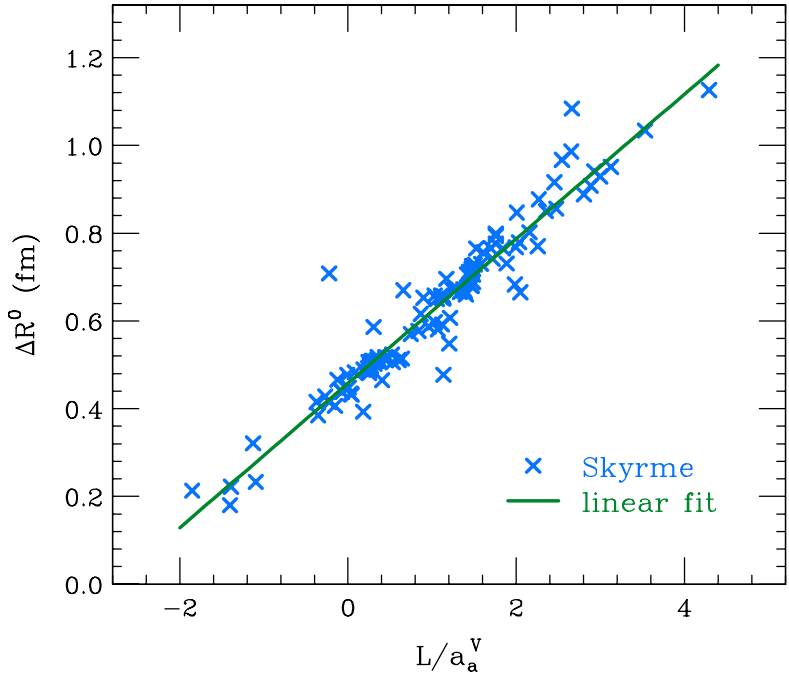}}
\caption{Displacement of isovector- relative to isoscalar-surface in semi-infinite nuclear-matter, plotted vs normalized slope-parameter of symmetry energy.  Symbols represent results for the Skyrme interactions of Table~\ref{tab:skypro}.  The line represents a linear fit to the Skyrme results.
}
\label{fig:lavdr}
\end{figure}

Rounding up the set of correlations between surface characteristics and features of symmetry energy in uniform matter, Fig.~\ref{fig:lavdr} shows the displacement of the isovector- relative to isoscalar-surface for the Skyrme interactions, as a function of scaled slope parameter.
Quality of this correlation does not change significantly when the shift is scaled with diffuseness, so we present the correlation without such a scaling, as possibly simpler to use.
With the correspondence between the displacement $\Delta R^0$ and the effective displacement $\Delta_e R$ proportional to $a_a^V/a_a^S$, cf.~Fig.~\ref{fig:drdr}, Fig.~\ref{fig:lavdr} is a counterpart of Fig.~\ref{fig:lavas}.  Linear fit to the Skyrme results, which is indicated in Fig.~\ref{fig:lavdr}, produces
\beq
\Delta R^0 \simeq \left(0.458 + 0.165 \, L/a_a^V \right) \, \text{fm} \, .
\eeq

\subsection{Some Context}

In discussing the isovector and isoscalar densities in the preceeding subsection, we have concentrated on the shift of these densities relative to each other and on their diffusenesses.  To the extent to which $\rho_a^0$ and $\rho_0$ might be approximated in terms of Fermi functions, these characteristics could suffice for describing the average sizes of asymmetry skins in nuclei.  The skin sizes have been, so far, the prime aspect of nuclear density distributions linked in the literature to the density dependence of symmetry energy.

Specifically, in the context of asymmetry skins, let us consider neutron and proton rms radii for a nucleus of mass $A$.  From \eqref{eq:rhonp}, we find for the mean squared proton radius:
\beq
\label{eq:r2n}
\langle r^2 \rangle_p = \langle r^2 \rangle + \frac{Z - N} {2Z} \left( \langle r^2 \rangle_a - \langle r^2 \rangle \right) \, ,
\eeq
where
\beq
\langle r^2 \rangle = \frac{1}{A} \int \text{d}V \, r^2 \, \rho( {\bf r}) 
 \, ,
\eeq
and
\beq
\langle r^2 \rangle_a = \frac{\int \text{d}V \, r^2 \, \rho_a( {\bf r})}
{ \int \text{d}V \, \rho_a( {\bf r})} \equiv \frac{1}{N-Z} \int \text{d}V \, r^2 \, \rho_{np}( {\bf r}) \, .  
\eeq
With the $N \leftrightarrow Z$ interchange on the r.h.s.\ of \eqref{eq:r2n} yielding the neutron instead of the proton mean squared-radius, the difference between nucleon squared-radii may be represented as
\beq
\langle r^2 \rangle_n - \langle r^2 \rangle_p = \frac{A \, (N-Z)}{2 N Z} \left( \langle r^2 \rangle_a - \langle r^2 \rangle \right) \, .
\eeq
From the above, we finally get for the size of asymmetry skin
\beq
\label{eq:rmsnp}
\langle r^2 \rangle_n^{1/2} - \langle r^2 \rangle_p^{1/2} \simeq \frac{A \, (N-Z)}{4 N Z} \,
\frac{\langle r^2 \rangle_a - \langle r^2 \rangle}{ \langle r^2 \rangle^{1/2} } \, .
\eeq
It turns than out that the size of asymmetry skin directly probes the difference in mean squared radii for the isovector and isoscalar densities.  When a density can be approximated in the Fermi form, the mean squared radius can be expressed in terms of the $\rho_0/2$-position combined with diffuseness.  Looking at Figs.~\ref{fig:density} and~\ref{fig:densvec}, it is apparent that isoscalar densities are more likely to be well described in terms of a Fermi shape than isovector, particularly for the most extreme symmetry energies, characterized by either very low or very high slope-parameters~$L$.

Outside of the scope of this paper, analyses of data point to a symmetry energy in uniform matter that is characterized by a slope in the upper range of the values possible for Skyrme interactions, especially for the slope scaled with $a_a^V$.  Thus, e.g.\ the analysis of excitation energies of isobaric analog states \cite{Danielewicz:2004et,Danielewicz:2007pf} yields independent values of $a_a^V$ and~$a_a^S$.  While the volume symmetry coefficient from this type of analysis, $a_a^V \simeq \text{(31.5--33.5)} \, \text{MeV}$, comes out quite in the middle of values found for the Skyrme interactions, the surface symmetry coefficient, $a_a^S \simeq \text{(9.5--12)} \, \text{MeV}$, comes out right at the lower end of values encountered for the Skyrme interactions.  The coefficient ratio from that analysis is in the range $a_a^V/a_a^S \simeq \text{(2.8--3.3)}$.  That ratio produces the effective surface displacement in the range of $\Delta_e R = (r_0/3) \, (a_a^V/a_a^S) \simeq  \text{(1.06--1.26)} \, \text{fm}$.  Moreover, Figs.~\ref{fig:drdr} and~\ref{fig:lavas} yield the respective ranges of $\Delta R^0 \simeq \text{(0.85--1.05)} \, \text{fm}$ and $L/a_a^V \simeq \text{(2.4--3.4)}$ or $L \simeq \text{(78--111)} \, \text{MeV}$.  The analysis \cite{Danielewicz:2004et,Danielewicz:2007pf} is relatively model-independent, provided curvature effects play little role for heavier nuclei.  If~the latter were not the case, though, a bit softer symmetry energy would need to be deduced.

The large displacement for the isovector surface, inferred above, would give rise to large predicted asymmetry skins, cf.~Eq.~\eqref{eq:rmsnp}, towards the top of those that appear possible within the Skyrme-Hartree-Fock descriptions \cite{Danielewicz:2004et,PhysRevC.64.027302}.  Notably, still higher sizes of asymmetry skins can be encountered in the relativistic mean-field descriptions of nuclear systems, along with higher values of the $L$-parameter than encountered for the Skyrme parameterizations \cite{PhysRevC.64.027302}. Aiming at constraints beyond those in~\cite{Danielewicz:2004et,Danielewicz:2007pf}, on the symmetry energy in uniform matter, we intend to develop strategies for the constraints, further in this series, exploiting available experimental information on nucleonic densities, along the lines of developments of the present paper.  Otherwise, the isovector density, as a fundamental quantity, deserves attention on its own when analyzing data on nucleonic densities.

\section{Conclusions}
\label{sec:conc}

In this paper, we have considered nuclear energy in the macroscopic limit for a nucleus.  Further, we have considered the Hohenberg-Kohn functional for a nuclear system, in terms of proton and neutron densities.  Finally, we have carried out Skyrme-Hartree-Fock calculations of half-infinite particle-stable nuclear-matter.  In each case, we have concentrated on the role of neutron-proton asymmetry in the system and on the symmetry energy.

In discussing nuclear energy in the macroscopic limit, we have shown that surface symmetry-energy emerges, as an unavoidable ingredient of the net nuclear energy, from simultaneous considerations of nuclear surface and symmetry energies.  A consequence of this conclusion is that the net nuclear surface and volume energies combine in the similar manner as energies of connected capacitors in electrostatics.  The role of charge flowing and distributing itself between the capacitors, is taken over by the asymmetry that distributes itself between the nuclear interior and surface, in proportion to volume and surfaces capacitances.  The capacitances for the interior and surface are proportional, respectively, to the volume and surface area and are inversely proportional to the volume and surface symmetry coefficients.

When considering a continuous limit of the Hohenberg-Kohn functional, we have broken up the nuclear part of the functional into the functional for symmetric matter and a~symmetry term that is bilinear, in the lowest order, in the difference of neutron and proton densities.  The~kernel ${\mathcal S}$ in the symmetry term is generally nonlocal and depends on the net density.  In~the limit of weak nonuniformities, the kernel may be approximated in a local form as a~$\delta$-function multiplying symmetry energy in uniform matter, divided by local density.  We~have shown that, up to Coulomb corrections and terms of second order in asymmetry, the~net nucleonic density and shape of the neutron-proton density difference are invariant, in~the continuous limit, across an isobaric chain.  The neutron-proton density difference, in~particular, is~expressible in terms of the inverse operator ${\mathcal S}$ for the symmetric matter.  In~this context, we have introduced an isovector density, as a scaled neutron-proton density difference, and a~counterpart to the~net density, also termed isoscalar density.  The neutron and proton densities can be expressed as a combination of those nearly invariant isoscalar and isovector densities.  When the local approximation for ${\mathcal S}$ holds, the isovector density is proportional to the isoscalar density divided by the value of symmetry energy for uniform matter.  A generalized symmetry coefficient can be introduced for a nuclear system and can be expressed in terms of the inverse operator ${\mathcal S}$. The coefficient turns out, further, to be inversely proportional to the volume integral of the isovector density.  Contributions to the net capacitance of the system, for asymmetry, can be expressed in terms of the volume integral of the difference of isovector and isoscalar densities.

We have carried out Skyrme-Hartree-Fock calculations of symmetric and asymmetric semi-infinite nuclear matter, with a significantly higher accuracy than in the past.  The~accuracy is important when trying to extract subtle symmetry effects associated with the nuclear surface.  We have calculated symmetric surface and surface-symmetry characteristics for nearly all Skyrme parameterizations that have been proposed in the literature.  In~the calculations, we have verified the near-invariance of the isoscalar and isovector densities, first inferred within the considerations relying on the Hohenberg-Kohn functional.  We~have found that, up to the Friedel oscillations, the isovector density follows the expectation from the local approximation, down to about a quarter of normal density.  Within the WKBJ approximation, we have shown that the isovector density is expected to follow the local approximation within the classically allowed region for single-particle wavefunctions.  In the far-out forbidden region, on the other hand, the density fall-off is governed by separation energy.  We have extracted displacements of isovector- relative to isoscalar-density as well diffuseness values for the two densities.  By integrating differences of the two densities for symmetric matter, we have determined values of the surface symmetry coefficient for different Skyrme parameterizations.  Consistently with qualitative expectations, we found the displacements and volume-to-surface symmetry-coefficient ratios to be strongly correlated with the density dependence of symmetry energy in uniform matter.  The faster the symmetry energy drops at subnormal densities, the larger the relative displacement of the two densities and the larger the coefficient ratio.

One exciting possibility, emerging as a consequence of our investigations, is that, due to the invariance of two fundamental densities, isoscalar and isovector, features of the symmetry energy could be investigated by studying systematics of proton distributions alone, without reference to neutron distributions.  Such an investigation requires, however, a careful separation and/or circumvention of shell, pairing, Coulomb and deformation effects on the proton distribution for finite nuclei.  The obvious difficulty of such separation may be compensated by the rather extensive knowledge of proton densities.

\acknowledgements
The authors are grateful to J.\ Rikovska Stone for supplying them with values of force constants for the majority of Skyrme interactions employed in this work.  They further thank H.~S.~Kohler, W.~Nazarewicz and J.\ M.\ Pearson for comments related to the paper.  This work was supported by the National Science Foundation under Grants PHY-0551164, PHY-0555893, PHY-0606007 and PHY-0800026.

\newpage

\bibliography{ss07,skyrme}

\bibliographystyle{my}  

\end{document}